\documentclass[aps,prd,preprint,superscriptaddress]{revtex4}

\usepackage{graphicx}
\usepackage{amsmath,amssymb}
\usepackage{bm}
\usepackage[caption=false]{subfig}
\usepackage{color}
\usepackage{setspace}
\usepackage{soul}
\usepackage{times}

\def\p{\partial}

\def\=:{=\hspace{-.7em}\raisebox{1.1ex}{.}\hspace{.1em}\raisebox{-0.2ex}{.} }

\newcommand {\beq}{\begin{eqnarray}}
\newcommand {\eeq}{\end{eqnarray}}
\newcommand {\non}{\nonumber\\}

\newcommand {\1}[1]{\frac{1}{#1}}

\newcommand {\tr}{{\rm tr}\,}

\begin{document}

\preprint{NORDITA-2014-124}

\title{Baryonic torii: Toroidal baryons in a generalized Skyrme model} 


\author{Sven Bjarke Gudnason}
\affiliation{
Nordita, KTH Royal Institute of Technology and Stockholm University,
Roslagstullsbacken 23, SE-106 91 Stockholm, Sweden
}
\author{Muneto Nitta}
\affiliation{Department of Physics, and Research and Education Center for Natural 
Sciences, Keio University, Hiyoshi 4-1-1, Yokohama, Kanagawa 223-8521, Japan
}



\date{\today}
\begin{abstract}
We study a Skyrme-type model with a potential term motivated by 
Bose-Einstein condensates (BECs), which we call the BEC Skyrme
model. 
We consider two flavors of the model, the first is the Skyrme model
and the second has a sixth-order derivative term instead of the Skyrme
term; both with the added BEC-motivated potential.
The model contains toroidally shaped Skyrmions and they are
characterized by two integers $P$ and $Q$, representing the winding
numbers of two complex scalar fields along the toroidal and poloidal
cycles of the torus, respectively.
The baryon number is $B=PQ$. 
We find stable Skyrmion solutions for $P=1,2,3,4,5$ with 
$Q=1$, while for $P=6$ and $Q=1$ it is only metastable.
We further find that configurations with higher $Q>1$ are all unstable 
and split into $Q$ configurations with $Q=1$. 
Finally we discover a phase transition, possibly of first order, in
the mass parameter of the potential under study. 
\end{abstract}


\maketitle

\section{Introduction}

Half a century has passed since Skyrme proposed \cite{Skyrme:1962vh}
that Skyrmions characterized by the topological charge
$\pi_3(S^3)\simeq\mathbb{Z}$ describe nucleons in the pion effective
field theory or the chiral Lagrangian \cite{Adkins:1983ya}, where the 
Skyrme term, i.e.~quartic in derivatives, is needed to stabilize
Skyrmions against shrinkage. 
Although nucleons are now known to be bound states of quarks, the idea
of the Skyrme model is still attractive. 
In fact, the Skyrme model is still valid as a low-energy description
of QCD, has only a small number of parameters and is, for instance,
used also in holographic QCD \cite{Sakai:2004cn,Hata:2007mb}. 

Meanwhile in condensed matter physics, considerable efforts have been
made recently to realize stable 3-dimensional Skyrmions in
two-component Bose-Einstein condensates (BECs)
\cite{Ruostekoski:2001fc,3D-skyrmions,Kawakami:2012zw,Nitta:2012hy} 
(see Ref.~\cite{Kasamatsu:2005} for a review of 
two-component BECs).  
In Ref.~\cite{Nitta:2012hy}, the creation of a Skyrmion is proposed to 
be a consequence of the annihilation of a brane and an anti-brane
\cite{Takeuchi:2012ee}. 
At strong coupling, these systems reduce to the SU(2) principal chiral 
model, but the existence of Skyrmions is elusive due to the lack of
the Skyrme term (or an even higher-order derivative term) \footnote{It
  has also been proposed that a stable 3-dimensional Skyrmion can
  exist as a ground state in the SU(2)-symmetric case, by introducing
  ``artificial'' non-Abelian gauge fields \cite{Kawakami:2012zw}.}. 
One interesting feature in these systems is that a potential term,
breaking the SU(2) symmetry is present, which deforms the (would-be)
Skyrmion to the shape of a torus \cite{Ruostekoski:2001fc}.  
Consequently, the Skyrmion can be interpreted 
\cite{Ruostekoski:2001fc,Nitta:2012hy,Metlitski:2003gj} 
as a vorton \cite{Davis:1988jq,Vilenkin:2000,Radu:2008pp,Garaud:2013iba}, 
that is, a vortex ring in the first component with the second
component flowing inside said ring.

In this paper, we consider a Skyrme-like model with a potential term
in the form $V = m^2|\phi_1|^2|\phi_2|^2$ which was introduced in our
previous papers \cite{Gudnason:2014hsa,Gudnason:2014gla} 
and is motivated by two-component BECs
\cite{Ruostekoski:2001fc,3D-skyrmions,Nitta:2012hy}, 
where we use a notation of two complex scalar fields $\phi_1(x)$ and
$\phi_2(x)$ with the constraint $|\phi_1|^2 + |\phi_2|^2=1$ along the 
lines of two-component BECs. 
For higher-derivative terms needed to stabilize Skyrmion, 
we consider either the conventional fourth-order derivative term,
i.e.~the Skyrme term or a sixth-order derivative term, which is 
the baryon charge density squared 
(see,
e.g.~Refs.~\cite{Adam:2010fg,Gudnason:2013qba,Gudnason:2014gla}); for
a short-term notation we will call the first case the 2+4 model and the
second case the 2+6 model. 
We construct stable Skyrmions which were elusive in two-component BECs 
in the absence of the Skyrme term or other higher-order derivative
terms, and find that they take the shape of a torus as two-component
BECs. 
We find that the most general solutions are characterized by two
integers $P$ and $Q$, representing the winding numbers of 
the scalar fields $\phi_1$ and $\phi_2$ along the toroidal and
poloidal cycles of the torus, respectively, and show that the baryon
number or the Skyrmion number of $\pi_3(S^3)\simeq\mathbb{Z}$ is
$B=PQ$ (which is also known as the linking number). 
We explicitly construct stable Skyrmion solutions with $P=1,2,3,4,5$
and $Q=1$, yielding the baryon numbers $B=1,2,3,4,5$. We also
construct the $P=6$, $Q=1$ solution and find that it is metastable,
i.e.~is energetically prone to decay into two $B=3$ objects. 
This turns out to be the case for both the 2+4 and the 2+6 model.  

The energy and baryon charge distributions of the configuration of
$P=1$ are spherically symmetric in the 2+4 model, whereas in the
2+6 model it is a deformed ball (with a hint of a torus-like shape).
The configurations with $P>1$ are all of toroidal shapes (for both
models) when the mass is bigger than a certain critical mass.
This is in contrast to the conventional Skyrmions (i.e.~without our
BEC-motivated potential) for which the configuration of $B=1$ is
spherically symmetric, that of $B=2$ is toroidal, and those of $B>1$
have energy distributions with some point symmetry. 
We compare our $B=2$ solutions in the 2+4 and 2+6 models to those of
the conventional model (i.e.~without the BEC-motivated potential), and
find that the energy distribution of the solution in the 2+6 model is
a surface of a torus while the energy distributions of the solutions 
in the 2+4 model and the conventional model are solid torii,
i.e.~filled torii. 

Although the classification of our solutions is given by the integers
$P$ and $Q$, we find that configurations with $Q>1$ are unstable, that
is, a configuration with $(P,Q)$ decays into $Q$ copies of the $(P,1)$
configuration.

We also note that our configurations can be identified as global
analogues of vortons \cite{Davis:1988jq,Vilenkin:2000,Radu:2008pp}, 
that is, twisted closed global vortex strings as in two-component
BECs \footnote{Strictly speaking, there is a supercurrent or a
  superflow due to the trapped field along the ring of the vorton. 
  This can be achieved by rotating the phase of the trapped field 
  linearly in time as $\phi_2 \sim e^{i z + i \alpha t}$ with $z$
  being the coordinate along the string.
  In the case of BECs, such a time dependence is automatic in the
  presence of the phase gradient along the string, because of the
  first derivative in time in the non-relativistic Lagrangian.}. 
While vortices in this model are global vortices so that straight
vortices have divergent energy per unit length, a closed string has
finite energy because of cancellation of vorticity. 
A vortex in the field $\phi_1$ traps the field $\phi_2$ in its core
and has the U(1) phase modulus of $\phi_2$.  
The integers $Q$ and $P$ are identified with the winding numbers of the
vortex of the $\phi_1$ field and of the $\phi_2$ field along the ring
inside the vortex core, respectively. 
The identification of the Skyrmions with global vortex rings also
explains why configurations with higher $Q>1$ are unstable. 
This is because $Q$ is the winding number of the vortex in the field
$\phi_1$, and a global vortex with higher winding is unstable to decay 
as two global vortices repel each other. 

Finally we discover a first-order phase transition between the
configuration (local minimum) where the Skyrmions have a (discrete)
point symmetry and the toroidal configuration (another local minimum)
at some critical mass, $m_{\rm critical}$.  
For concreteness we carry out this investigation at $B=3$ where the
Skyrmion has tetrahedral symmetry for $m<m_{\rm critical}$ and has
axial symmetry (i.e.~it is a torus) for $m>m_{\rm critical}$. For
$m<m_{\rm critical}$ the toroidal state is metastable and for
$m>m_{\rm critical}$ the tetrahedral state is metastable. For
sufficiently large $m\sim 2m_{\rm critical}$, the tetrahedral solution
becomes unstable and thus for large $m$ only the torus exists.

This paper is organized as follows. 
In Sec.~\ref{sec:model}, we present our model and explain the
symmetries and topological structures of the model. 
In Sec.~\ref{sec:wall-vortex}, we construct a domain wall and a global
vortex which serve as constituents of the torus. 
Finally, in Sec.~\ref{sec:toroidal-wall}, we construct toroidal
Skyrmions which are the strings wrapped up on a circle and we further
study their stability.
The phase transition between the Skyrmions with point symmetry and
with axial symmetry is studied in Sec.~\ref{sec:transition}.
Sec.~\ref{sec:summary} is devoted to a summary and discussions. 
In Appendix \ref{app:stringsplitting}, 
we show that solutions with $P=1,2$ and 
$Q=2$ are unstable to decay into 
two configurations of $P=1,2$ and $Q=1$.
In Appendix \ref{app:comparison}, we compare our $B=2$ solutions in
the 2+4 and 2+6 models and that in the conventional models
(i.e.~without the BEC-motivated potential).

\section{A Skyrme-like model with BEC-motivated
  potential \label{sec:model}} 

We consider the SU(2) principal chiral model with the addition of the
Skyrme term and a sixth-order derivative term in $d=3+1$ dimensions. 
In terms of the SU(2)-valued field $U(x)\in$ SU(2), the Lagrangian
which we are considering is given by 
\beq
\mathcal{L} = \frac{f_\pi^2}{16} 
\tr (\p_{\mu}U^{\dagger} \p^{\mu} U) 
+ \mathcal{L}_4
+ \mathcal{L}_6
- V(U),
\eeq
where we use the mostly-negative metric and the higher-derivative
terms are given by
\begin{align}
\mathcal{L}_4 &= \frac{\kappa}{32 e^2} 
\tr (\big[U^\dag \p_{\mu} U, U^\dag \p_{\nu} U\big]^2),\\ 
\mathcal{L}_6 &= \frac{c_6}{36 e^4 f_\pi^2} 
\left(\epsilon^{\mu\nu\rho\sigma}\tr\big[U^\dag\p_\nu U U^\dag \p_\rho
  U U^\dag \p_\sigma U\big]\right)^2.
\end{align}
The symmetry of the Lagrangian for $V=0$ is 
$\tilde G =$ SU(2)$_{\rm L} \times $SU(2)$_{\rm R}$ acting on $U$ as 
$U \to U'= g_{\rm L} U g_{\rm R}^\dag$.
The requirement of a finite-energy configuration, however,
spontaneously breaks this symmetry down to $\tilde H \simeq$
SU(2)$_{\rm L+R}$, which in turn acts as $U \to U'= g U g^\dag$ so
that the target space is 
$\tilde G/\tilde H \simeq$ SU(2)$_{\rm L-R}$.
The conventional potential term, i.e.~the pion mass term, is 
$V = m_{\pi}^2\tr (2{\bf 1}_2 - U - U^\dag)$,
which breaks the symmetry $\tilde G$ to SU(2)$_{\rm L+R}$
\emph{explicitly}. 

In this paper, it will prove convenient to use the following notation
where we express the field $U$ in terms of two complex scalar fields, 
$\phi^{\rm T} = (\phi_1(x),\phi_2(x))$, as
\beq
U = 
\begin{pmatrix}
\phi_1 & -\phi_2^*\\
\phi_2 & \phi_1^*
\end{pmatrix},
\eeq
subject to the constraint
\beq
\det U = |\phi_1|^2 + |\phi_2|^2 = 1.
\eeq
We further rescale the lengths to be in units of $2/(e f_\pi)$ and
energy to be in units of $f_\pi/(2e)$, for which we can write the
static Lagrangian density as
\begin{align}
-\mathcal{L} 
&= \1{2} \p_i\phi^\dag \p_i\phi
+ \frac{\kappa}{4} \left[(\p_i \phi^\dag \p_i\phi)^2  
  -\1{4}(\p_i\phi^\dag \p_j\phi + \p_j\phi^\dag \p_i\phi)^2\right] 
+ \frac{c_6}{4} \left(\epsilon^{i j k}
  \phi^\dag\p_i\phi\p_j\phi^\dag\p_k\phi\right)^2 \non
&\phantom{=\ }
+ V(\phi,\phi^*).
\end{align}
The full symmetry $\tilde G$ is not manifest in terms of $\phi$, 
where only SU(2)$_{\rm L}$ is manifest but SU(2)$_{\rm R}$ is not. 
The U(1) subgroup generated by $\sigma_3$ in SU(2)$_{\rm R}$, however, 
is manifest and acts on $\phi$ as $\phi \to e^{i\alpha} \phi$, 
constituting a U(2) group with SU(2)$_{\rm L}$.

The target space (the vacuum manifold with $m=0$) $M\simeq$ SU(2)
$\simeq S^3$ has a nontrivial homotopy group  
\beq
\pi_3(M) = \mathbb{Z}, 
\eeq
which admits Skyrmions as usual. 
The baryon number (the Skyrme charge) of $B \in \pi_3(S^3)$ is defined
as 
\beq
B &=& -\1{24\pi^2} \int d^3x \; \epsilon^{ijk} 
\tr \left( U^\dag\p_i U U^\dag\p_j U U^\dag\p_k U\right) \non
&=& \1{24\pi^2} \int d^3x \; \epsilon^{ijk} 
\tr \left( U^\dag\p_i U\p_j U^\dag\p_k U\right) \non
&=& \1{4\pi^2} \int d^3x \; \epsilon^{ijk} 
\phi^\dag \p_i\phi \p_j\phi^\dag \p_k \phi .
\eeq

Instead of the conventional potential term, we consider here a
potential term motivated by two-component Bose-Einstein condensates
(BECs), given by 
\beq
 V(\phi,\phi^*) 
= \frac{m^2}{8} \left[1 - (\phi^\dagger \sigma_3 \phi)^2 \right] 
= \frac{1}{2} m^2 |\phi_1|^2 |\phi_2|^2;
 \label{eq:potential}
\eeq
see the Appendix of Ref.~\cite{Gudnason:2014hsa} for a relation to
BECs. 
With this potential, the full symmetry $\tilde G$ is explicitly broken
down to 
\beq
G = \mathrm{U}(1) \times \mathrm{O}(2) \simeq 
  \mathrm{U}(1)_0 \times [\mathrm{U}(1)_3 \rtimes (\mathbb{Z}_2)_{1,2}].
\eeq
Here, each group is defined as
\beq
 \mathrm{U}(1)_0:&& \quad \phi \to e^{i \alpha} \phi,  \\
 \mathrm{U}(1)_3:&& \quad \phi \to e^{i \beta \sigma_3} \phi,   \\
 (\mathbb{Z}_2)_{1,2}:&& \quad e^{i (\pi/2) \sigma_{1,2}} \phi 
\eeq
where U(1)$_3$ acts on $\mathbb{Z}_2$ so that they are a semi-direct
product denoted by $\rtimes$.
The vacua of the potential in Eq.~\eqref{eq:potential} are 
\begin{align}
\begin{split}
\odot\ : \quad   \phi^{\rm T} = (e^{i\alpha},0), \\
\otimes\ : \quad \phi^{\rm T} = (0,e^{i\beta}), 
\end{split}
\label{eq:vacua}
\end{align}
and the unbroken symmetry $H$ is
\beq
 H_{\odot} = \mathrm{U}(1)_{0-3}:  && \quad 
 \phi \to e^{i \alpha}  e^{-i \alpha \sigma_3} \phi, \non  
 H_{\otimes} = \mathrm{U}(1)_{0+3}:  && \quad  
  \phi \to e^{i \alpha}  e^{+i \alpha \sigma_3} \phi,
\eeq
for the $\odot$ and the $\otimes$ vacuum of Eq.~(\ref{eq:vacua}),
respectively.  
Therefore, the vacuum manifold (or the moduli space of vacua) is given
by 
\beq
 \mathcal{M} \simeq G/H = 
 \frac{\mathrm{U}(1)_0 \times [\mathrm{U}(1)_3 \rtimes (\mathbb{Z}_2)_{1,2}]}{\mathrm{U}(1)_{0 \pm 3}}
 \simeq \mathrm{SO}(2)_{0\mp 3} \rtimes (\mathbb{Z}_2)_{1,2} = \mathrm{O}(2).
\eeq
The nontrivial homotopy groups of the vacuum manifold are
\beq
 \pi_0(\mathcal{M}) = \mathbb{Z}_2, \quad
 \pi_1(\mathcal{M}) = \mathbb{Z}, 
\eeq
admitting domain walls and vortices, respectively.

By means of the Hopf map $\vec{n} = \phi^\dagger \vec{\sigma} \phi$,  
the principal chiral SU(2) model can be mapped to 
the O(3) nonlinear sigma model with $\vec{n}^2 =1$ or equivalently 
the ${\mathbb C}P^1$ model.
The potential term in Eq.~(\ref{eq:potential}) is 
mapped to $V=\frac{m^2}{8} (1-n_3^2)$, 
which is referred to as the Ising-type potential in ferromagnets 
\cite{Kobayashi:2014xua}.
The ${\mathbb C}P^1$ model with the same potential is
often called the massive ${\mathbb C}P^1$ model 
\cite{Abraham:1992vb,Arai:2002xa,Nitta:2012xq,Nitta:2012kj}. 
This map can be obtained by 
coupling a U(1) gauge field to $\phi$ with 
common U(1) charges  
and subsequently taking the strong gauge coupling limit 
$e \to \infty$.

\section{Domain walls and vortices}\label{sec:wall-vortex}

In this section, we will review the constituents which will be used in
the next section in modified or compactified forms.

\subsection{Domain walls}
In $d=1+1$ dimensions, a (n anti-)kink solution interpolating between 
the two vacua in Eq.~(\ref{eq:vacua}) is given by 
\beq
\phi^{\rm T} = \1{\sqrt{1 + e^{\pm 2m(x-X)}}}
 (e^{i \alpha} , e^{\pm m(x-X) + i \beta }),
\label{eq:kink_sol} 
\eeq
with $X\in\mathbb{R}$ being the translational modulus of the kink. 
Here $\alpha$ and $\beta$ are not moduli of the kink but moduli of the
vacua in Eq.~(\ref{eq:vacua}). 
Note that this solution is (statically) exact in the form given above,
even in the presence of the Skyrme or sixth-order derivative term
(this can easily be understood as the Skyrme (sixth-order derivative)
term is nonzero only when a solution nontrivially depends on two
(three) spatial coordinates). 
Once waves on top of this static solution are considered, the
higher-order derivative terms must be taken into account; see
e.g.~Ref.~\cite{Kudryavtsev:1997nw}.

In the static case, the kink can trivially be extended
to a domain line in $d=2+1$ dimensions and to a domain wall in $d=3+1$ 
dimensions, with a one- and two-dimensional world volume,
respectively. 

By the Hopf map, the solution \eqref{eq:kink_sol} is mapped to a kink 
in the massive $\mathbb{C}P^1$ model
\cite{Abraham:1992vb,Arai:2002xa,Nitta:2012wi}. 
In that case, the phase difference $\beta - \alpha$ becomes a modulus
of the kink. 

In the $(3+1)$-dimensional case, we can think of our toroidal objects
in Sec.~\ref{sec:toroidal-wall} as a domain wall wrapped up on a torus
with its $S^1$ moduli twisted in both world-volume directions. 
It will, however, prove convenient to take a different point of view,
as we shall see, namely to consider first a vortex string which is
then wrapped up on a circle. In the next subsection we therefore
review the (global) vortex.

\subsection{Vortices}
In $d=2+1$ dimensions the model allows for global vortices.
The vortices of $\phi_1$ trap or localize $\phi_2$ in their cores and
they carry a U(1) modulus being the phase of $\phi_2$. 

\begin{figure}[!htb]
\mbox{
\includegraphics[width=0.45\linewidth]{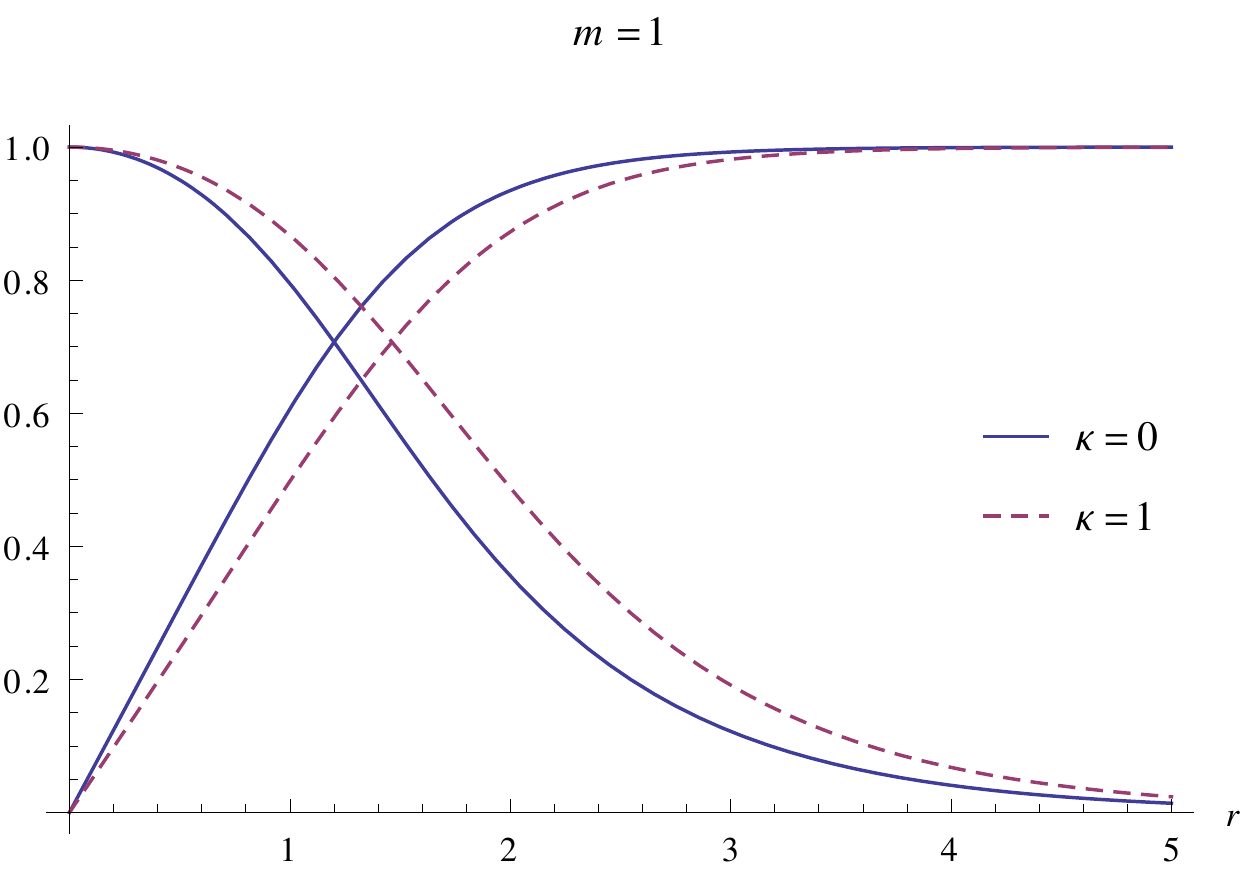}\quad
\includegraphics[width=0.45\linewidth]{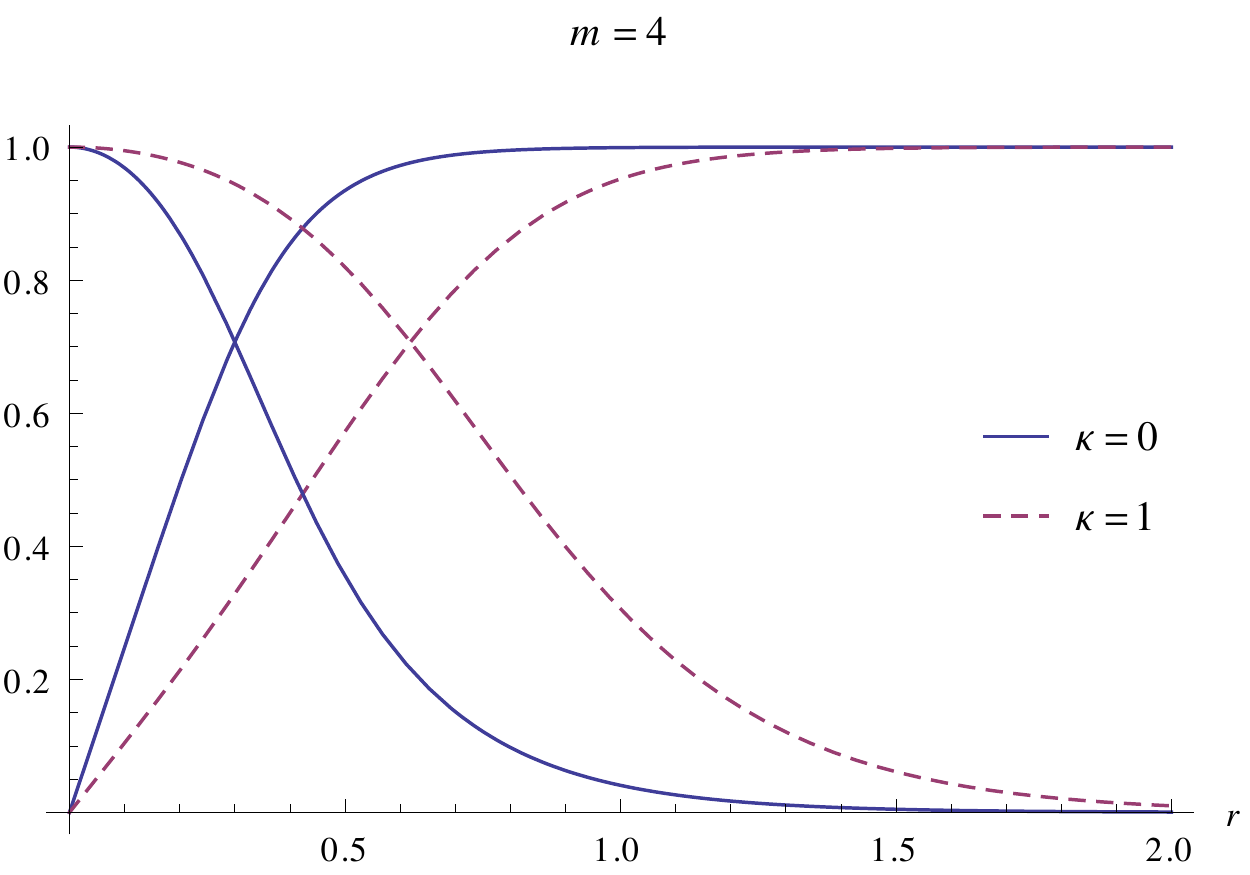}}
\mbox{
\includegraphics[width=0.45\linewidth]{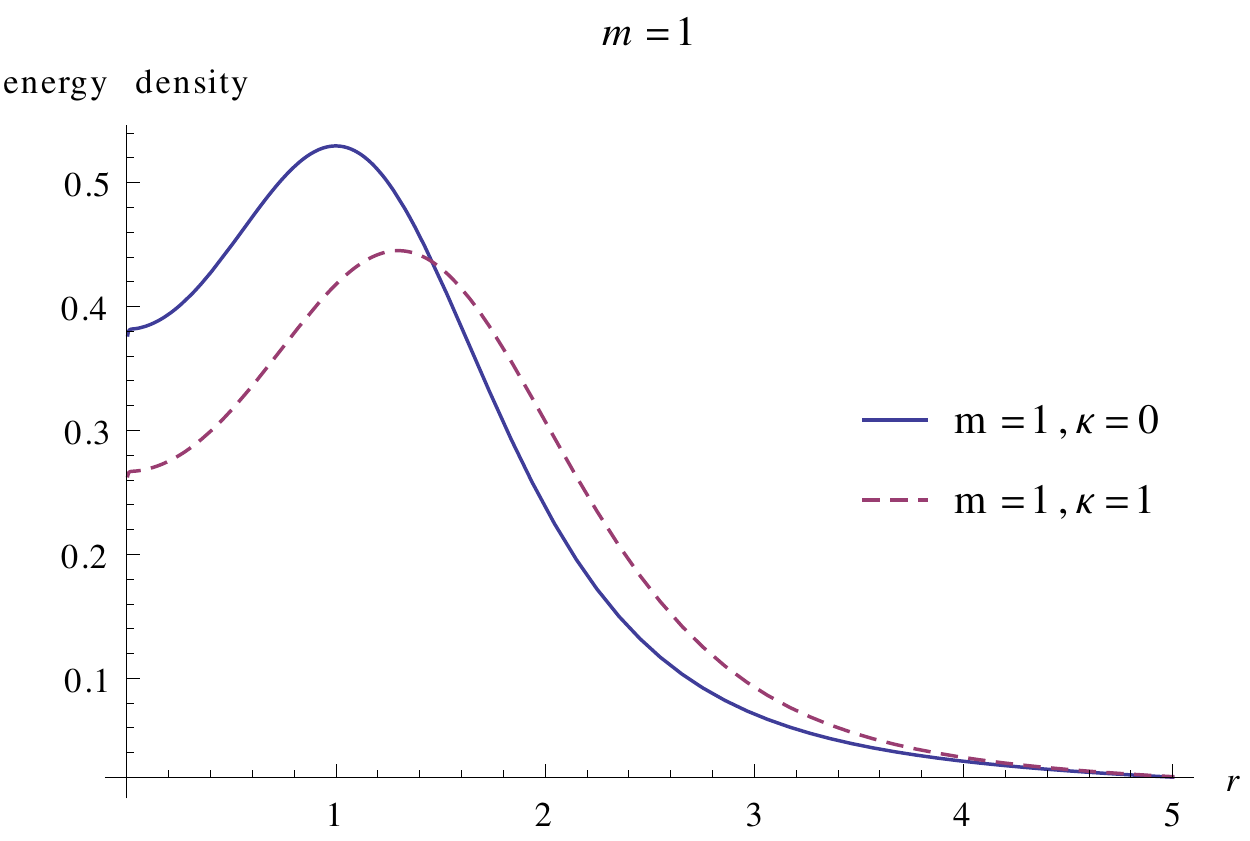}\quad
\includegraphics[width=0.45\linewidth]{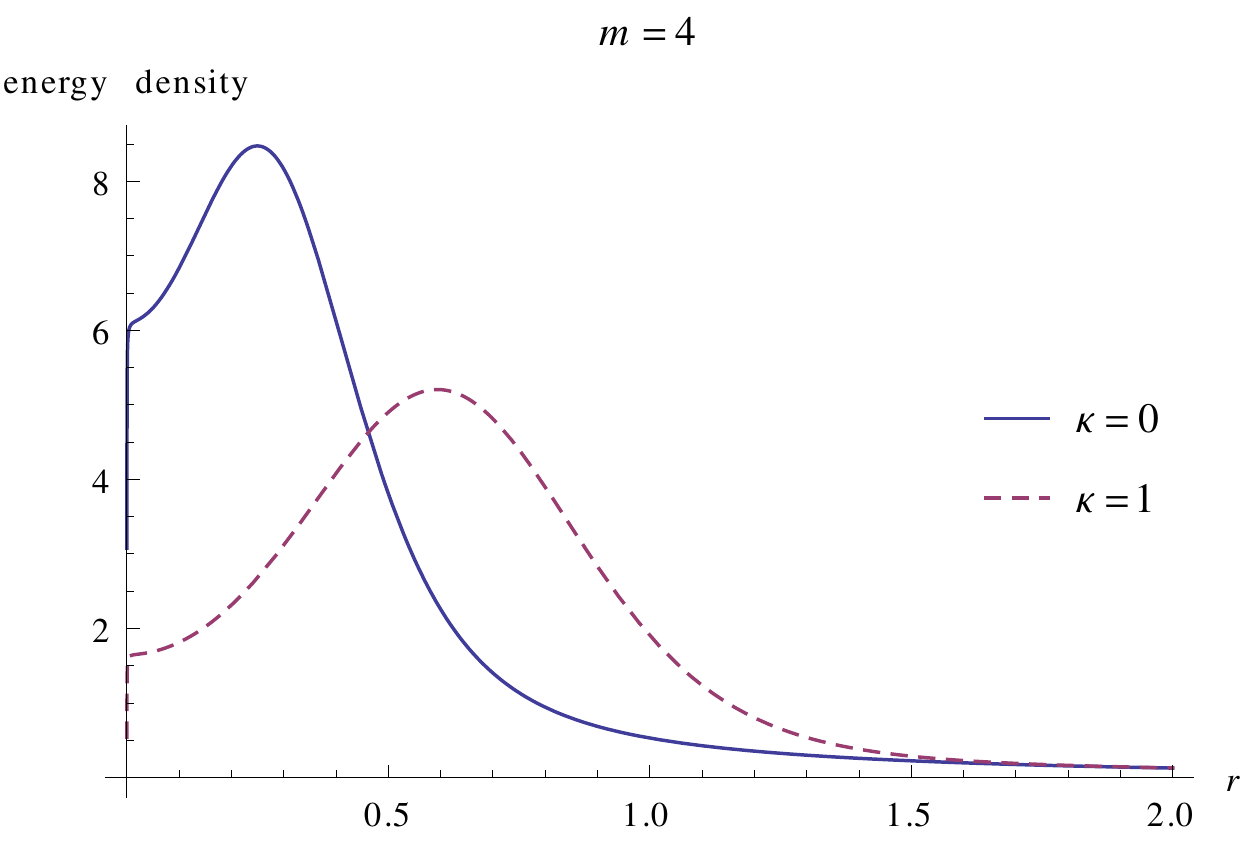}}
\caption{Vortex profiles and energy densities for solutions without
  the Skyrme term $\kappa=0$ (blue curve) and with the Skyrme term
  $\kappa=1$ (dotted red curve) for $m=1$ (left panels) and $m=4$
  (right panels).}
\label{fig:vortex}
\end{figure}

We will now review the global vortex in the nonlinear sigma model
with the potential \eqref{eq:potential}, see \cite{Gudnason:2014hsa}. 
The vortex can be constructed using the following Ansatz
\beq
\phi^{\rm T} = \left(\sin f(r) e^{i\varphi + i \alpha}, 
  \cos f(r) e^{i \beta}\right),
\eeq
where $r\in[0,\infty),\varphi\in[0,2\pi)$ are polar coordinates in a
plane.
The constant, $\alpha$, can be absorbed by 
a redefinition of the coordinate $\varphi$, 
while the constant $\beta$ is a U(1) modulus.
This simplifies the Lagrangian density to \cite{Gudnason:2014hsa} 
\beq
-\mathcal{L} = 
\frac{1}{2}f_r^2
+\frac{1}{2r^2}\sin^2 f
+\frac{\kappa}{2r^2}\sin^2(f) f_r^2
+\frac{1}{8} m^2 \sin^2(2f),
\eeq
for which the equation of motion reads \cite{Gudnason:2014hsa}
\beq
f_{rr} + \frac{1}{r} f_r - \frac{1}{2r^2}\sin 2f
+\frac{\kappa}{r^2}\sin^2 f\left(f_{rr} - \frac{1}{r} f_r\right)
+\frac{\kappa}{2r^2}\sin(2f) f_r^2
-\frac{1}{4} m^2 \sin 4f = 0.
\eeq
The boundary conditions for the vortex system are given by
\beq
f(0) = 0, \qquad
f(\infty) = \frac{\pi}{2}.
\eeq
We show numerical solutions in Fig.~\ref{fig:vortex} for $m=1,4$ and 
$\kappa=0,1$. 
By the Hopf map, they can (topologically) be mapped to lumps. 

In $d=3+1$ dimensions, these vortices are extended to vortex strings
or cosmic strings. They are global analogues of Witten's
superconducting strings \cite{Witten:1984eb}. 
We may call them superflowing cosmic strings. 
Once extended to $(3+1)$-dimensional spacetime, the strings bear a
U(1) modulus, which we can parametrize as
\beq
\phi^{\rm T} = \left(\sin f(r) e^{i\varphi}, 
\cos f(r) e^{i\zeta(z)}\right),
\eeq
In the next section we will compactify these strings on a circle which
requires a nontrivial twist of the modulus $\zeta$.

\section{Toroidal Skyrmions in $3+1$ dimensions
\label{sec:toroidal-wall}}

In this section we will consider a closed vortex string, i.e.~the
vortex string wound up on a circle and thus forming a torus-like
object. 
Such a closed vortex string is unstable unless its U(1) modulus is
twisted along the string (viz.~it is topologically trivial otherwise). 

\begin{figure}[!tbh]
\begin{center}
\includegraphics[width=0.5\linewidth]{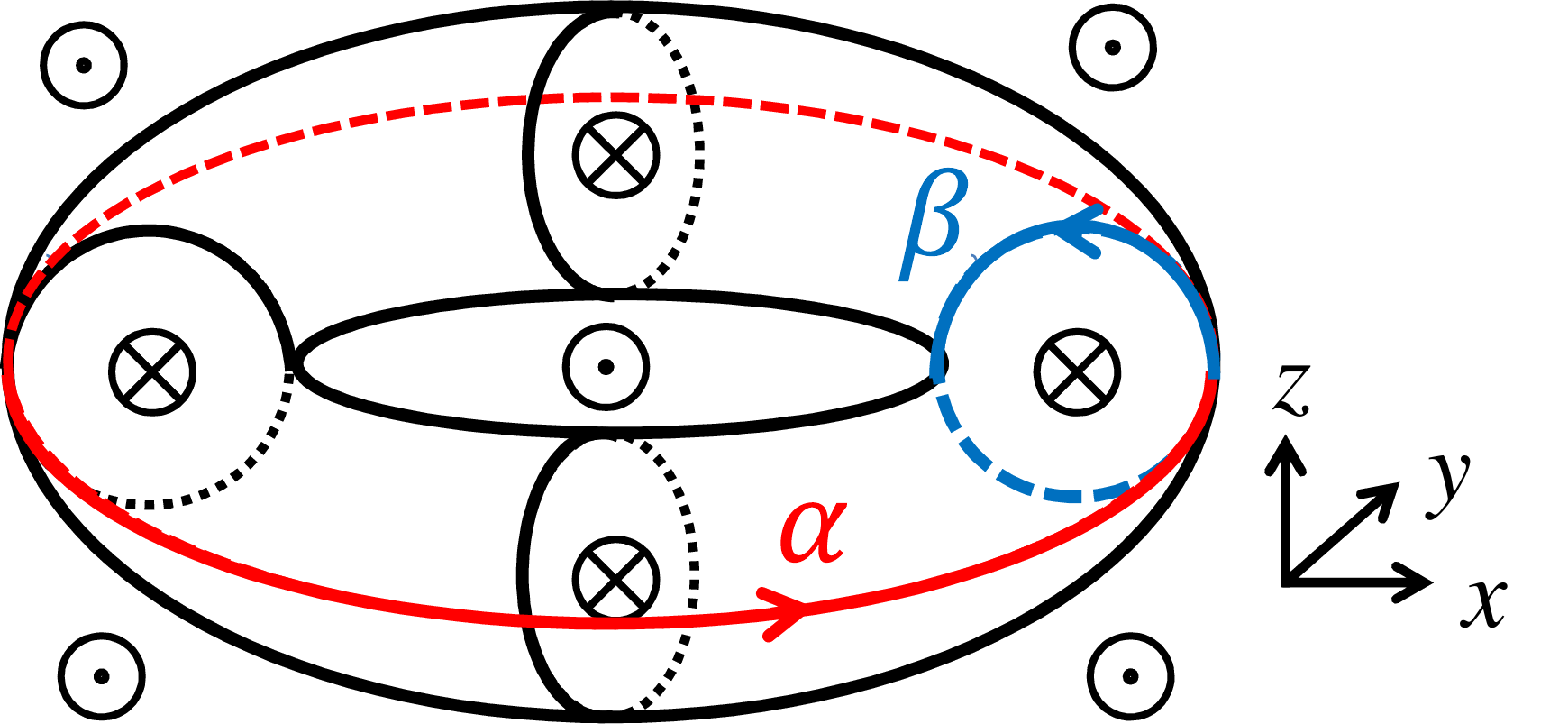}
\end{center}
\caption{The two cycles of the torus. 
The toroidal and poloidal cycles are 
denoted by $\alpha$ and $\beta$, respectively. 
The $\odot$ and $\otimes$ denote the vacua 
in Eq.~(\ref{eq:vacua}), respectively.  
The U(1) modulus is twisted $P$ and $Q$ times 
along the cycles $\alpha$ and $\beta$,
respectively. \label{fig:cycles}} 
\end{figure}

In the final configuration, the U(1) modulus is twisted $P$ times
along the toroidal ($\alpha$) cycle of the torus and the global string
winds $Q$ times ``along'' the poloidal ($\beta$) cycle of the torus;
see Fig.~\ref{fig:cycles}. 

The torus-shaped solution requires us to study the full partial
differential equation (PDE) numerically, for which we will use the
relaxation method on a cubic square lattice. Because of the
topological nature of the objects we study, it is sufficient to employ 
Neumann conditions on the boundary of the lattice whereas the initial
condition is very important. 
For the initial configuration we will use the following Ansatz
\begin{align}
\phi^{\rm T} = \left(\sin\left[
    \cos^{-1}\{\sin f(r) \sin\theta\} \right] e^{i Q \tan^{-1}
    (\tan f(r)\cos\theta)}, \cos\left[
    \cos^{-1}\{\sin f(r) \sin\theta\} \right] e^{i P \phi}
\right), \label{eq:ansatz}
\end{align}
where $r\in[0,\infty)$, $\theta\in[0,\pi]$, $\varphi\in[0,2\pi)$ and
$f(r)$ is an appropriately chosen monotonically decreasing function
satisfying the boundary conditions 
\begin{align}
f(r \to 0) \to \pi, \quad
f(r \to \infty) \to 0.
\end{align}
The baryon number (Skyrme charge) of $\pi_3(S^3)\simeq\mathbb{Z}$ for
the configuration given in Eq.~\eqref{eq:ansatz} is 
\begin{align}
\begin{split}
B &= \frac{1}{4\pi^2} \int d^3x\: \epsilon^{i j k} 
\phi^\dag \p_i \phi \p_j \phi^\dag \p_k \phi \\
&= -\frac{1}{2 \pi^2} \int_0^{\infty} dr\: \int_0^\pi d\theta\: \int_0^{2 \pi} d\phi\; \sin\theta P Q f^\prime(r) \sin^2 f(r) \\
&= -\frac{P Q}{\pi} \int_0^\infty dr\: \p_r \left[f(r) - \sin f(r) \cos f(r) \right] \\
&= P Q.
\end{split}
\end{align}
Although we seemingly have two quantum numbers to dial in the
configuration, it will prove convenient to think about the winding
number $Q$ as that of the global vortex. This may suggest that $Q>1$
will be unstable as global vortices repel with a force $\sim 1/d$,
where $d$ (here) is the separation distance between two strings. 
We confirm this expectation by numerically solving the equations and
find for a wide range of parameters that for $Q>1$, the
relaxation method always splits up the object into $Q$ individual
strings; each with a $P$-wound U(1) phase. For details, see Appendix
\ref{app:stringsplitting}.

We can therefore study the numerical solutions with baryon number
$B=P$, for which the Ansatz \eqref{eq:ansatz} reduces to
\beq
\phi^{\rm T} = \left(\cos f(r) + i\sin f(r)\cos\theta,
\sin f(r) \sin\theta e^{i P \phi} \right).
\label{eq:torus_reduced}
\eeq
This is exactly the axially symmetric generalization of the hedgehog
Ansatz and this is just what we need (note that for Skyrmions without
our BEC-motivated potential, this Ansatz is only appropriate for
$B=1,2$ while for $B>2$ the axial symmetry no longer yields the 
minimum-energy configuration). 
We will study two cases in turn; in the first we turn on only the
fourth-order derivative term, i.e.~$\kappa=1$ and $c_6=0$ while in the
second case we switch off the fourth-order but use the sixth-order
derivative term, i.e.~$\kappa=0$ and $c_6=1$. We will call them the
2+4 model and the 2+6 model, respectively. 

\begin{figure}[!htp]
\begin{center}
\captionsetup[subfloat]{labelformat=empty}
\mbox{
\subfloat[$(P,Q)=(1,1)$]{\includegraphics[width=0.32\linewidth]{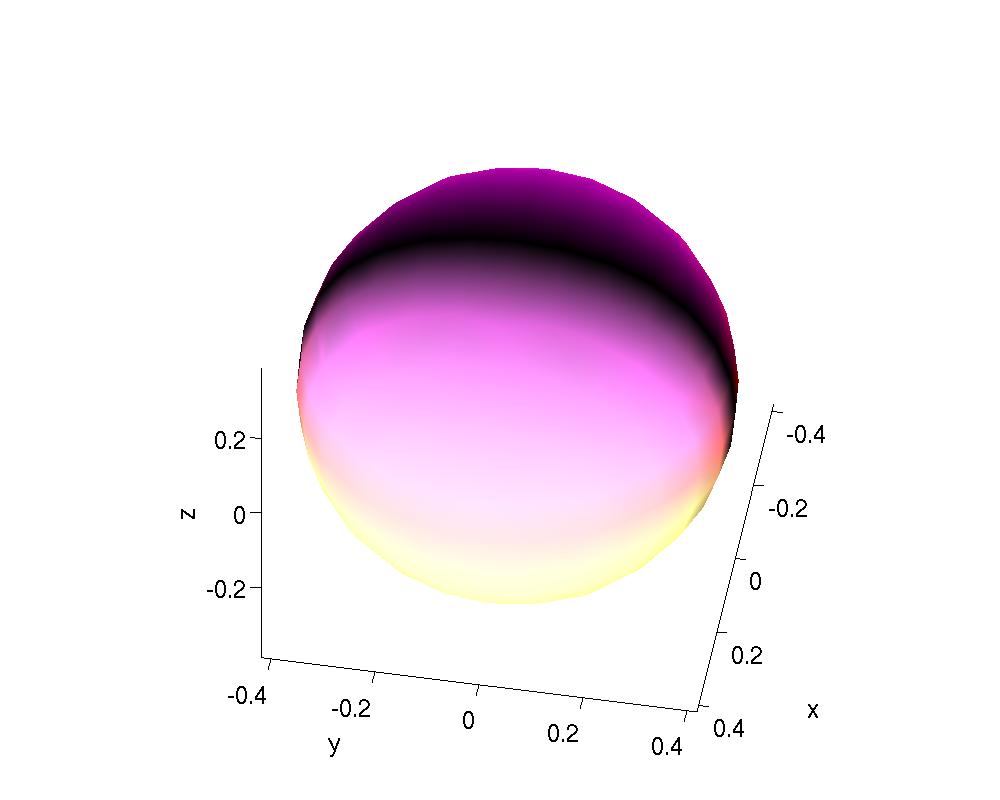}}
\subfloat[$(P,Q)=(2,1)$]{\includegraphics[width=0.32\linewidth]{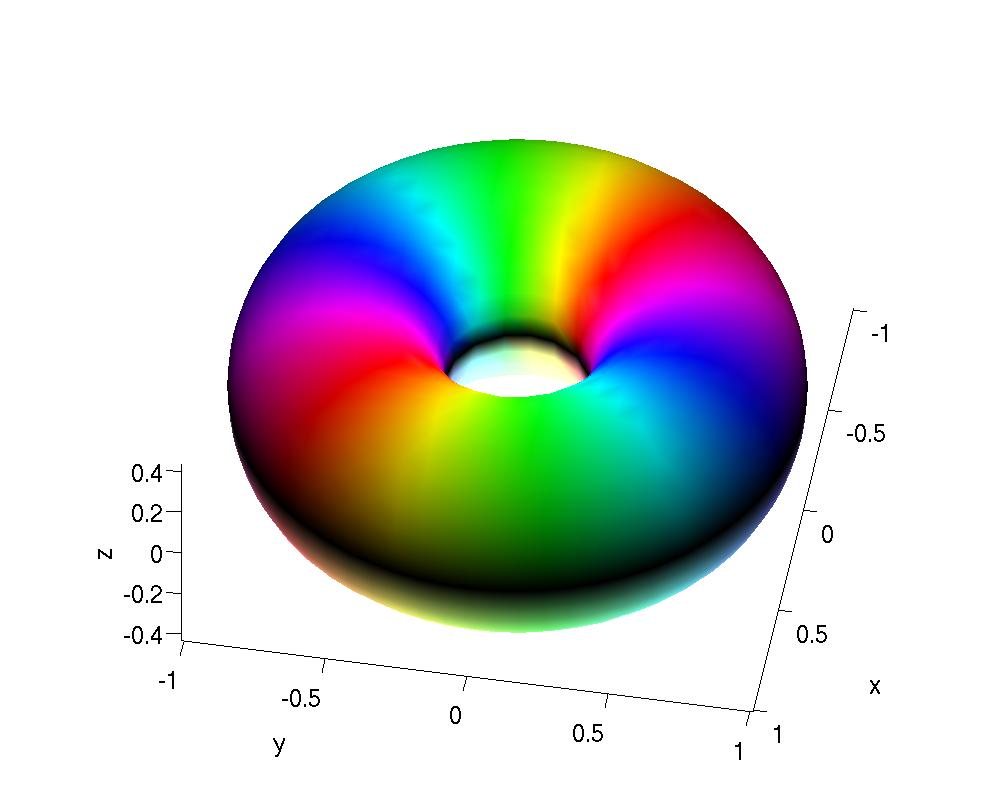}}
\subfloat[$(P,Q)=(3,1)$]{\includegraphics[width=0.32\linewidth]{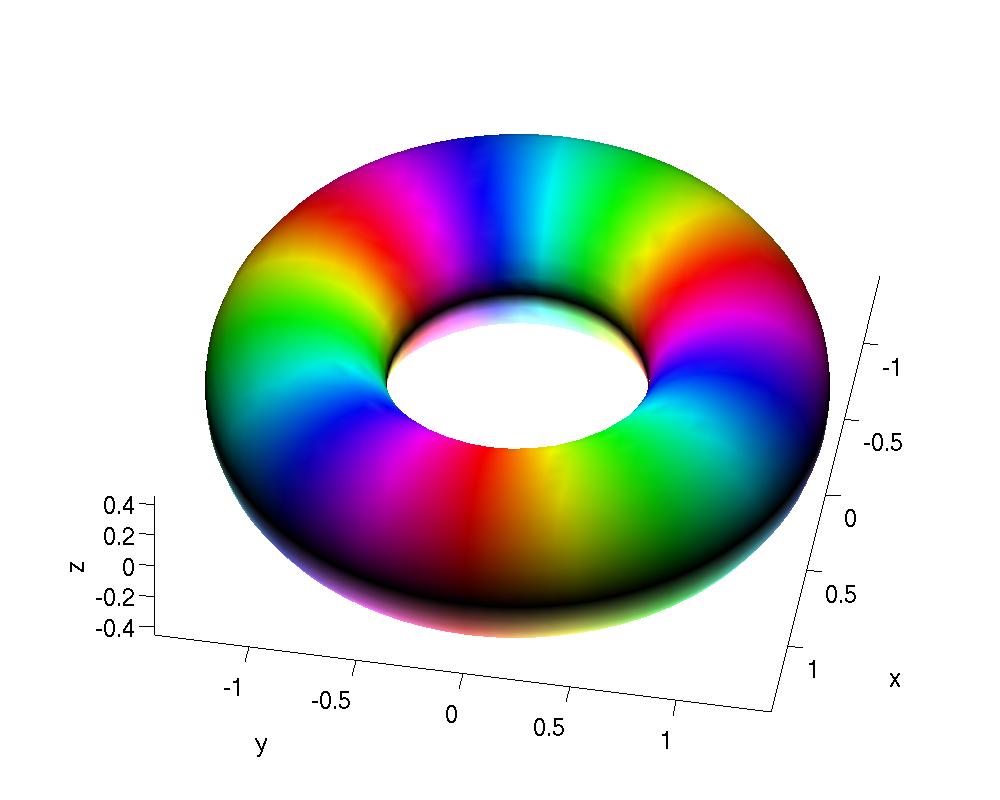}}}
\mbox{
\subfloat[$(P,Q)=(4,1)$]{\includegraphics[width=0.32\linewidth]{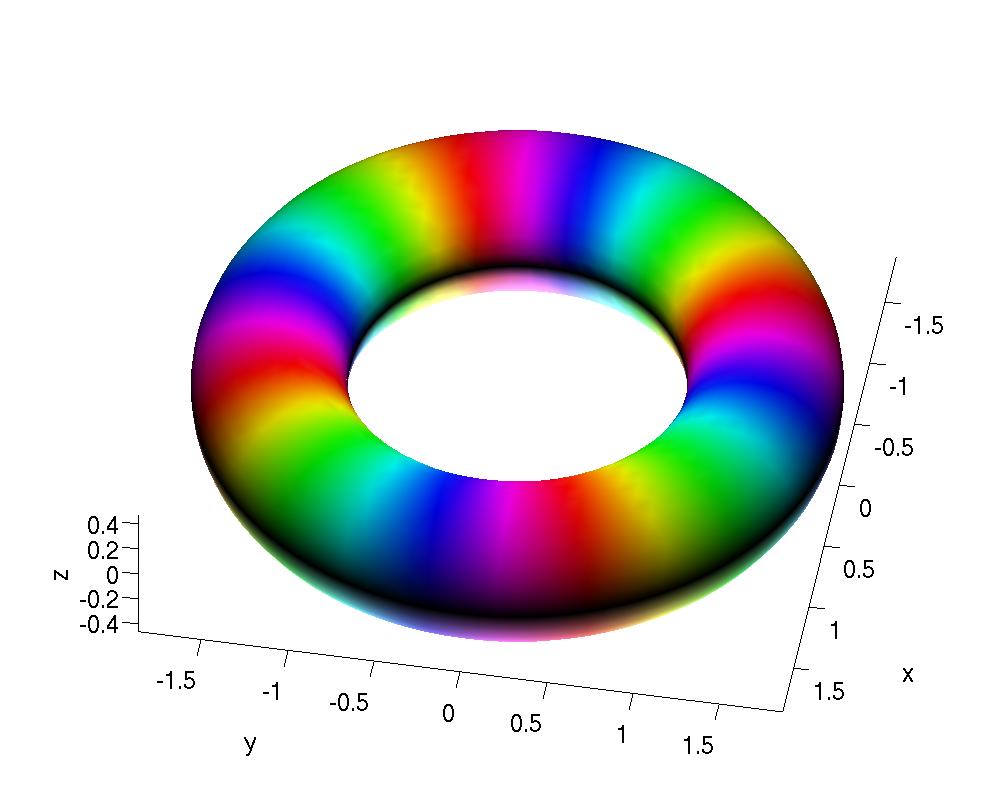}}
\subfloat[$(P,Q)=(5,1)$]{\includegraphics[width=0.32\linewidth]{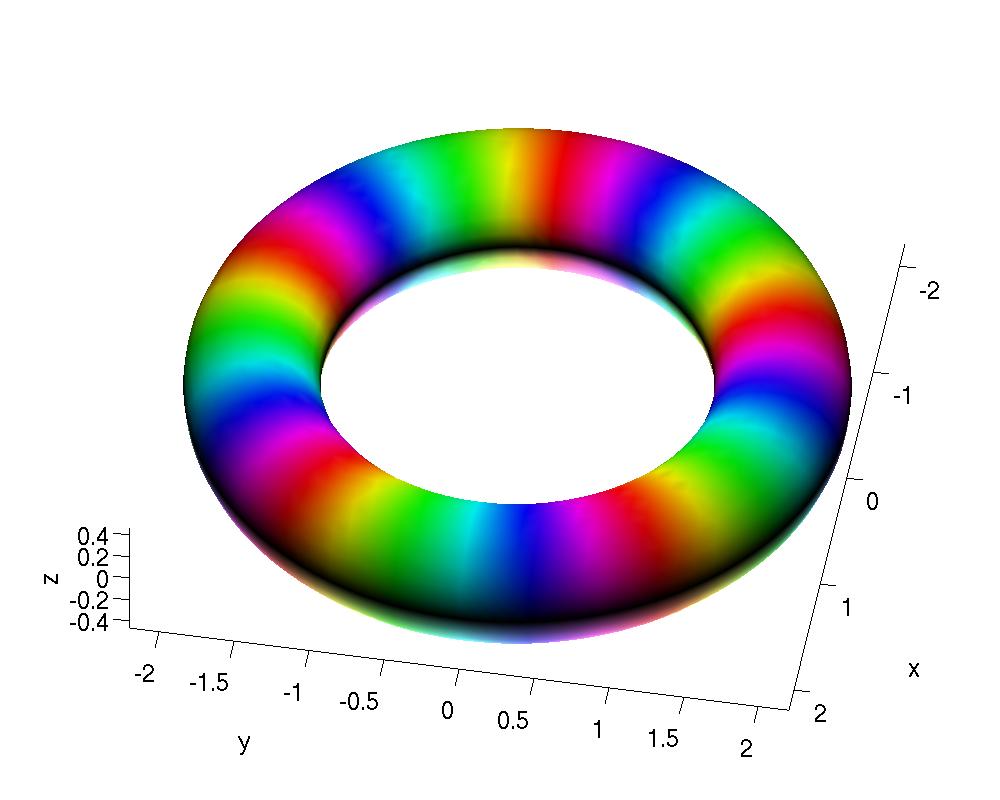}}
\subfloat[$(P,Q)=(6,1)$]{\includegraphics[width=0.32\linewidth]{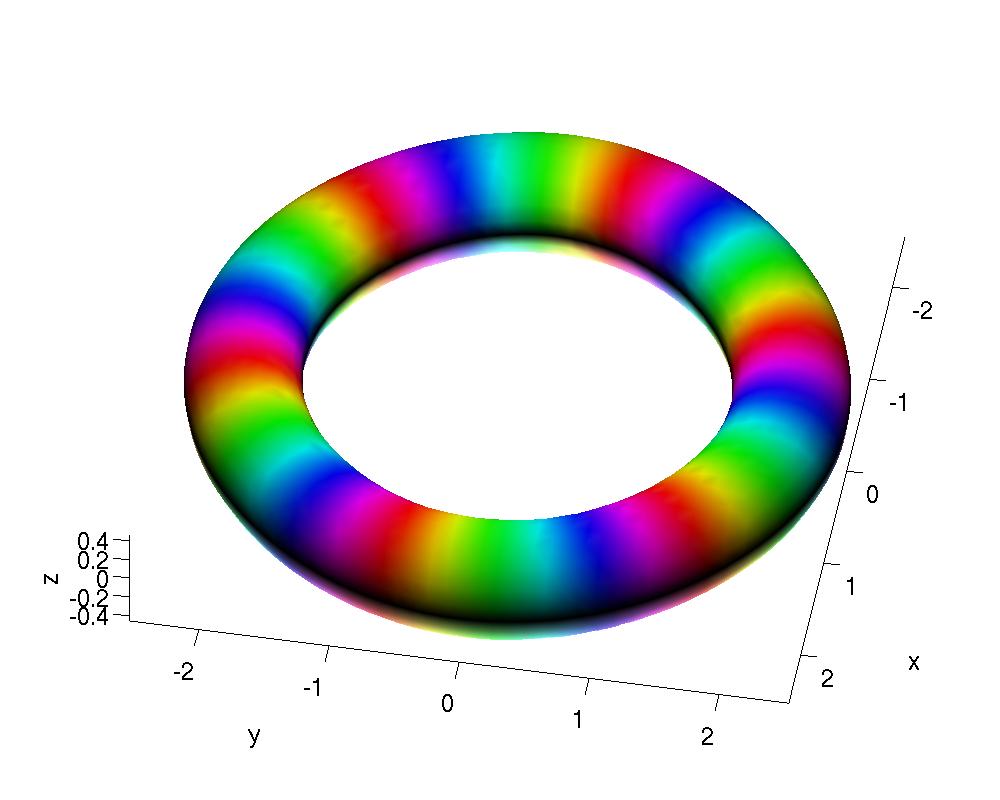}}}
\mbox{
\includegraphics[width=0.5\linewidth]{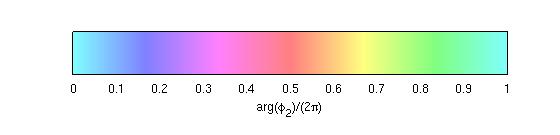}}
\caption{
Isosurfaces showing the solutions for the 2+4 model, i.e.~for
$\kappa=1$ and $c_6=0$, at constant baryon charge density equal to half
its maximum value. 
The color represents the phase of the scalar field $\phi_2$ and the
lightness is given by $|\Im(\phi_1)|$. 
The calculations are done on an $81^3$ cubic lattice with the
relaxation method. 
}
\label{fig:t4}
\end{center}
\end{figure}

\begin{figure}[!htp]
\begin{center}
\captionsetup[subfloat]{labelformat=empty}
\mbox{
\subfloat[$(P,Q)=(1,1)$]{\includegraphics[width=0.49\linewidth]{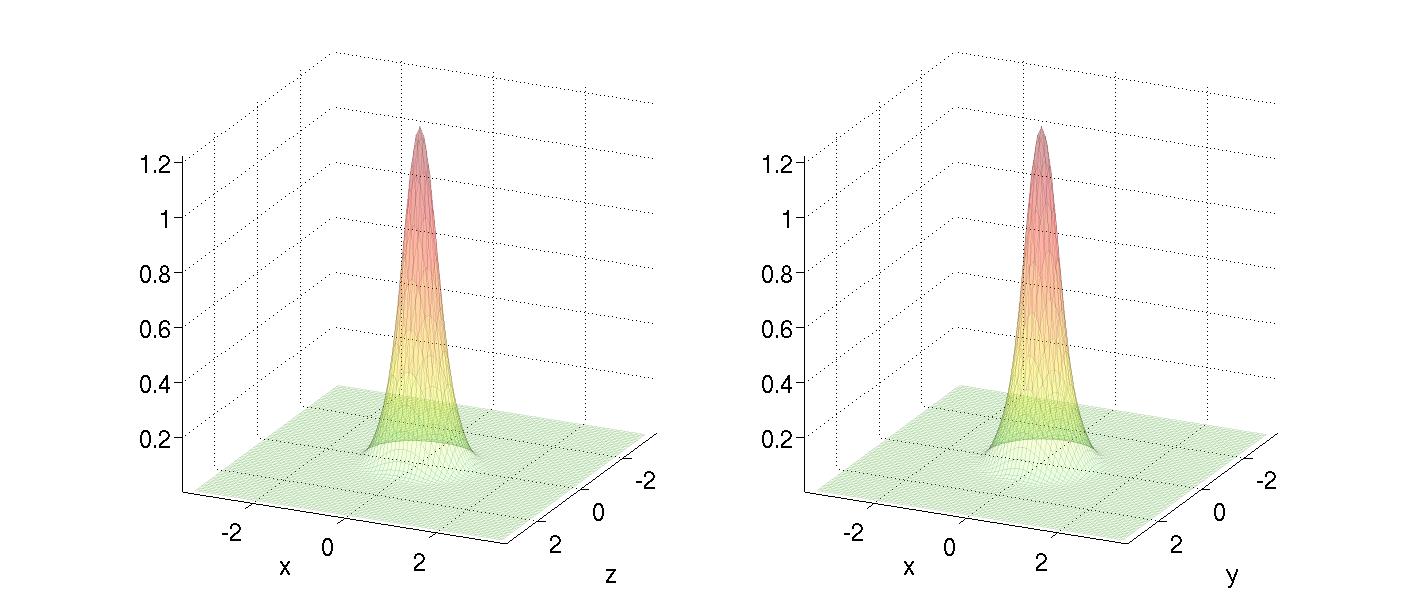}}
\subfloat[$(P,Q)=(2,1)$]{\includegraphics[width=0.49\linewidth]{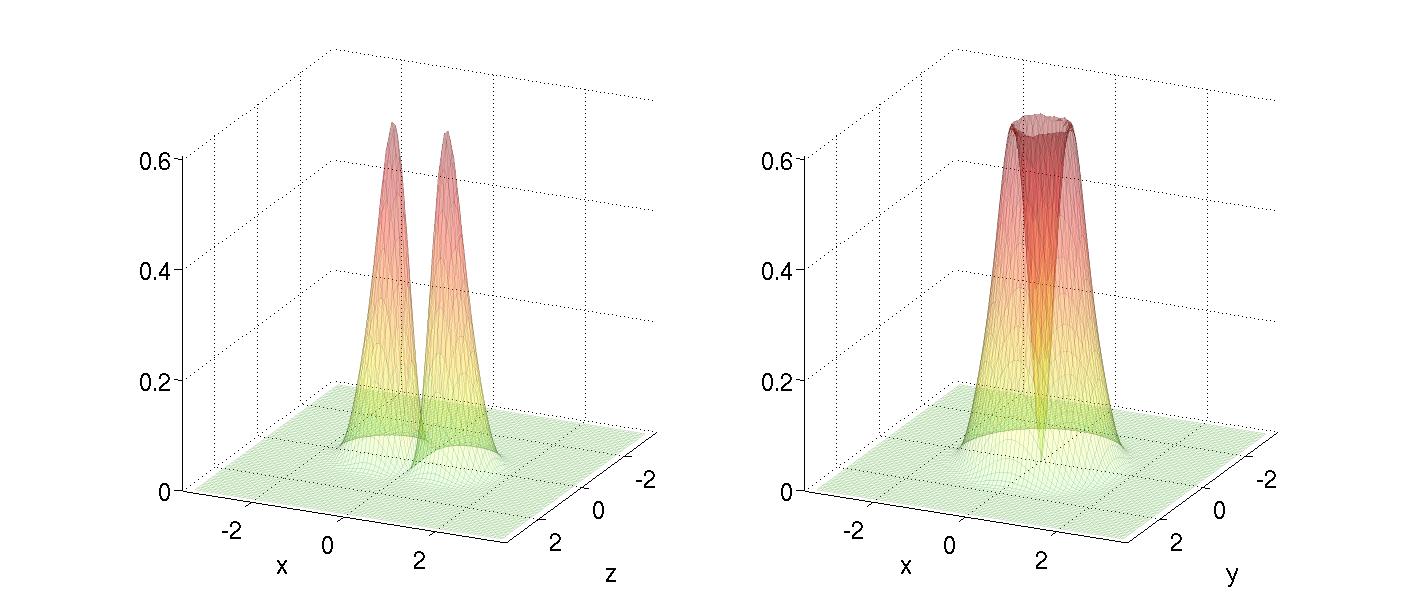}}}
\mbox{
\subfloat[$(P,Q)=(3,1)$]{\includegraphics[width=0.49\linewidth]{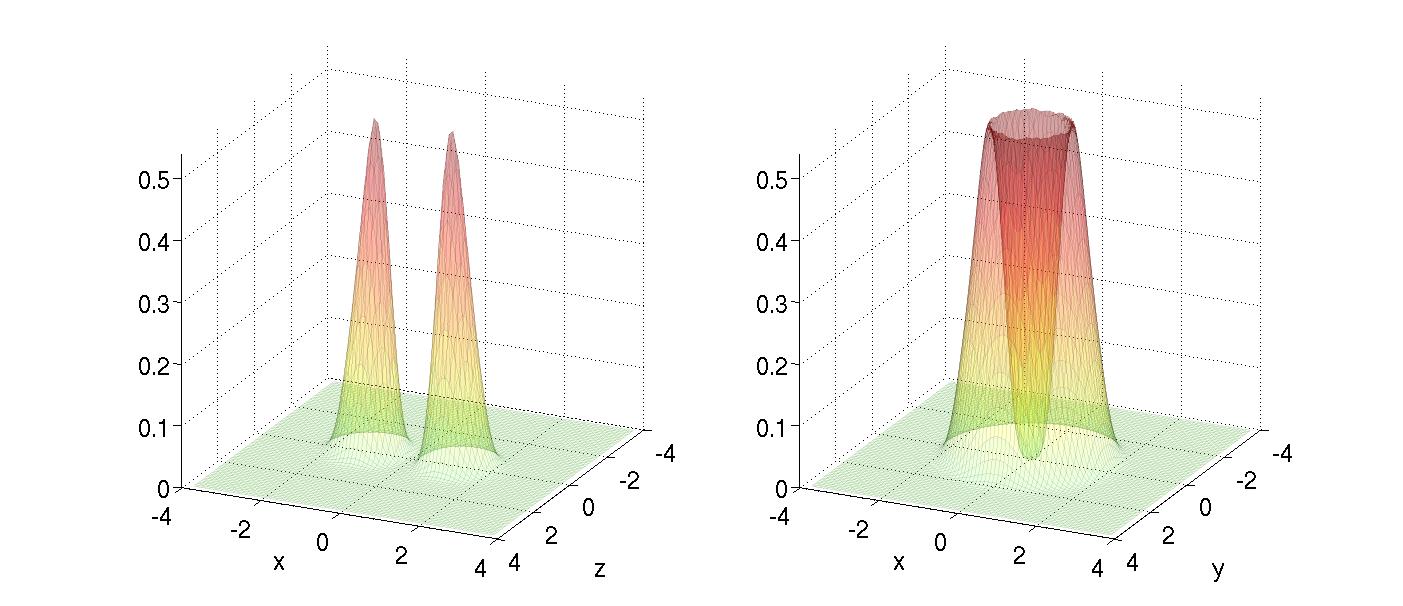}}
\subfloat[$(P,Q)=(4,1)$]{\includegraphics[width=0.49\linewidth]{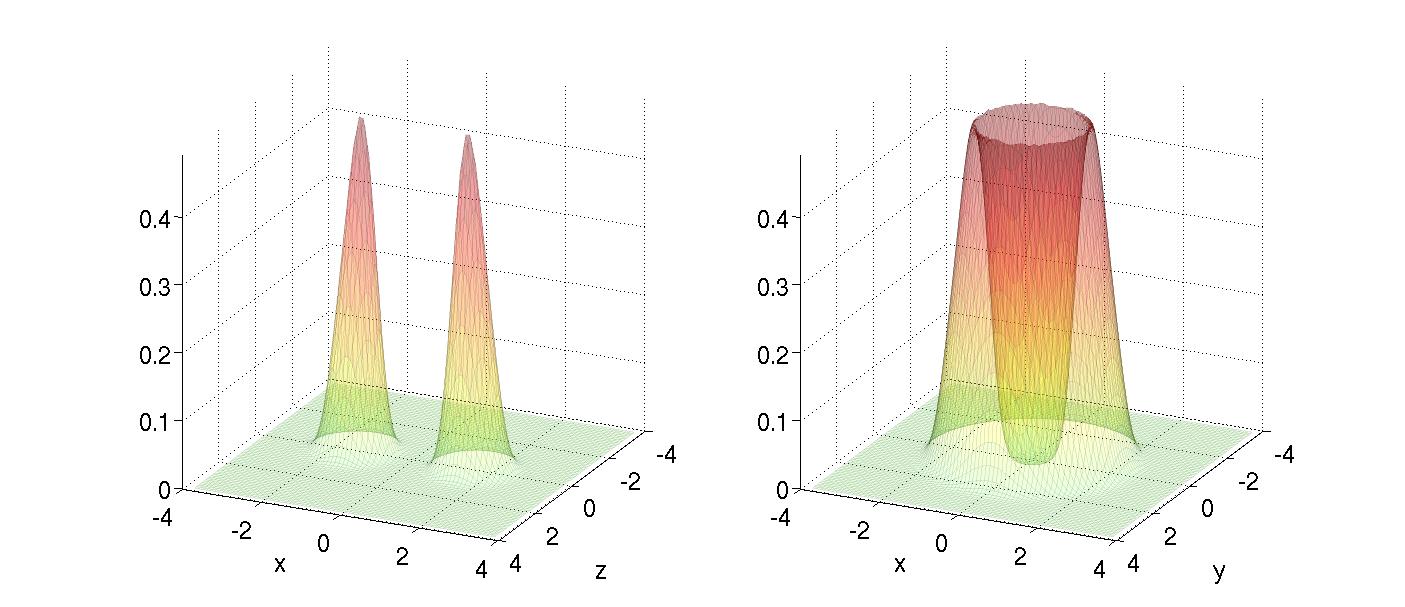}}}
\mbox{
\subfloat[$(P,Q)=(5,1)$]{\includegraphics[width=0.49\linewidth]{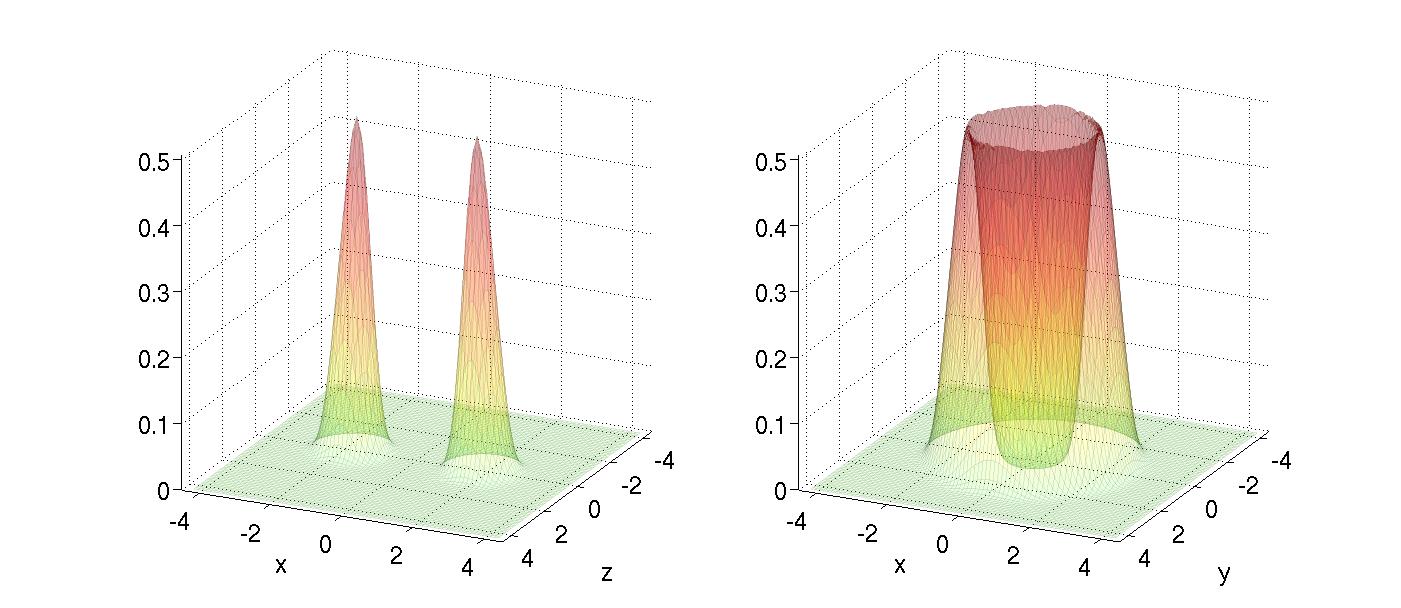}}
\subfloat[$(P,Q)=(6,1)$]{\includegraphics[width=0.49\linewidth]{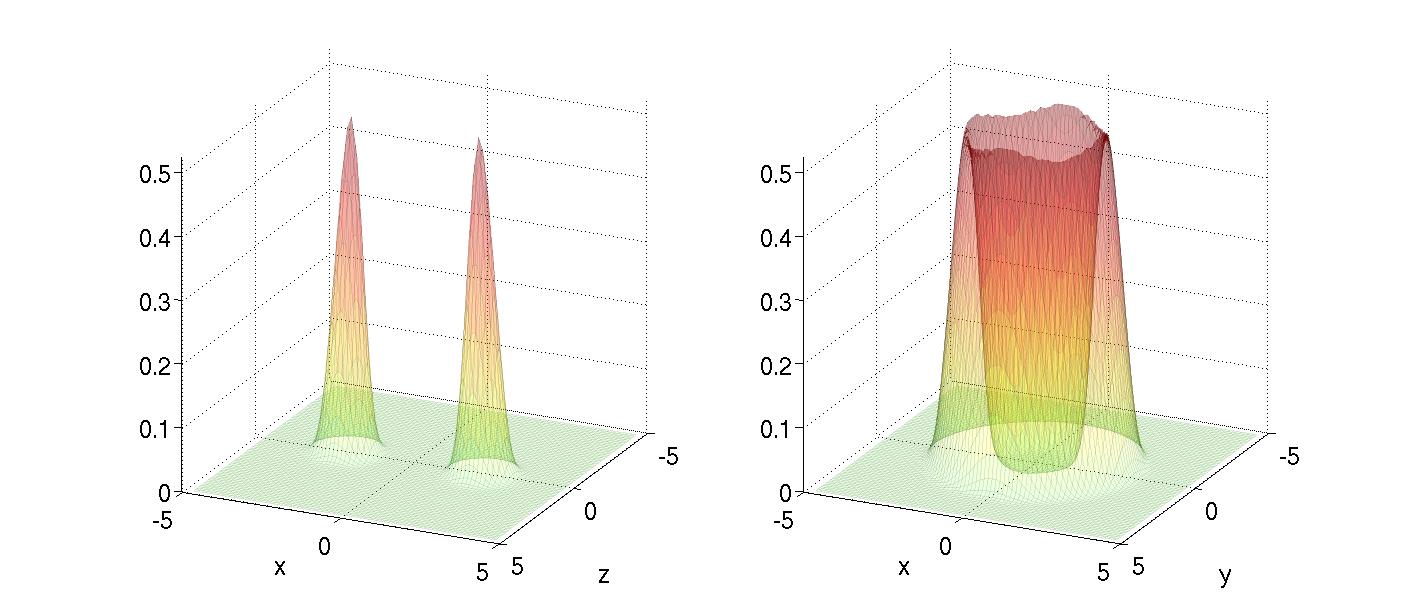}}}
\caption{
Baryon charge density for solutions in the 2+4 model, i.e.~with
$\kappa=1$ and $c_6=0$, at $xz$ slices (for $y=0$) and $xy$ slices (for
$z=0$). $yz$ slices are omitted as they are identical to the $xz$
slices by rotational symmetry of the torus. The calculations are done
on an $81^3$ cubic lattice with the relaxation method. 
}
\label{fig:t4_baryonslice}
\end{center}
\end{figure}

\begin{figure}[!htp]
\begin{center}
\captionsetup[subfloat]{labelformat=empty}
\mbox{
\subfloat[$(P,Q)=(1,1)$]{\includegraphics[width=0.49\linewidth]{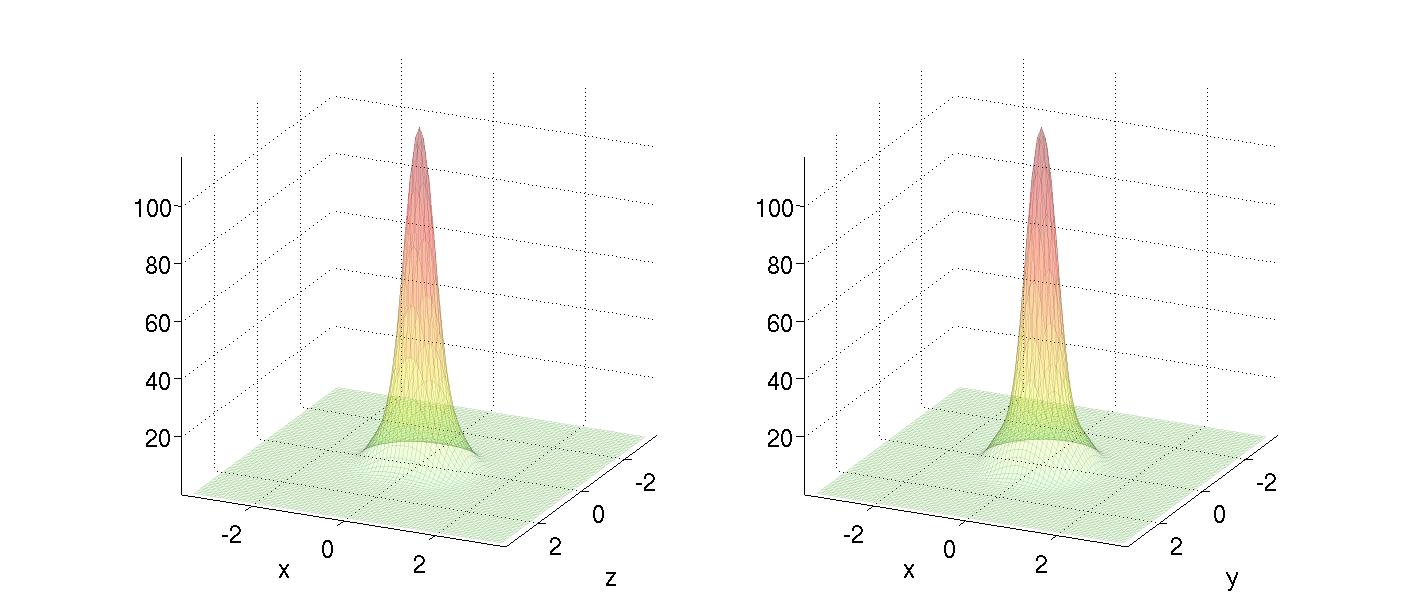}}
\subfloat[$(P,Q)=(2,1)$]{\includegraphics[width=0.49\linewidth]{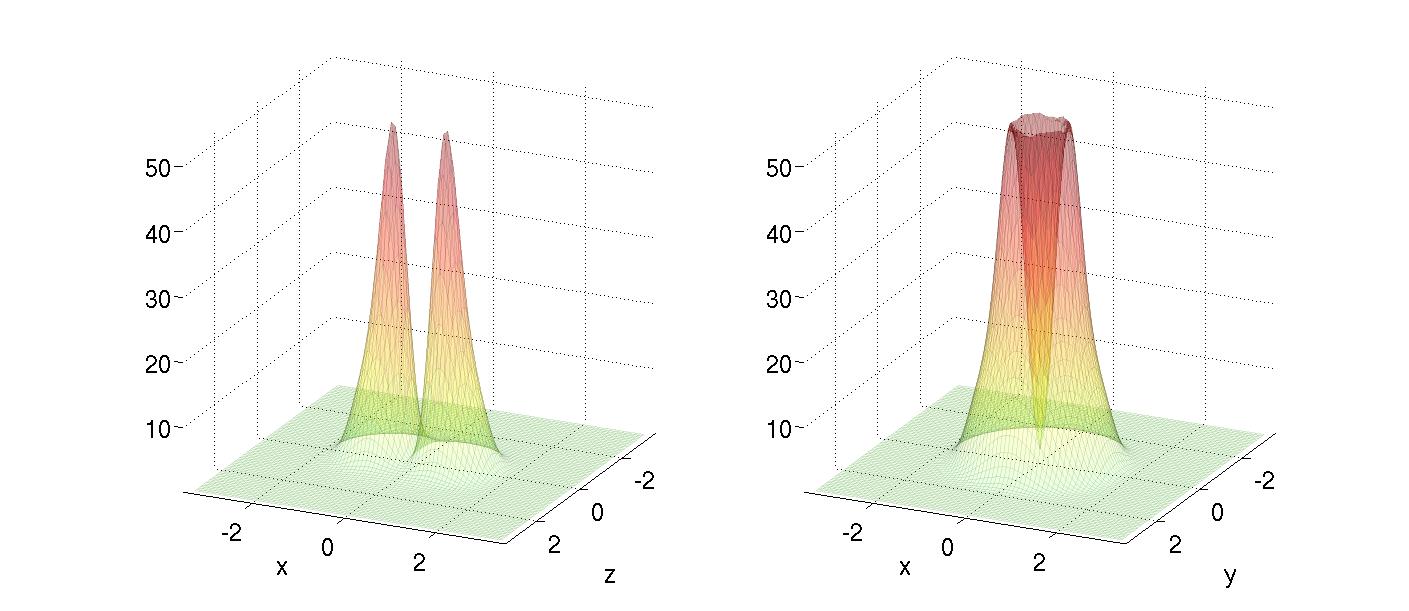}}}
\mbox{
\subfloat[$(P,Q)=(3,1)$]{\includegraphics[width=0.49\linewidth]{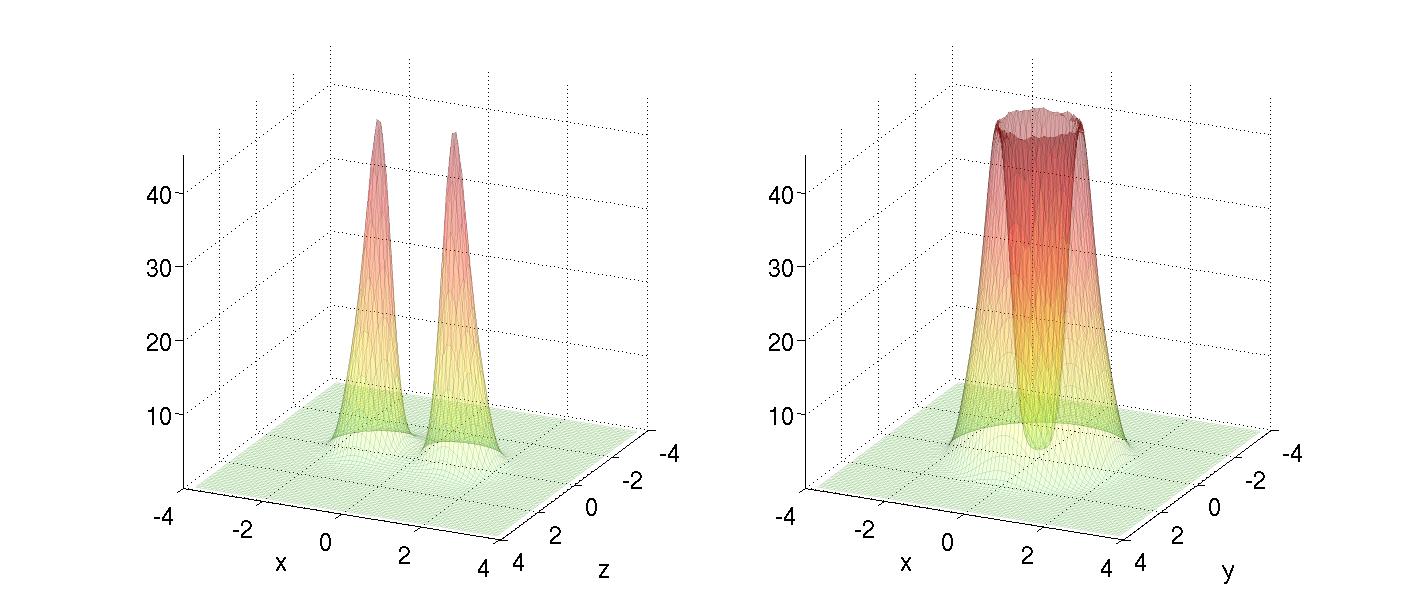}}
\subfloat[$(P,Q)=(4,1)$]{\includegraphics[width=0.49\linewidth]{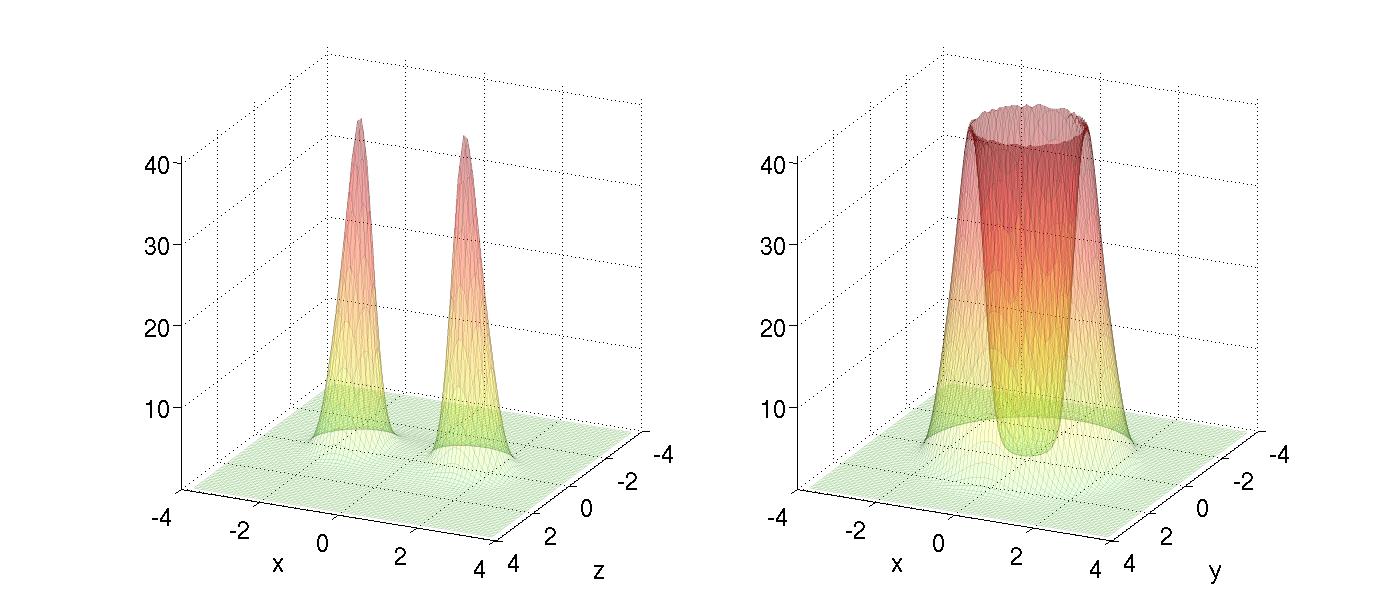}}}
\mbox{
\subfloat[$(P,Q)=(5,1)$]{\includegraphics[width=0.49\linewidth]{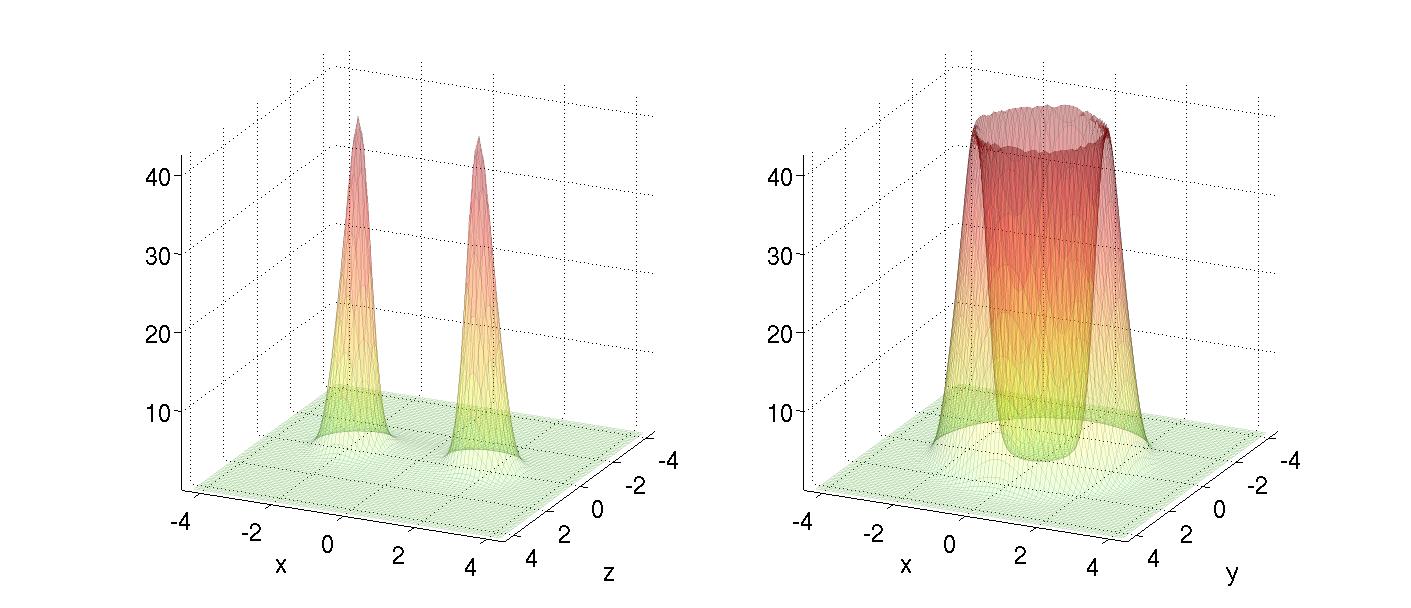}}
\subfloat[$(P,Q)=(6,1)$]{\includegraphics[width=0.49\linewidth]{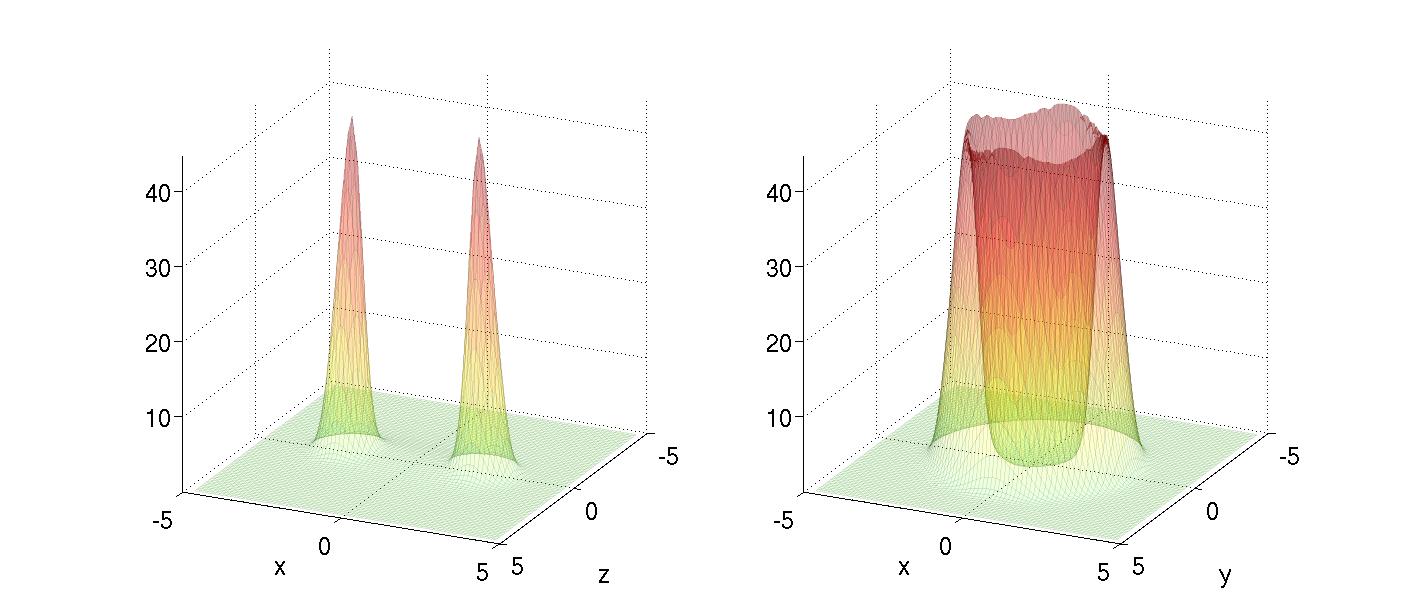}}}
\caption{
Energy density for solutions in the 2+4 model, i.e.~with $\kappa=1$
and $c_6=0$, at $xz$ slices (for $y=0$) and $xy$ slices (for
$z=0$). $yz$ slices are omitted as they are identical to the $xz$ 
slices by rotational symmetry of the torus. The calculations are done
on an $81^3$ cubic lattice with the relaxation method. 
}
\label{fig:t4_energyslice}
\end{center}
\end{figure}

In Figs.~\ref{fig:t4}, \ref{fig:t4_baryonslice} and
\ref{fig:t4_energyslice} we show solutions for case of the 2+4 model 
($\kappa=1$ and $c_6=0$) with mass $m=4$. In Fig.~\ref{fig:t4} is
shown the 3-dimensional isosurfaces at half the maximum value of the
baryon charge density. The color scheme used is chosen such that the
U(1) phase, $\mathop{\rm arg}\phi_2$, is mapped to the hue while the
lightness is given by the absolute value of the imaginary part of the
vortex condensate: $|\Im(\phi_1)|$. 
In Figs.~\ref{fig:t4_baryonslice} and \ref{fig:t4_energyslice} are
shown the baryon charge density and energy density, at two different
cross sections cutting through the origin of the torus, respectively. 
In this case, they are practically identical, which means that the
energy density is located where the baryon charge is.

As a check on our numerical precision, we calculate the baryon charge
density and integrate it numerically, see Table \ref{tab:EB_T4}. 
As already explained, our Skyrmionic torii are only stable for $Q=1$,
but to study whether they are stable for higher $P>1$, we need to
compare the energy of the configurations. In Table \ref{tab:EB_T4}, we
calculate the energy per $B=P$ and find that the energy drops for the
first four torii, viz.~$P=1,2,3,4$, but then it starts to increase
slightly. The increase is so small that the $P=5$ solution is still
energetically stable (also taking into account the numerical accuracy)
while $P=6$ is only metastable \footnote{The question of stability may
  also depend on the coefficients of the higher-derivative terms and
  the mass.}. That is, the energy of the $P=6$
solution is larger than two times that of the $P=3$ solution and hence
it is bound to decay. Here we have not studied the potential barrier
for the decay and thus cannot calculate its life time.

\begin{table}[!htp]
\begin{center}
\caption{Numerically integrated baryon charge and energy (mass) for
  the solutions in the 2+4 model. Stability is observed for the first
  five solutions whilst $P=6$ is only energetically metastable. }
\label{tab:EB_T4}
\begin{tabular}{c||cc}
$B$ & $B^{\rm numerical}$ & $E^{\rm numerical}/B$ \\
\hline\hline
$1$ & $0.9995$ & $93.3151\pm 0.0297$ \\
$2$ & $1.9994$ & $85.2782\pm 0.0223$ \\
$3$ & $2.9985$ & $84.0152\pm 0.0200$ \\
$4$ & $3.9981$ & $83.6919\pm 0.0516$ \\
$5$ & $4.9959$ & $84.1664\pm 0.0312$ \\
$6$ & $5.9921$ & $84.7335\pm 0.0204$
\end{tabular}
\end{center}
\end{table}

\begin{figure}[!htp]
\begin{center}
\captionsetup[subfloat]{labelformat=empty}
\mbox{
\subfloat[$(P,Q)=(1,1)$]{\includegraphics[width=0.32\linewidth]{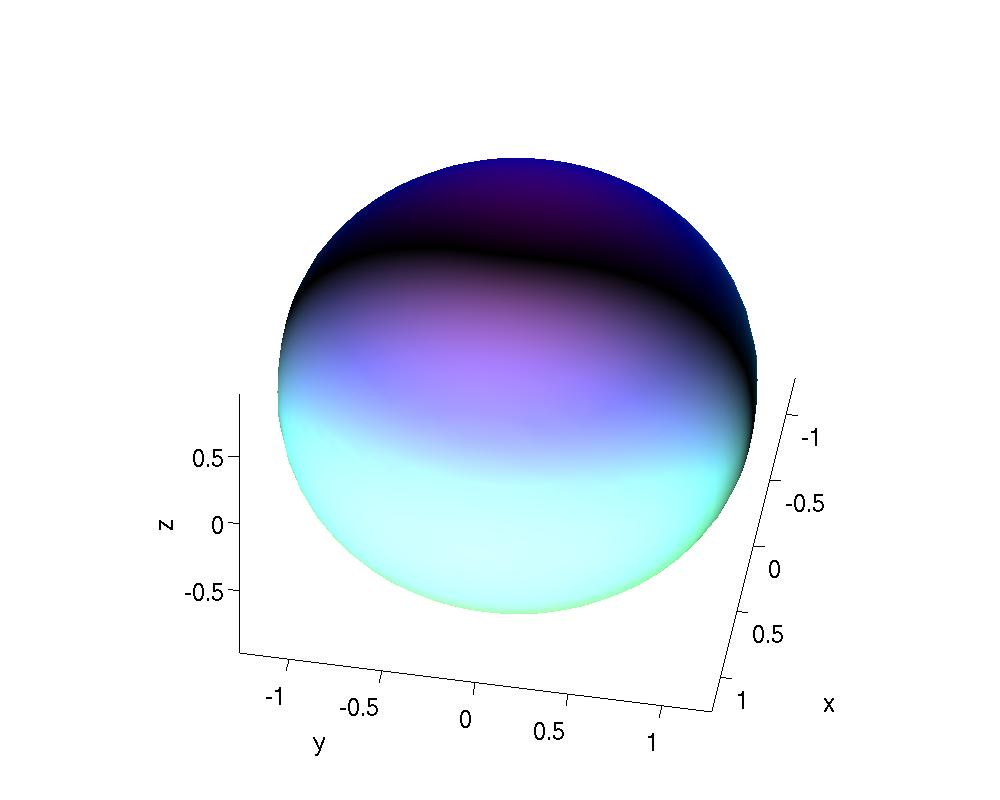}}
\subfloat[$(P,Q)=(2,1)$]{\includegraphics[width=0.32\linewidth]{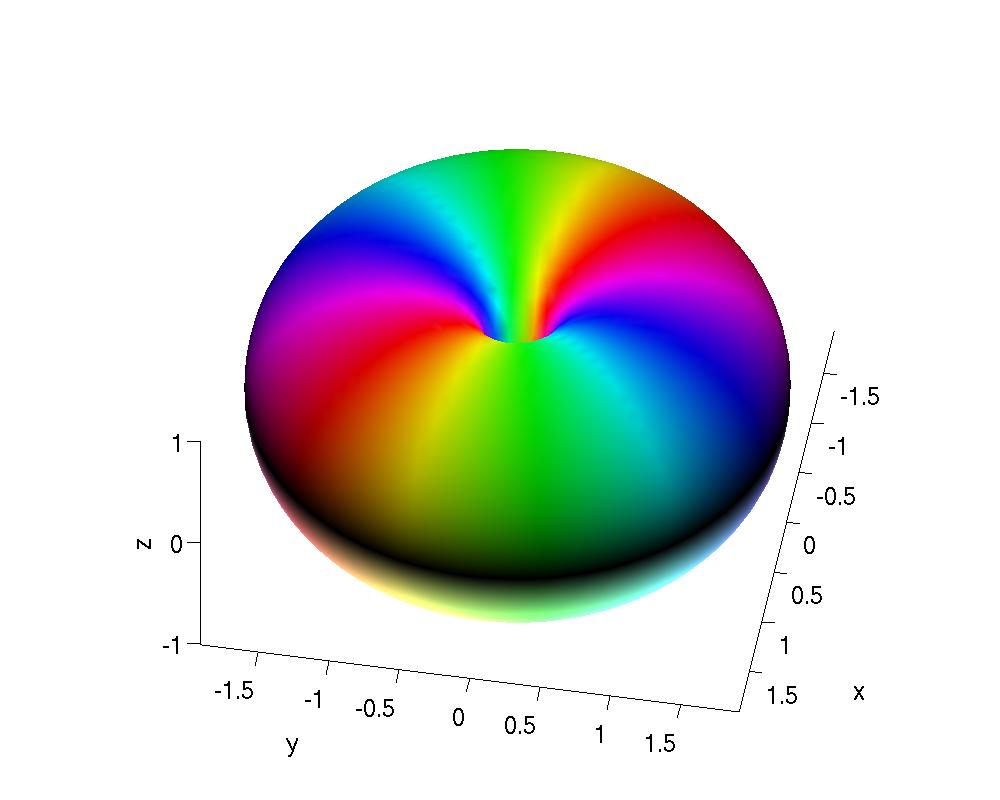}}
\subfloat[$(P,Q)=(3,1)$]{\includegraphics[width=0.32\linewidth]{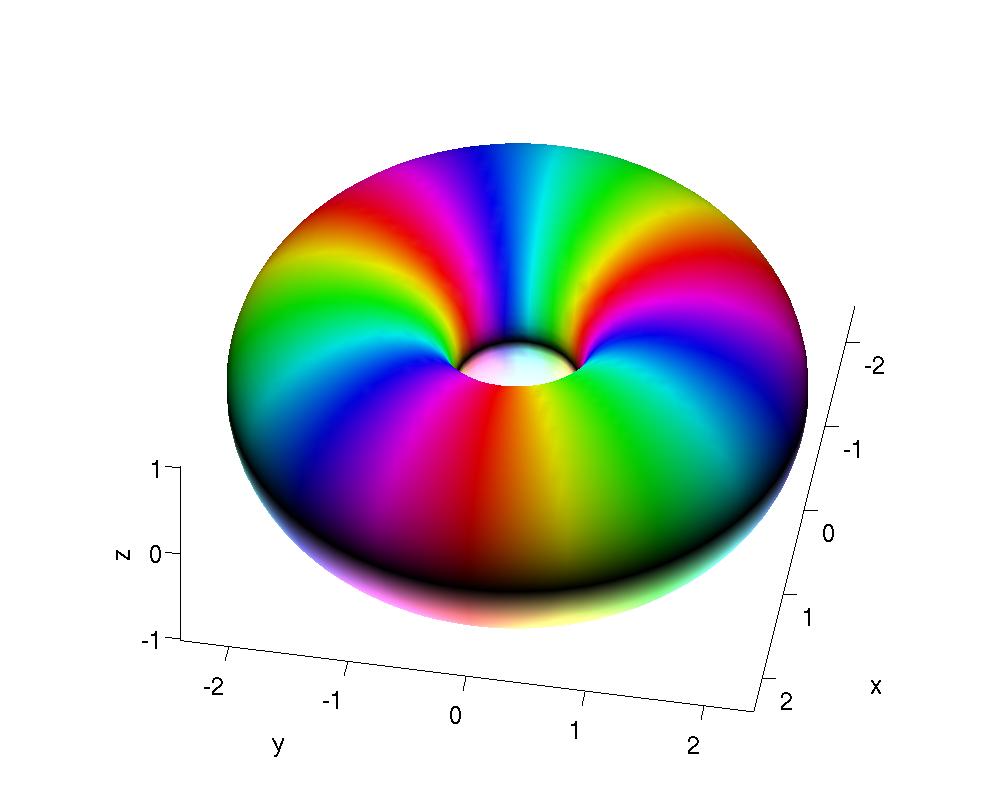}}}
\mbox{
\subfloat[$(P,Q)=(4,1)$]{\includegraphics[width=0.32\linewidth]{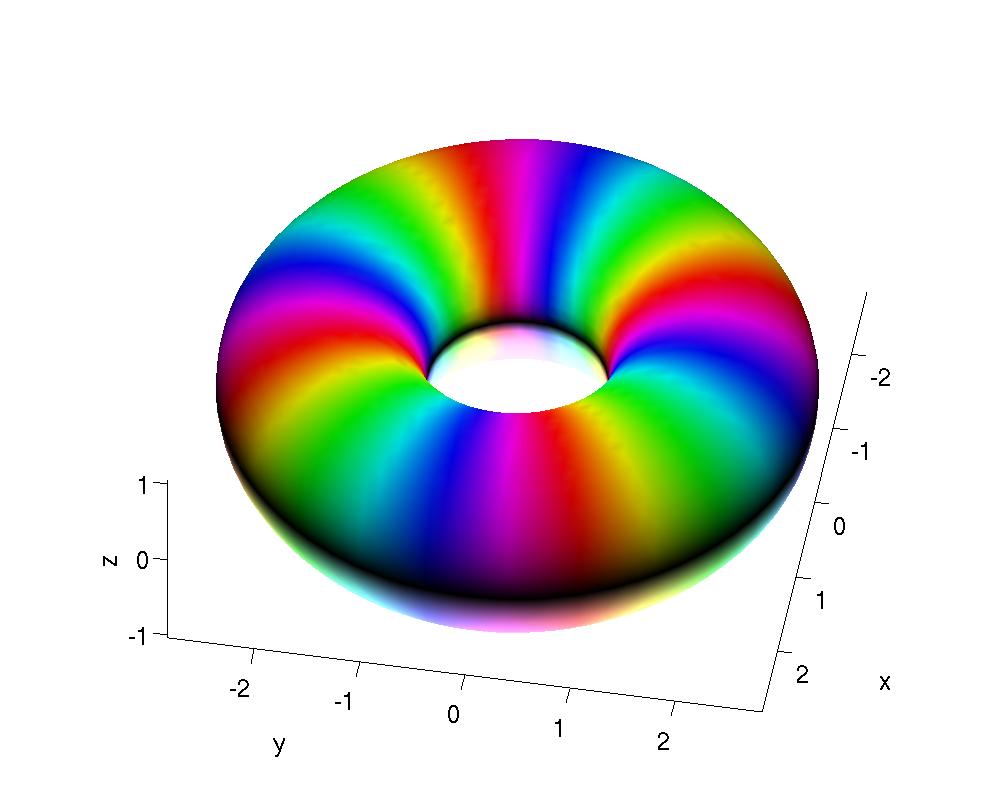}}
\subfloat[$(P,Q)=(5,1)$]{\includegraphics[width=0.32\linewidth]{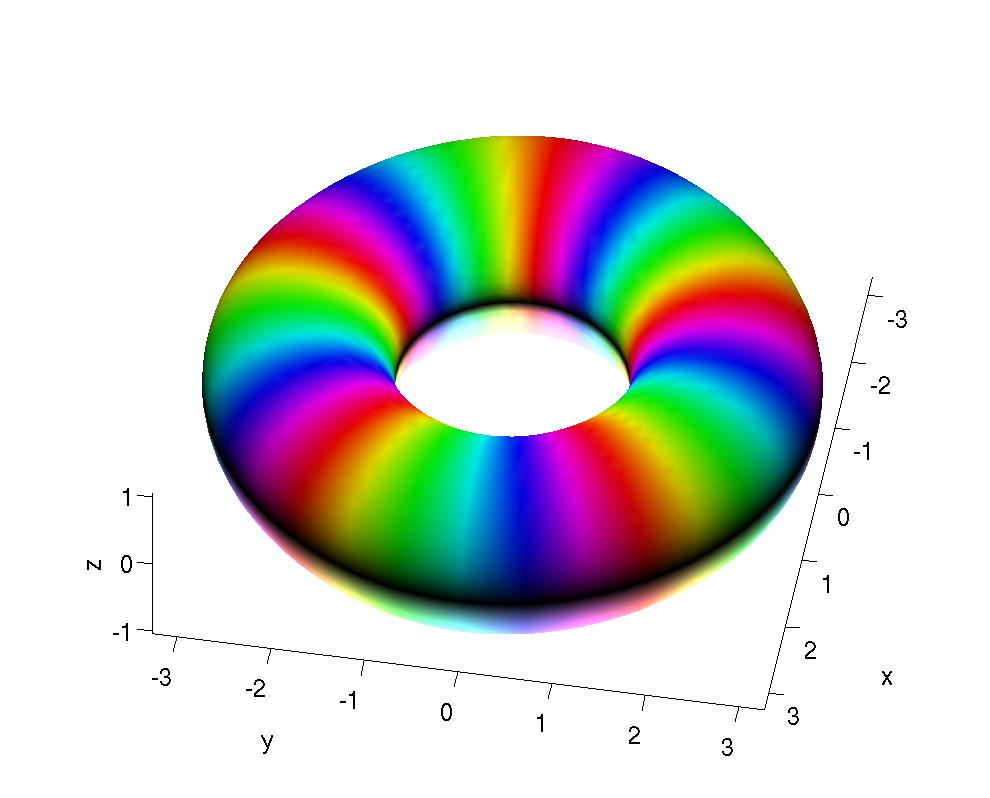}}
\subfloat[$(P,Q)=(6,1)$]{\includegraphics[width=0.32\linewidth]{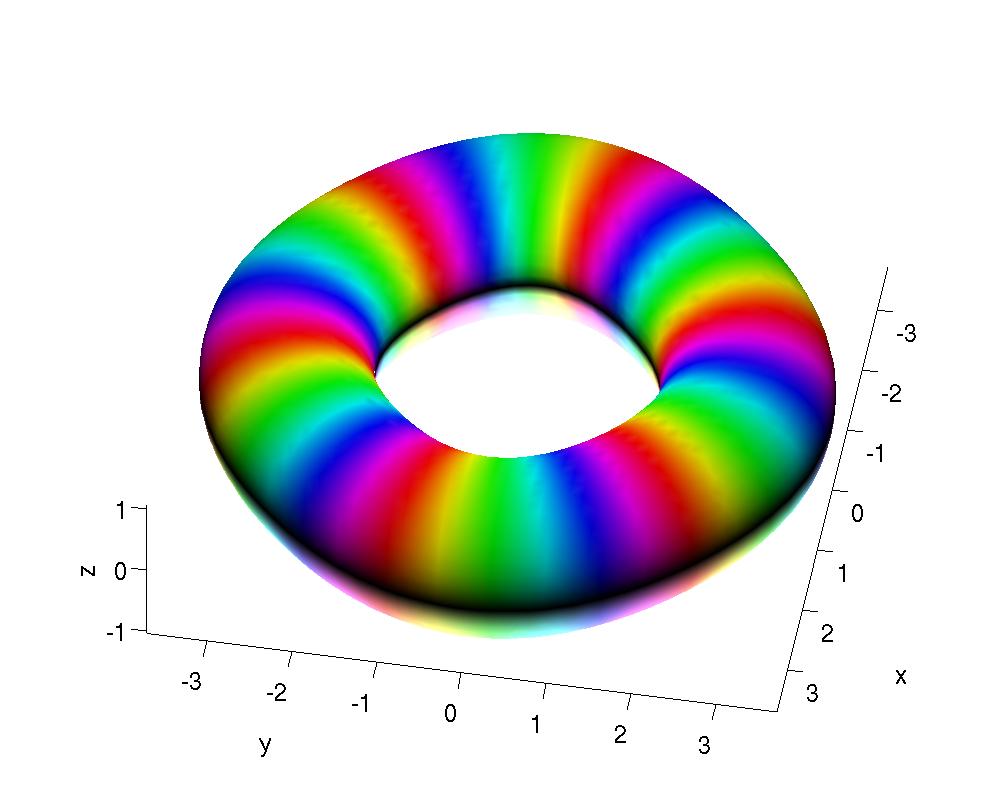}}}
\mbox{
\includegraphics[width=0.5\linewidth]{colorbar}}
\caption{
Isosurfaces showing the solutions for the 2+6 model, i.e.~for
$\kappa=0$ and $c_6=1$, at constant baryon charge density equal to half
its maximum value. 
The color represents the phase of the scalar field $\phi_2$ and the
lightness is given by $|\Im(\phi_1)|$. 
The calculations are done on an $81^3$ cubic lattice
with the relaxation method. 
}
\label{fig:t6}
\end{center}
\end{figure}

\begin{figure}[!htp]
\begin{center}
\captionsetup[subfloat]{labelformat=empty}
\mbox{
\subfloat[$(P,Q)=(1,1)$]{\includegraphics[width=0.49\linewidth]{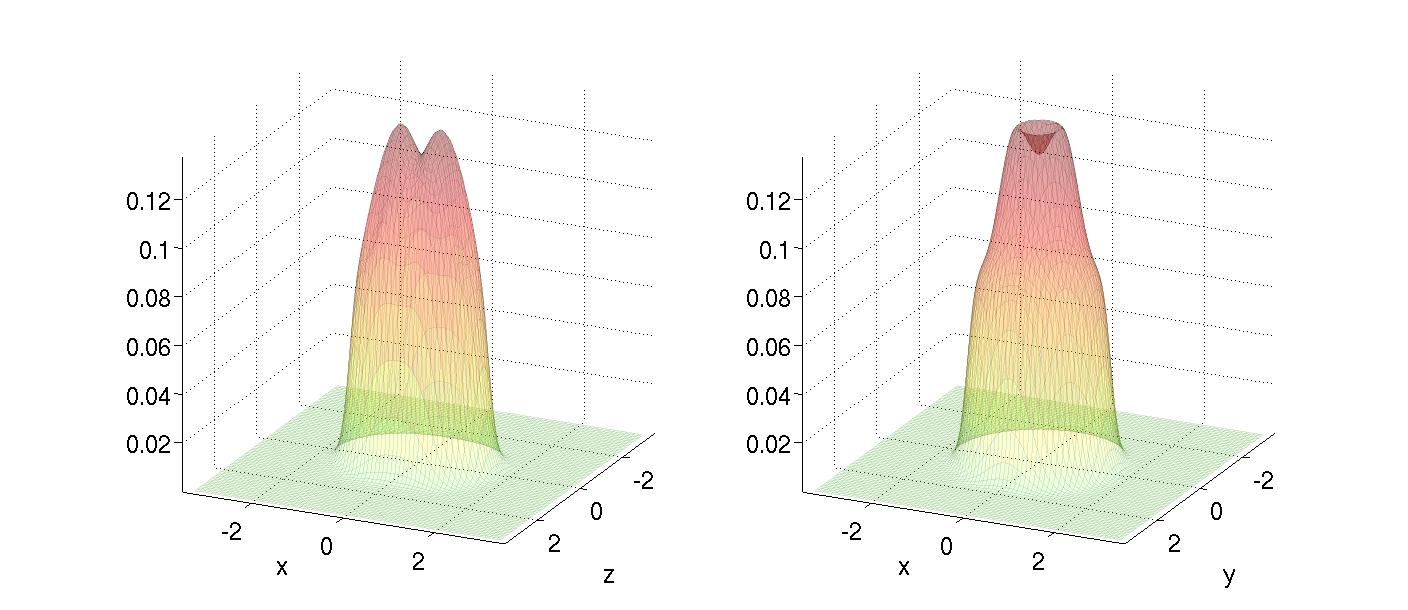}}
\subfloat[$(P,Q)=(2,1)$]{\includegraphics[width=0.49\linewidth]{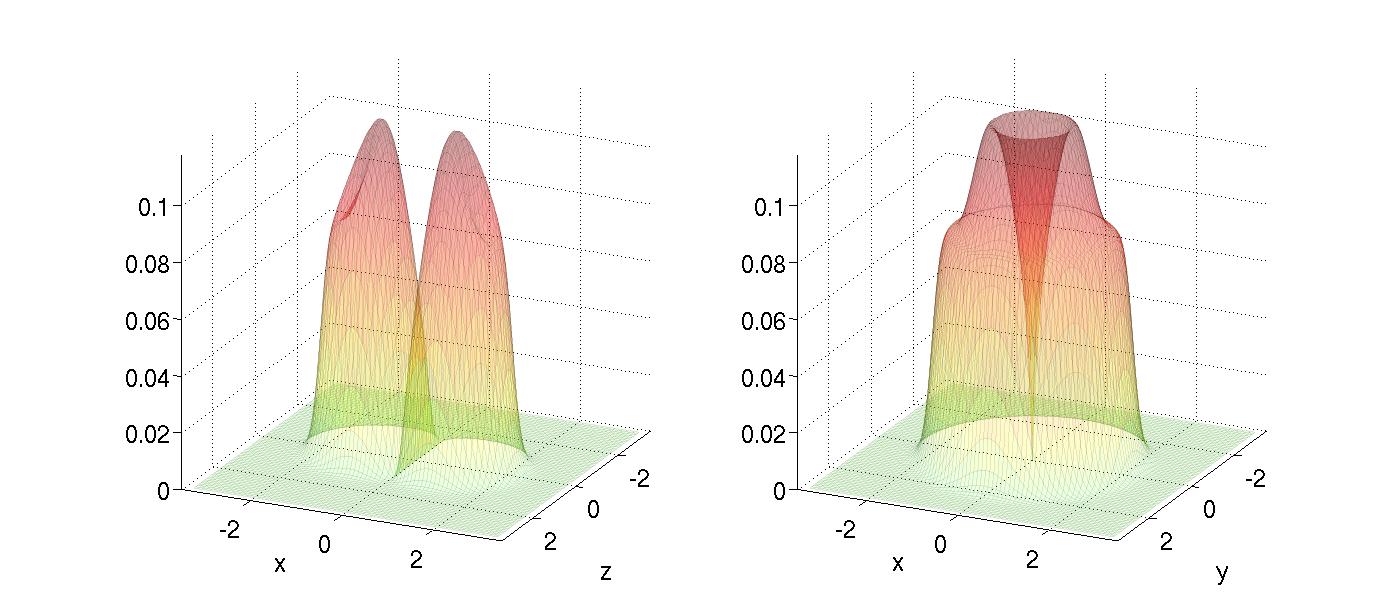}}}
\mbox{
\subfloat[$(P,Q)=(3,1)$]{\includegraphics[width=0.49\linewidth]{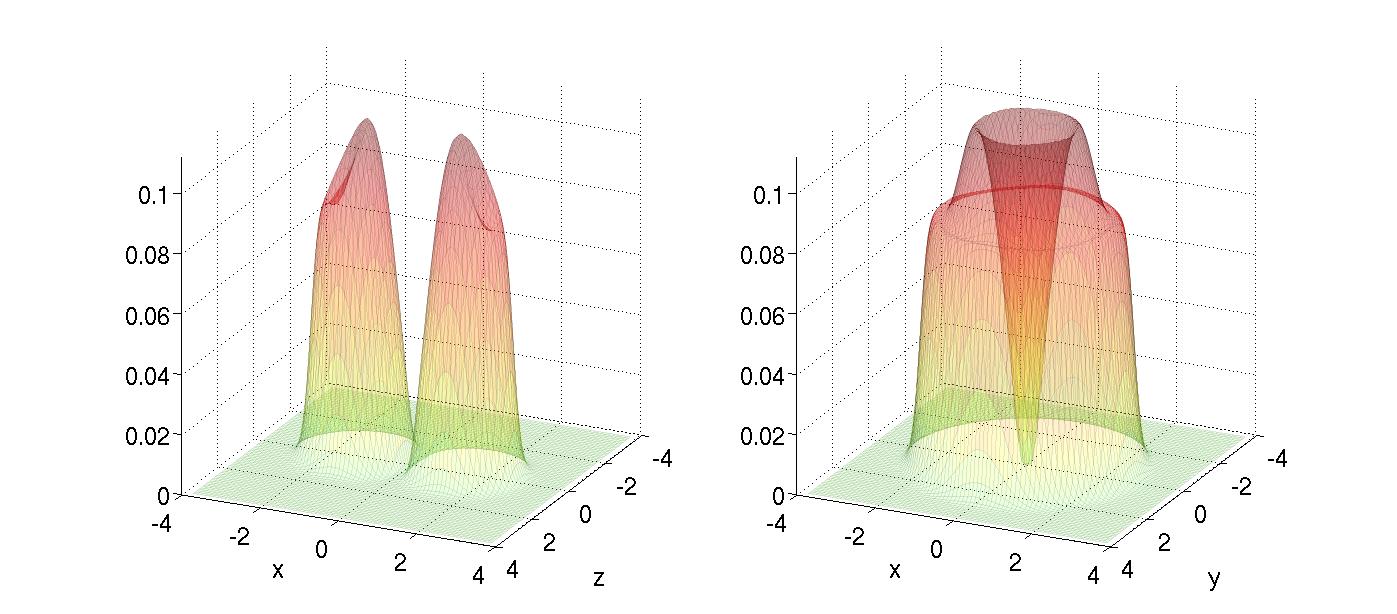}}
\subfloat[$(P,Q)=(4,1)$]{\includegraphics[width=0.49\linewidth]{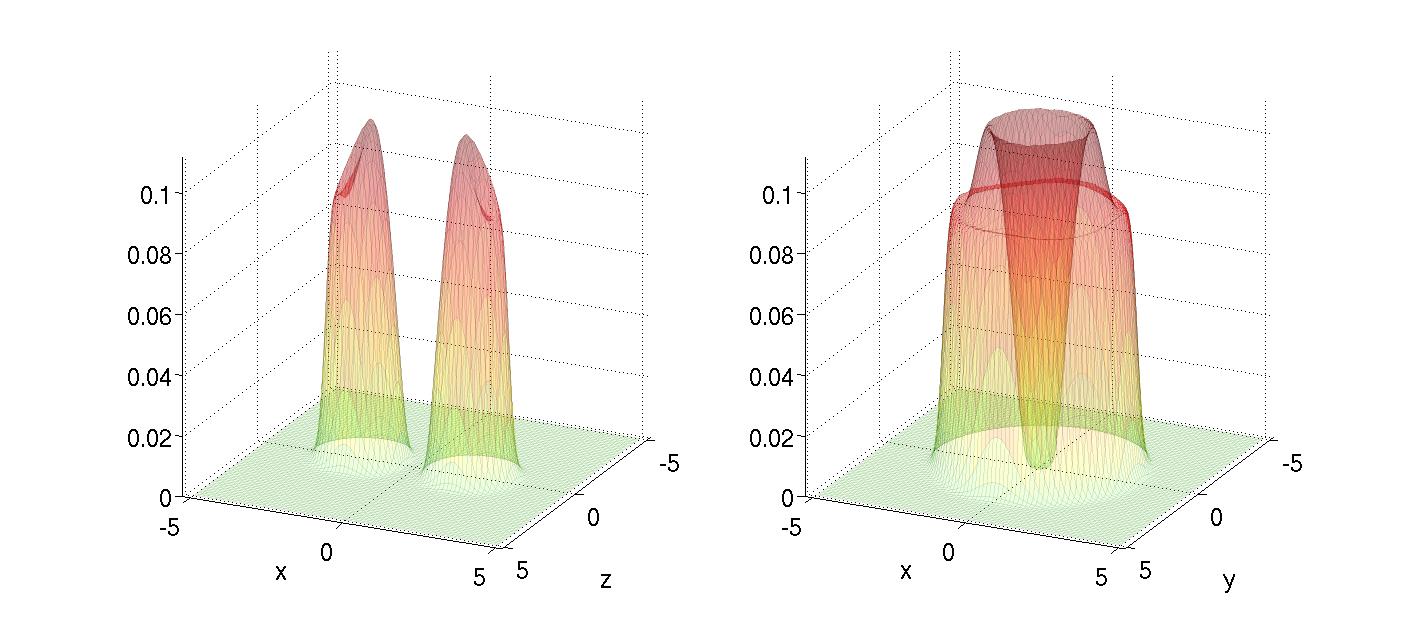}}}
\mbox{
\subfloat[$(P,Q)=(5,1)$]{\includegraphics[width=0.49\linewidth]{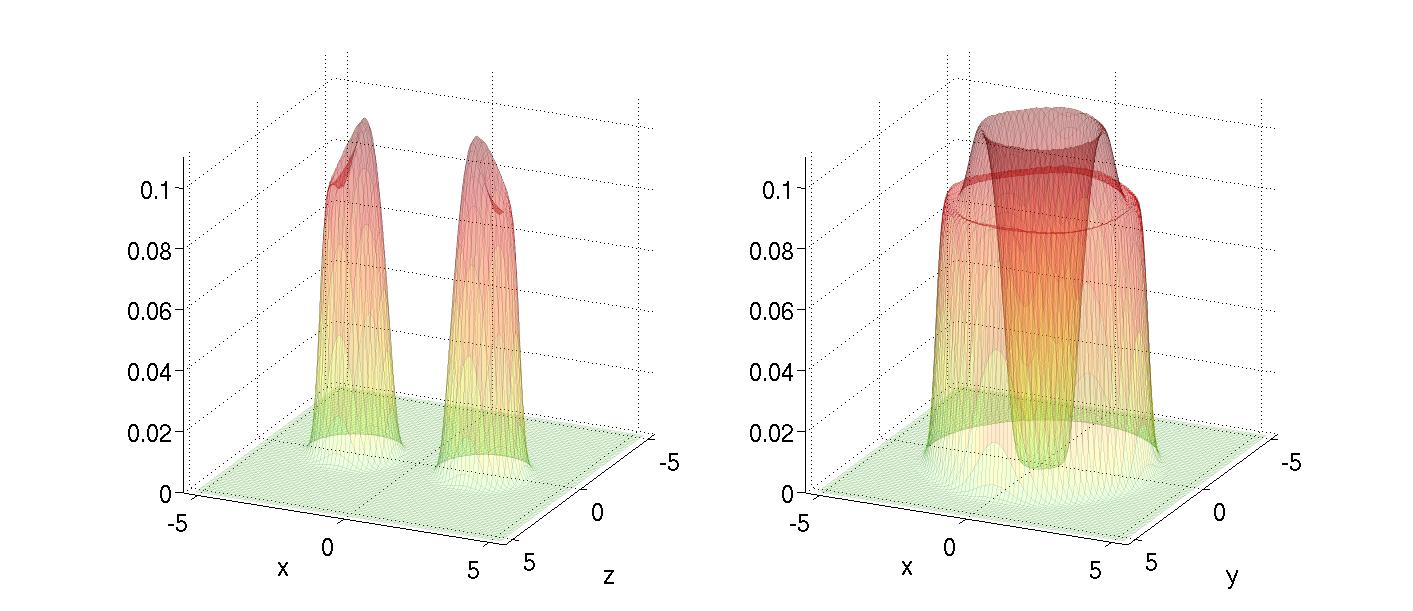}}
\subfloat[$(P,Q)=(6,1)$]{\includegraphics[width=0.49\linewidth]{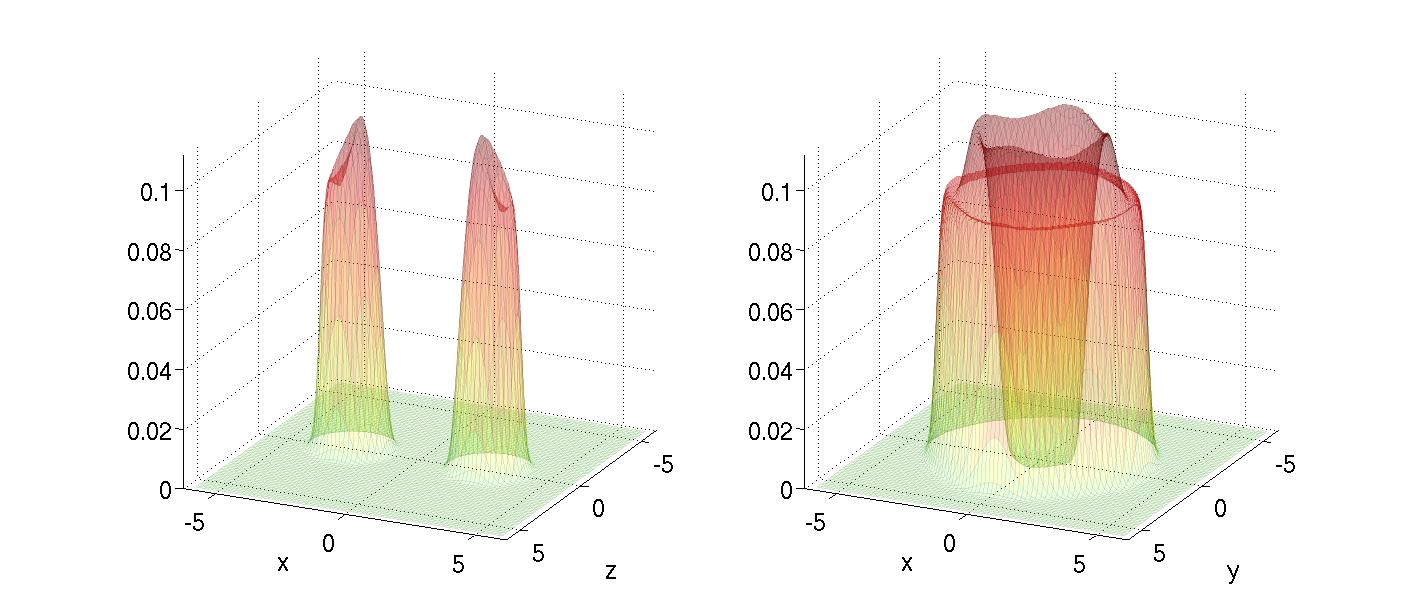}}}
\caption{
Baryon charge density for solutions in the 2+6 model, i.e.~with
$\kappa=0$ and $c_6=1$, at $xz$ slices (for $y=0$) and $xy$ slices (for
$z=0$). $yz$ slices are omitted as they are identical to the $xz$ 
slices by rotational symmetry of the torus. The calculations are done
on an $81^3$ cubic lattice with the relaxation method. 
}
\label{fig:t6_baryonslice}
\end{center}
\end{figure}

\begin{figure}[!htp]
\begin{center}
\captionsetup[subfloat]{labelformat=empty}
\mbox{
\subfloat[$(P,Q)=(1,1)$]{\includegraphics[width=0.49\linewidth]{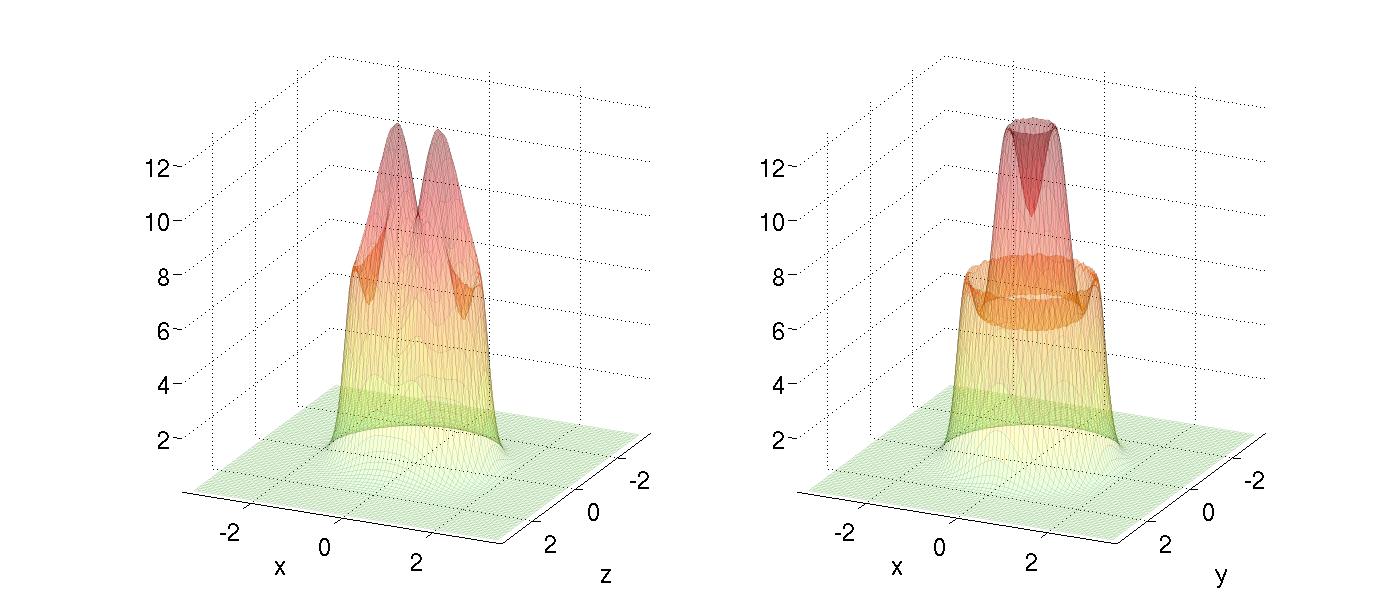}}
\subfloat[$(P,Q)=(2,1)$]{\includegraphics[width=0.49\linewidth]{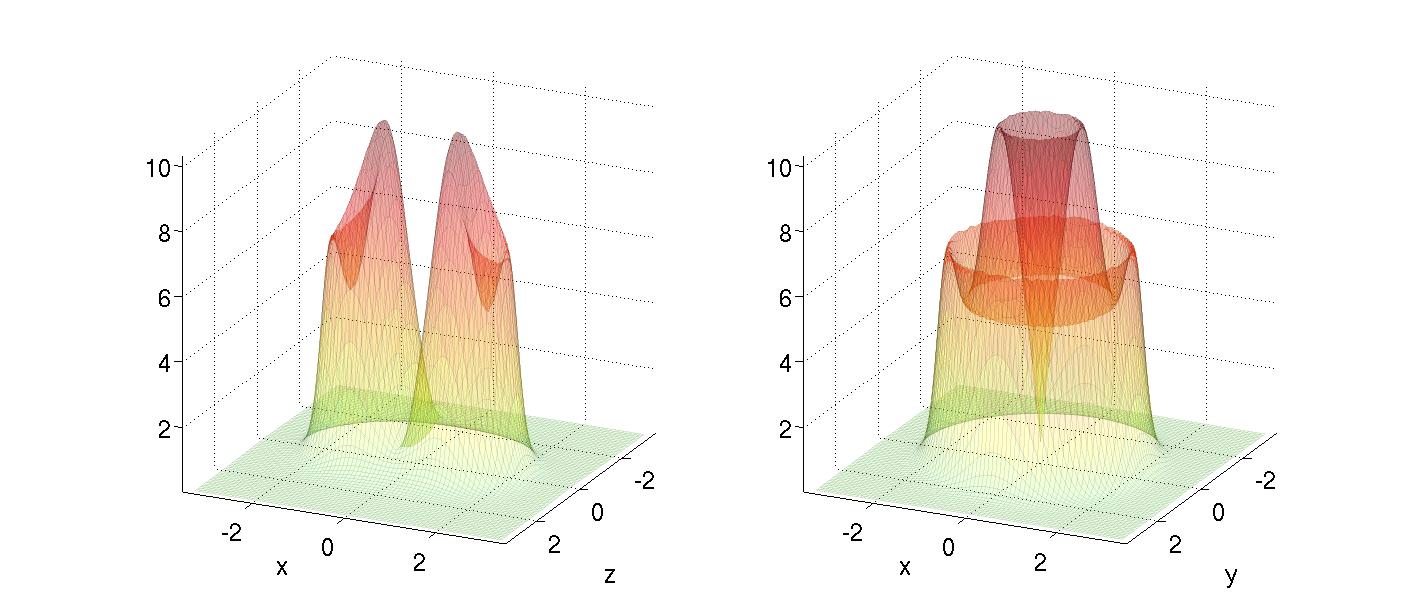}}}
\mbox{
\subfloat[$(P,Q)=(3,1)$]{\includegraphics[width=0.49\linewidth]{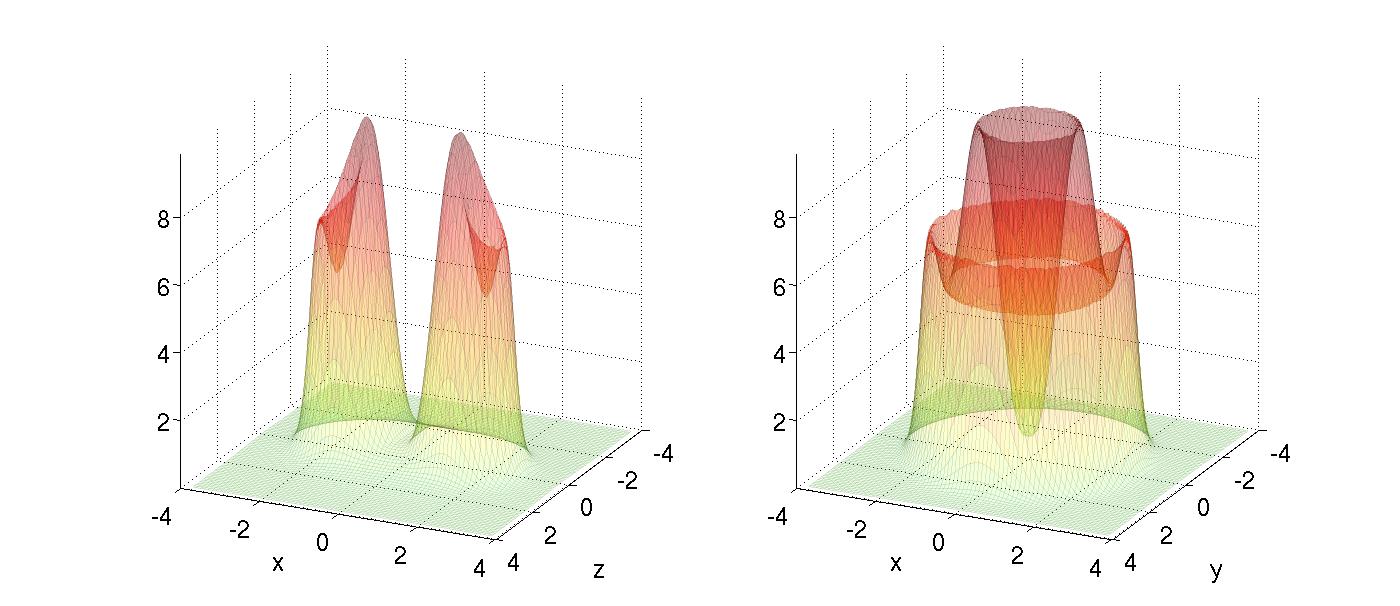}}
\subfloat[$(P,Q)=(4,1)$]{\includegraphics[width=0.49\linewidth]{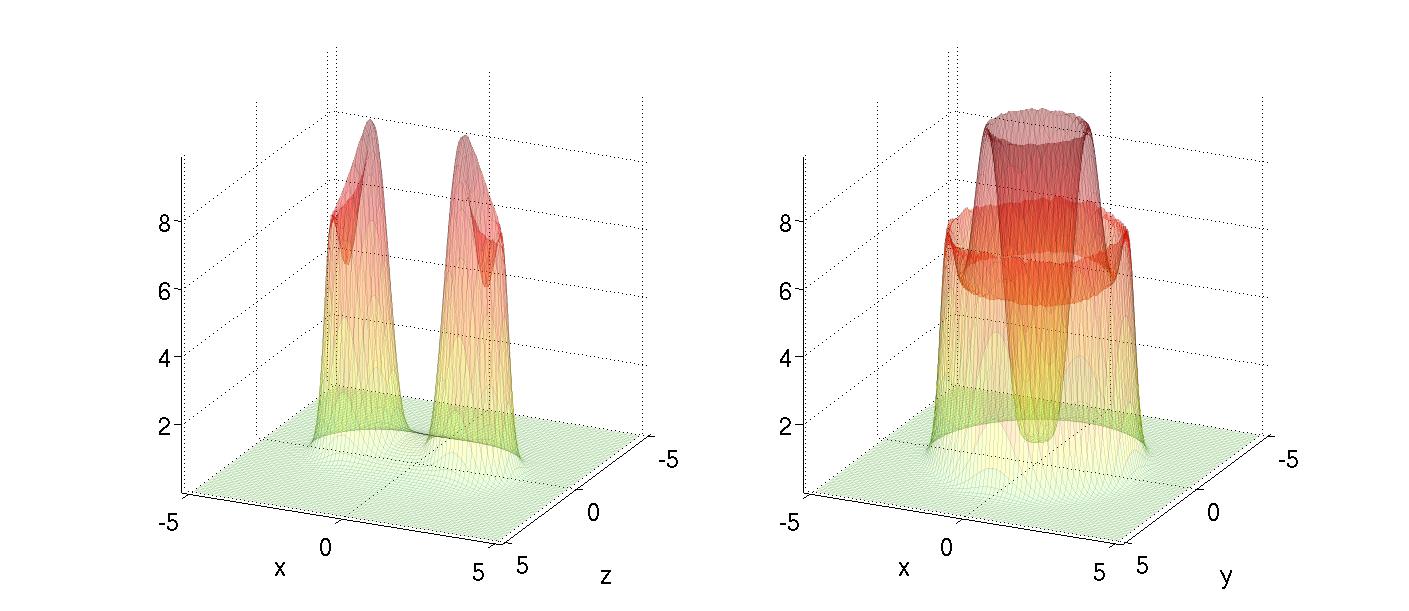}}}
\mbox{
\subfloat[$(P,Q)=(5,1)$]{\includegraphics[width=0.49\linewidth]{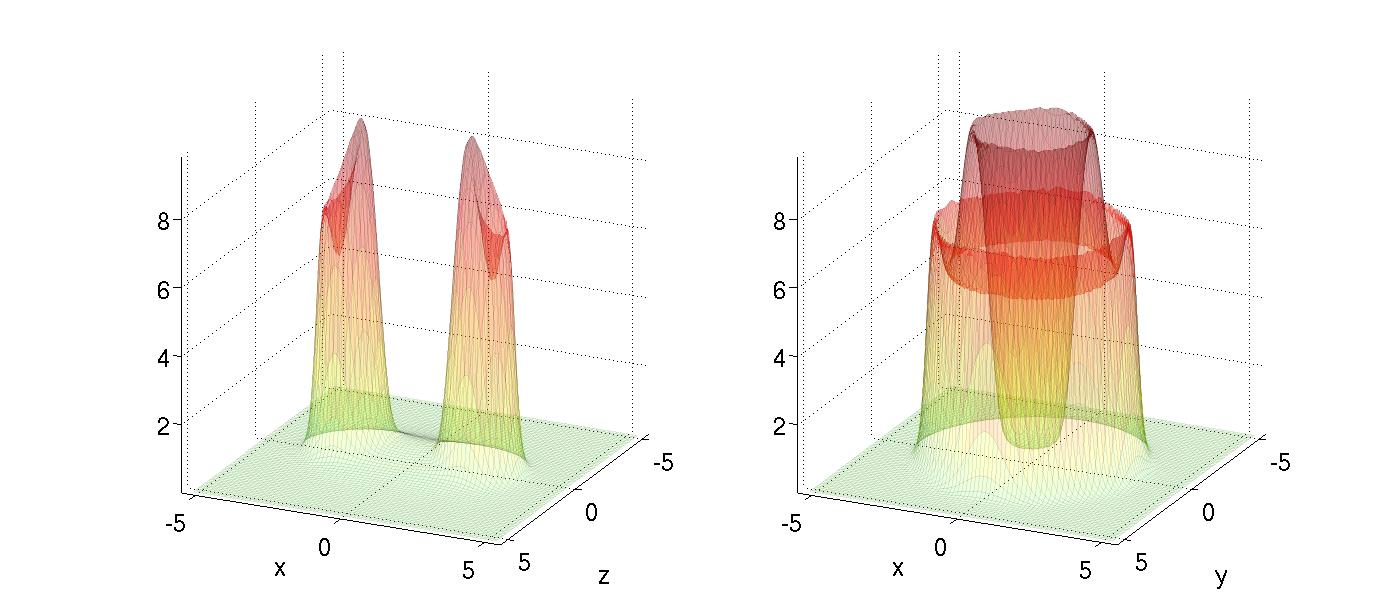}}
\subfloat[$(P,Q)=(6,1)$]{\includegraphics[width=0.49\linewidth]{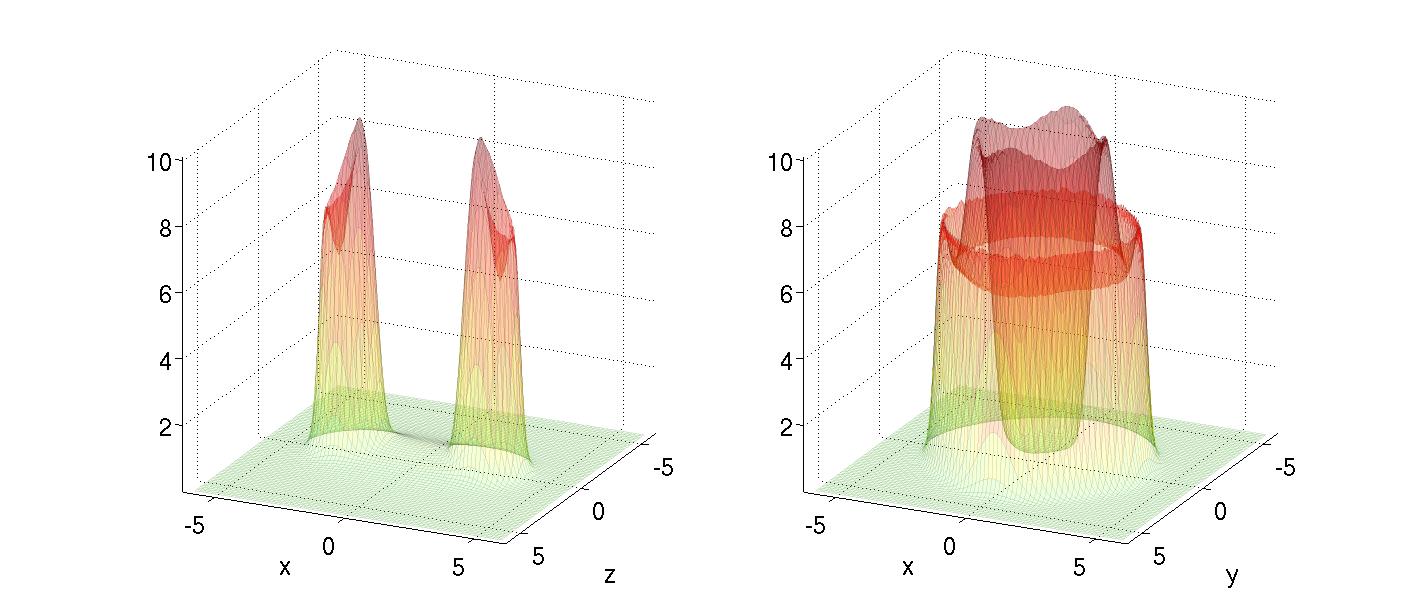}}}
\caption{
Energy density for solutions in the 2+6 model, i.e.~with $\kappa=0$
and $c_6=1$, at $xz$ slices (for $y=0$) and $xy$ slices (for
$z=0$). $yz$ slices are omitted as they are identical to the $xz$ 
slices by rotational symmetry of the torus. The calculations are done
on an $81^3$ cubic lattice with the relaxation method. 
}
\label{fig:t6_energyslice}
\end{center}
\end{figure}

Next we will turn to the case of the 2+6 model, i.e.~with only
sixth-order derivative terms ($\kappa=0$ and $c_6=1$) and again
with a mass of $m=4$. Numerical solutions are shown in
Figs.~\ref{fig:t6}, \ref{fig:t6_baryonslice} and
\ref{fig:t6_energyslice}. As in the previous case, we show the
3-dimensional isosurfaces of the baryon charge density at half the
maximum value in Fig.~\ref{fig:t6}. 
In Figs.~\ref{fig:t6_baryonslice} and \ref{fig:t6_energyslice} are
shown the baryon charge density and energy density, respectively, at
two different cross sections cutting the torus through the origin. 
Notice that the energy densities for these solutions are somewhat more 
complex than their respective baryon charge densities. This is one
difference between the 2+6 model and the 2+4 model. The second
difference is that in this case, the torus shape is vaguely visible 
already for $P=1$, whereas for the previous case $P=1$ has (unbroken)
spherical symmetry. 
Let us also comment on the circular shape of the torus for the
$(P,Q)=(6,1)$ solution along the toroidal direction in
Fig.~\ref{fig:t6}; this flattening out of the circle is not aligned
with the lattice, but is at almost 45 degrees to the lattice
axis. Since the small $P$ solutions do possess almost perfect circular
symmetry, we believe that this is not a lattice effect, but instead
signals metastability of the string: for high enough $B=P$ the string
wants to collapse and break up. The same effect can also be observed
in the $(P,Q)=(6,1)$ solution in Fig.~\ref{fig:t6_energyslice} on the
$xy$ slice where the energy density displays four distinct wave tops
around the toroidal cycle. 

We again check the numerical precision by numerically evaluating the
total baryon charge, see Table \ref{tab:EB_T6}. As for the stability
of the higher $P>1$ solutions, we numerically evaluate the energy
(mass) of the solutions and again find that the energy decreases as
$P$ is increased, for the first few solutions, but this time only for
the first three $P=1,2,3$ and then it starts to increase slightly. The
first five solutions are all energetically \emph{stable} while $P=6$
is only metastable.

\begin{table}[!htp]
\begin{center}
\caption{Numerically integrated baryon charge and energy (mass) for
  the solutions in the 2+6 model. Stability is observed for the first
  five solutions whilst $P=6$ is only energetically metastable. }
\label{tab:EB_T6}
\begin{tabular}{c||cr@{$\,\pm\,$}l}
$B$ & $B^{\rm numerical}$ & \multicolumn{2}{c}{$E^{\rm numerical}/B$} \\
\hline\hline
$1$ & $0.9999$ & $100.8613$ & $0.0410$ \\
$2$ & $1.9998$ & $89.7184$ & $0.0532$ \\
$3$ & $2.9995$ & $87.3095$ & $0.1871$ \\
$4$ & $3.9981$ & $87.5179$ & $0.0721$ \\
$5$ & $4.9970$ & $87.5560$ & $0.0901$ \\
$6$ & $5.9939$ & $88.1414$ & $0.1145$
\end{tabular}
\end{center}
\end{table}

\section{Transition to Toroidal Skyrmions
\label{sec:transition}}

In this section we study the transition from the normal Skyrmion of
higher charge (i.e.~with $m=0$) to the toroidal Skyrmion (i.e.~with
$m$ sufficiently large). For concreteness, we study the transition in
the normal Skyrme model ($\kappa=1$ and $c_6=0$) and for $B=3$ where the
transition is very visible (as opposed to for instance $B=1$ and
$B=2$). 
When the potential is turned off, the $B=3$ Skyrmion in the normal
Skyrme model is of tetrahedral shape \cite{Battye:1997qq}. 
Turning on the potential \eqref{eq:potential}, vorton-like Skyrmions
become the lowest-energy state for a sufficiently large mass
parameter, $m$. 
In order to find the critical mass necessary for obtaining torii or
global strings in the Skyrme model, we vary the mass parameter and
repeat the numerical calculation. 
We are using the relaxation method to find numerical solutions. One
weakness of this method is that it only finds the nearest \emph{local} 
minimal-energy solution, as opposed to the \emph{global} one. 
For this reason we make two series of numerical calculations: one
starting from the tetrahedral solution, whose initial guess is
\cite{Houghton:1997kg}
\beq
\mathbf{n} = \left\{
\frac{R + \bar{R}}{1+R\bar{R}}\sin f,
\frac{i(\bar{R}-R)}{1+R\bar{R}}\sin f,
\frac{1 - R\bar{R}}{1+R\bar{R}}\sin f,
\cos f\right\}, 
\eeq
where $R$ is the rational map Ansatz and for $B=3$ the tetrahedral
Ansatz is \cite{Houghton:1997kg}
\beq
R = \frac{z^3 - \sqrt{3}iz}{\sqrt{3}i z^2 - 1}, \qquad
z = \tan\left(\frac{\theta}{2}\right)e^{i\phi},
\label{eq:tetrahedral_ansatz}
\eeq
where $\theta,\phi$ are angles on the 2-sphere. 
The other series of numerical solutions use the initial guess provided
by the torus Ansatz of Eq.~\eqref{eq:torus_reduced}. 

\begin{figure}[!htp]
\begin{center}
\captionsetup[subfloat]{labelformat=empty}
\mbox{
\subfloat[$m=0$, $B^{\rm numerical}=2.9977$]{\includegraphics[width=0.33\linewidth]{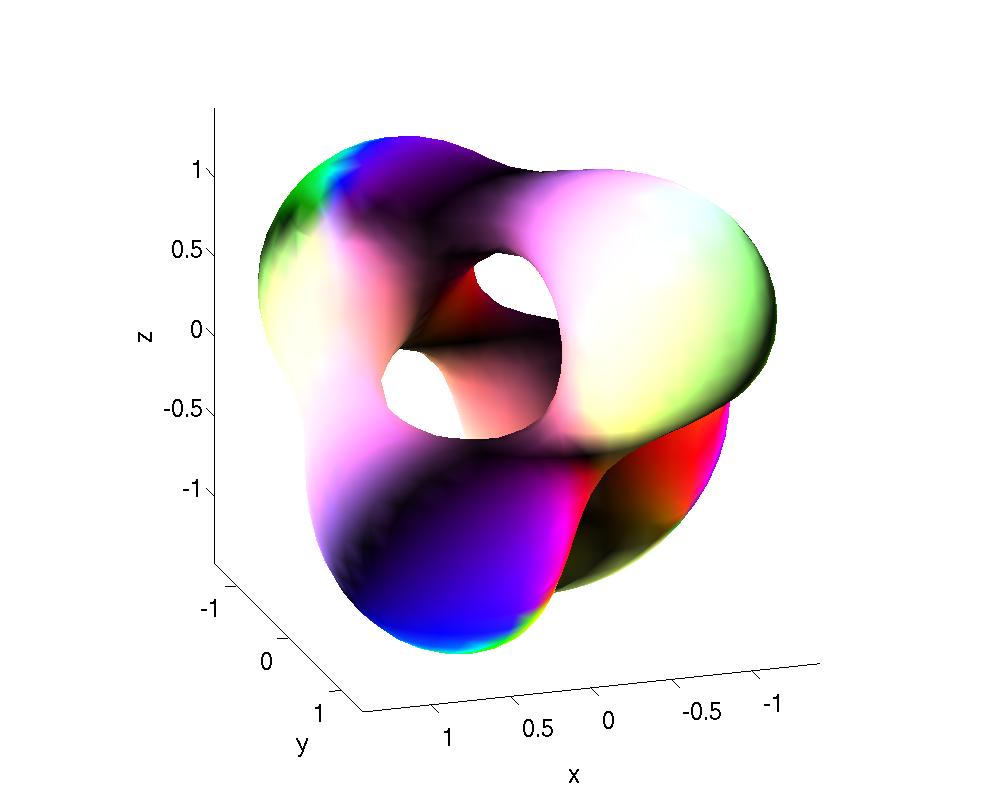}}
\subfloat[$m=1/2$, $B^{\rm numerical}=2.9983$]{\includegraphics[width=0.33\linewidth]{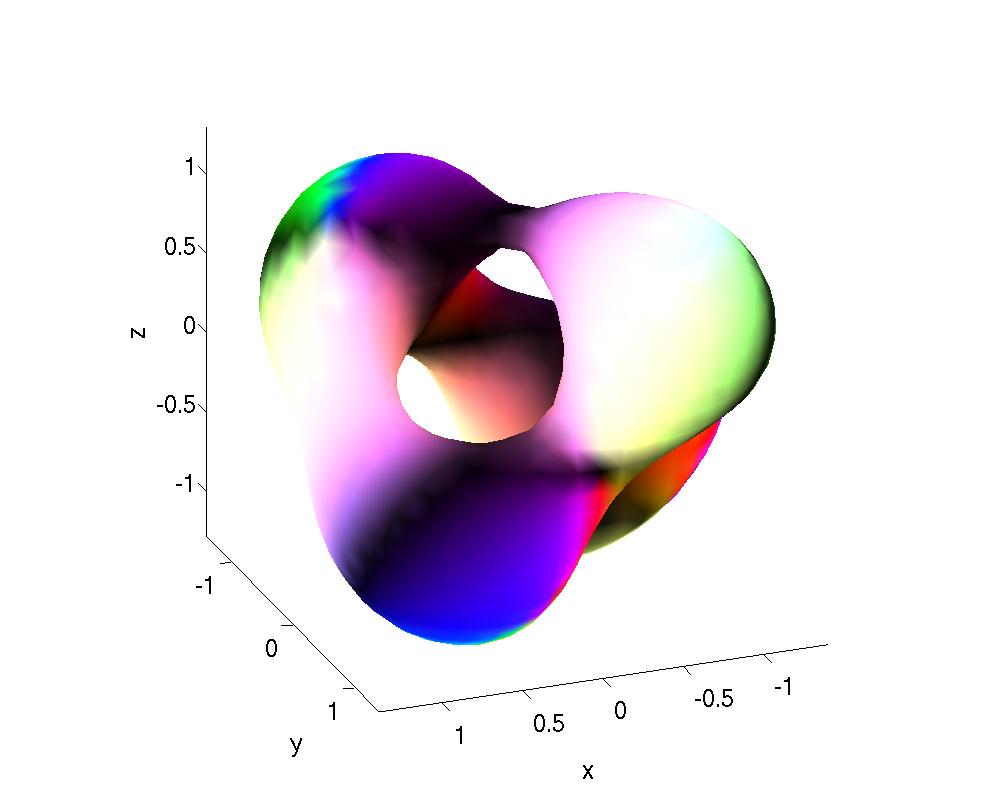}}
\subfloat[$m=1$, $B^{\rm numerical}=2.9986$]{\includegraphics[width=0.33\linewidth]{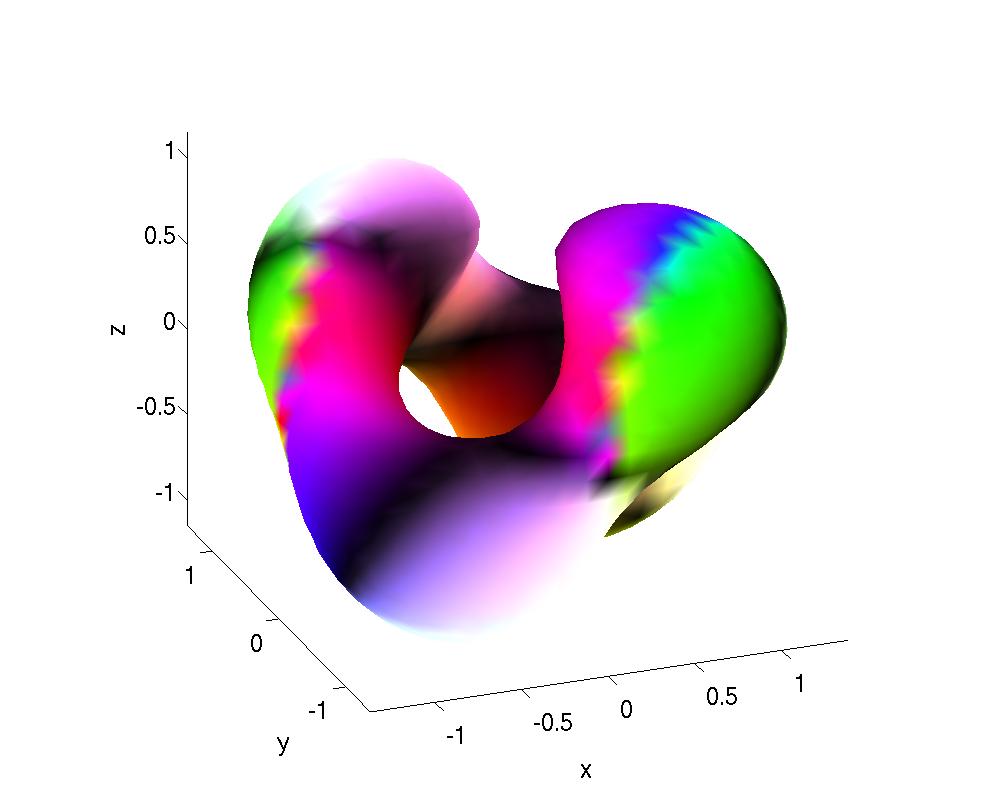}}}
\mbox{
\subfloat[$m=3/2$, $B^{\rm numerical}=2.9993$]{\includegraphics[width=0.33\linewidth]{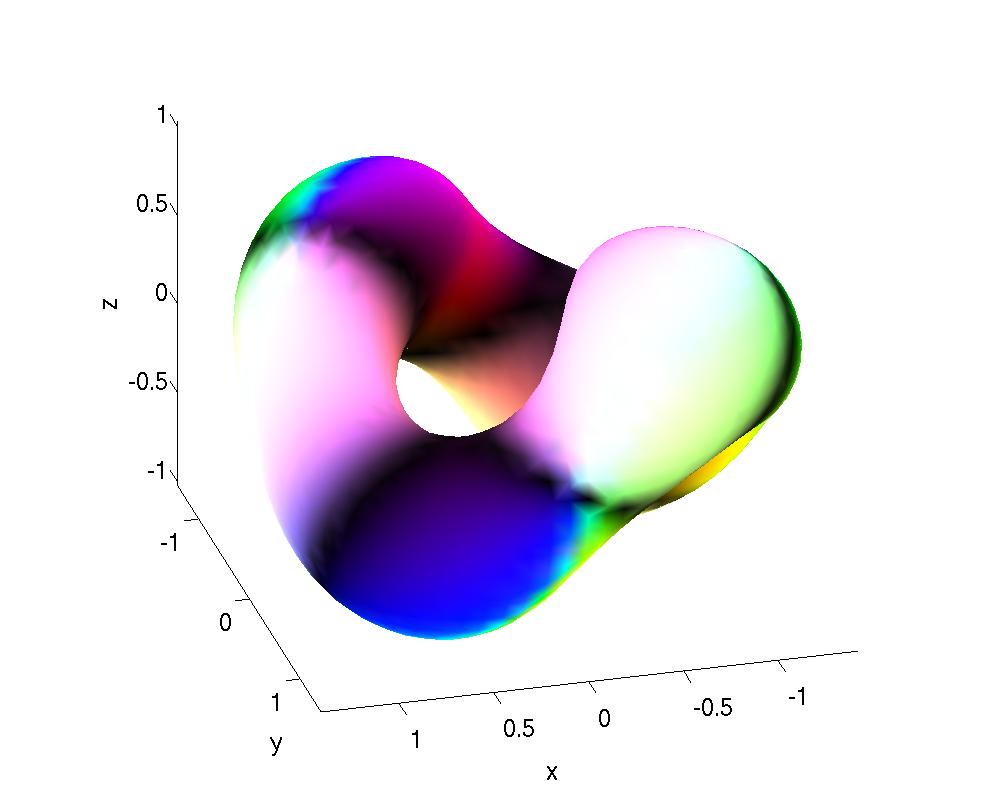}}
\subfloat[$m=2$, $B^{\rm numerical}=2.9986$]{\includegraphics[width=0.33\linewidth]{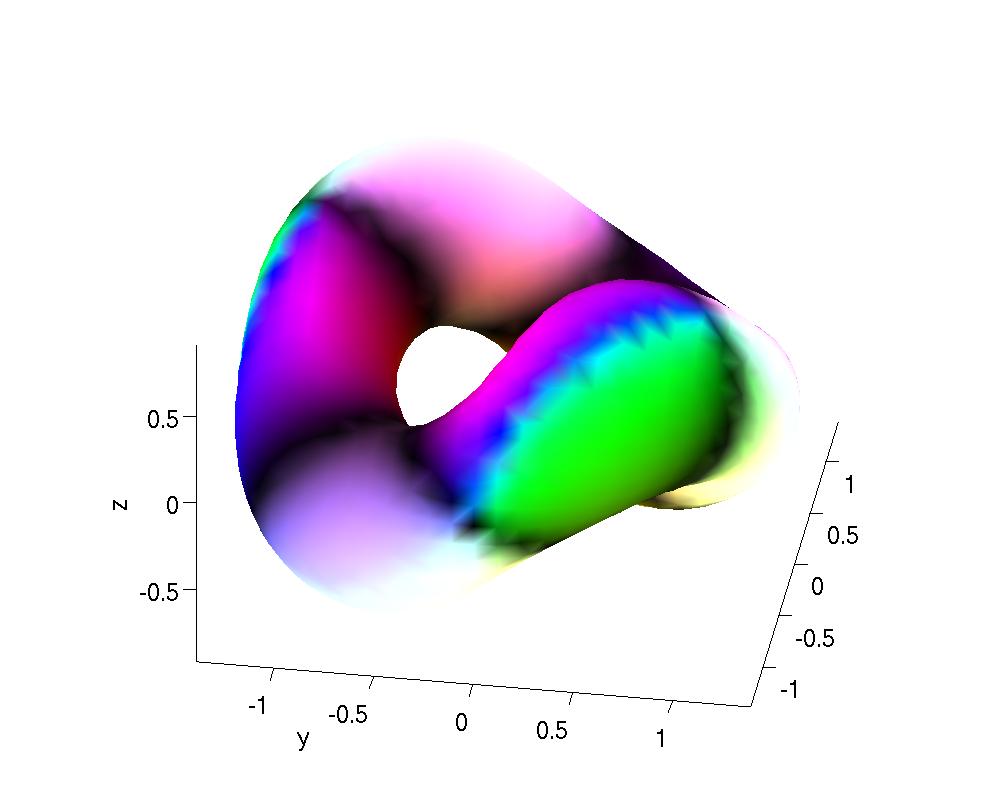}}
\subfloat[$m=3$, $B^{\rm numerical}=2.9982$]{\includegraphics[width=0.33\linewidth]{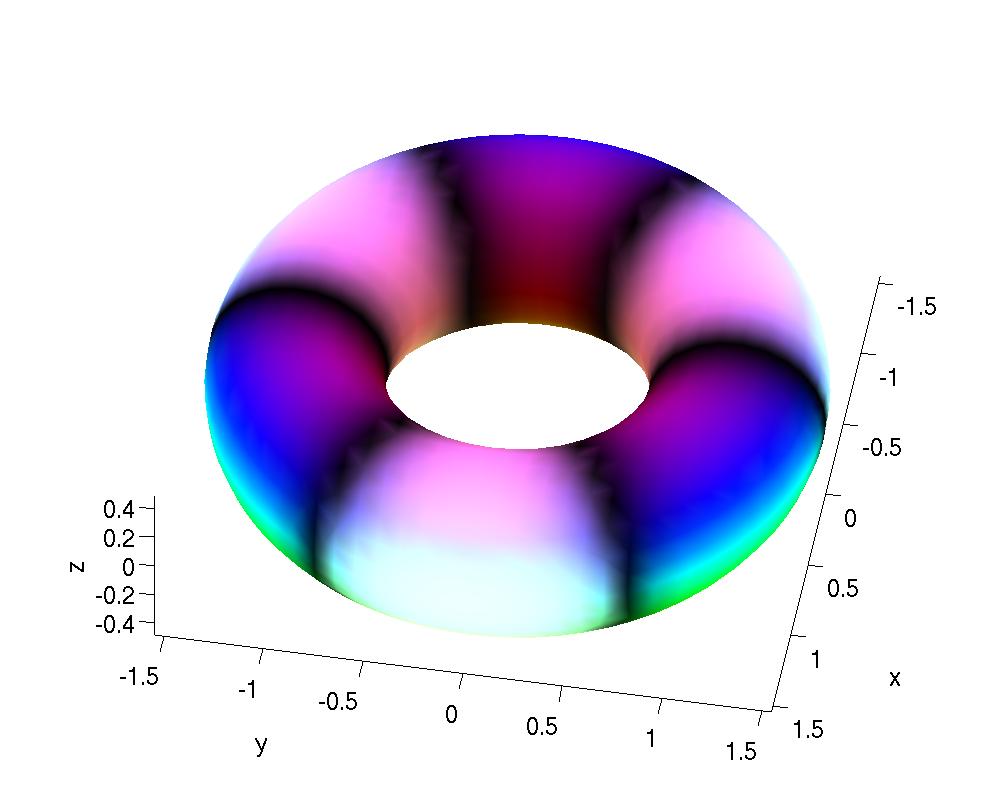}}}
\caption{Isosurfaces showing the solutions for the 2+4 model, 
$B=3$ and various values of the mass parameter, $m$, and the
tetrahedral Ansatz \eqref{eq:tetrahedral_ansatz} as initial guess for
the relaxation. 
The color represents the phase of the scalar field $\phi_2$ and the
lightness is given by $|\Im(\phi_1)|$. 
The calculations are done on an $81^3$ cubic lattice
with the relaxation method. }
\label{fig:T4B3_TEIC}
\end{center}
\end{figure}

\begin{figure}[!htp]
\begin{center}
\captionsetup[subfloat]{labelformat=empty}
\mbox{
\subfloat[$m=0$, $B^{\rm numerical}=2.9980$]{\includegraphics[width=0.33\linewidth]{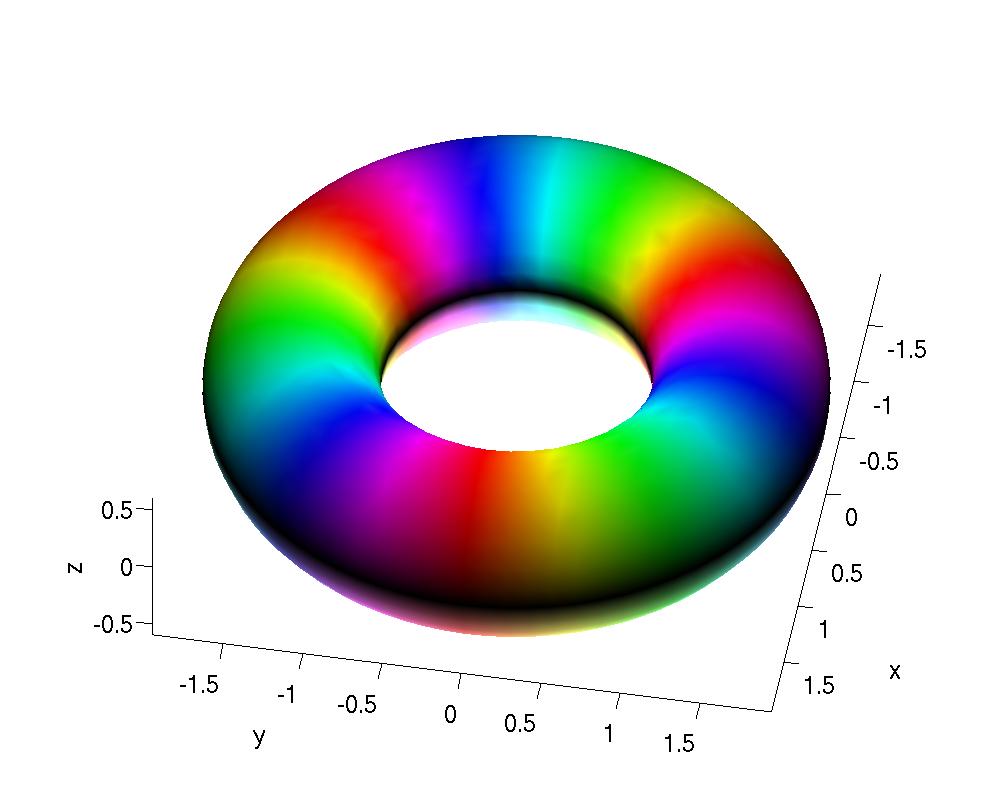}}
\subfloat[$m=1/2$, $B^{\rm numerical}=2.9983$]{\includegraphics[width=0.33\linewidth]{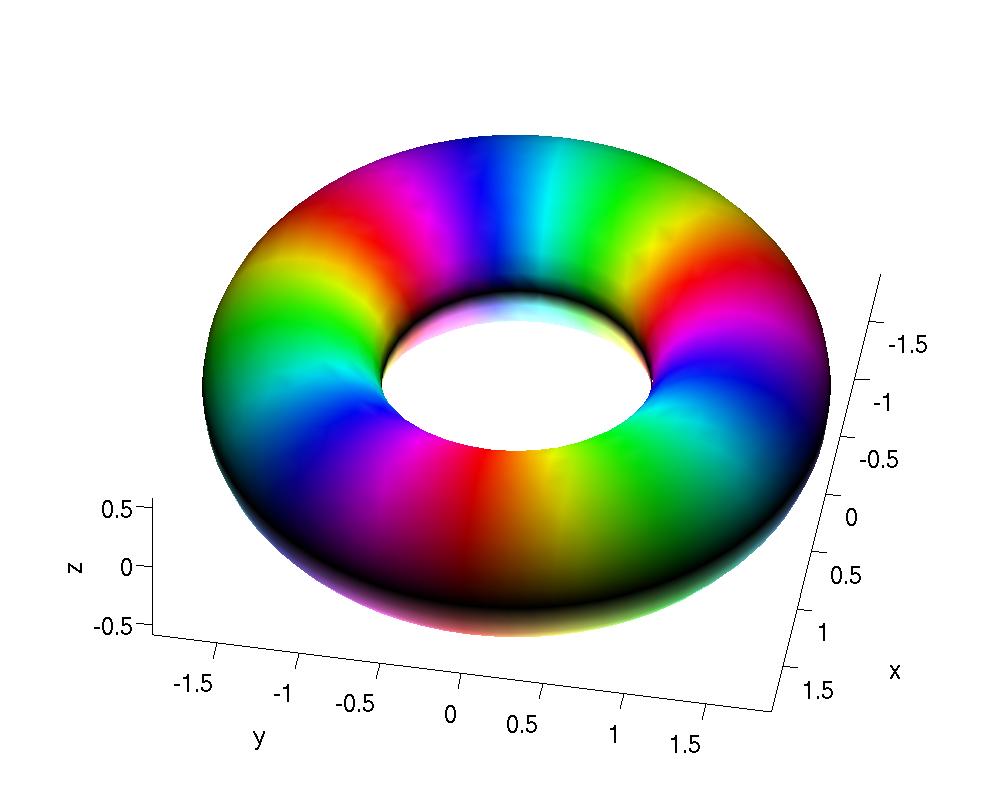}}
\subfloat[$m=1$, $B^{\rm numerical}=2.9987$]{\includegraphics[width=0.33\linewidth]{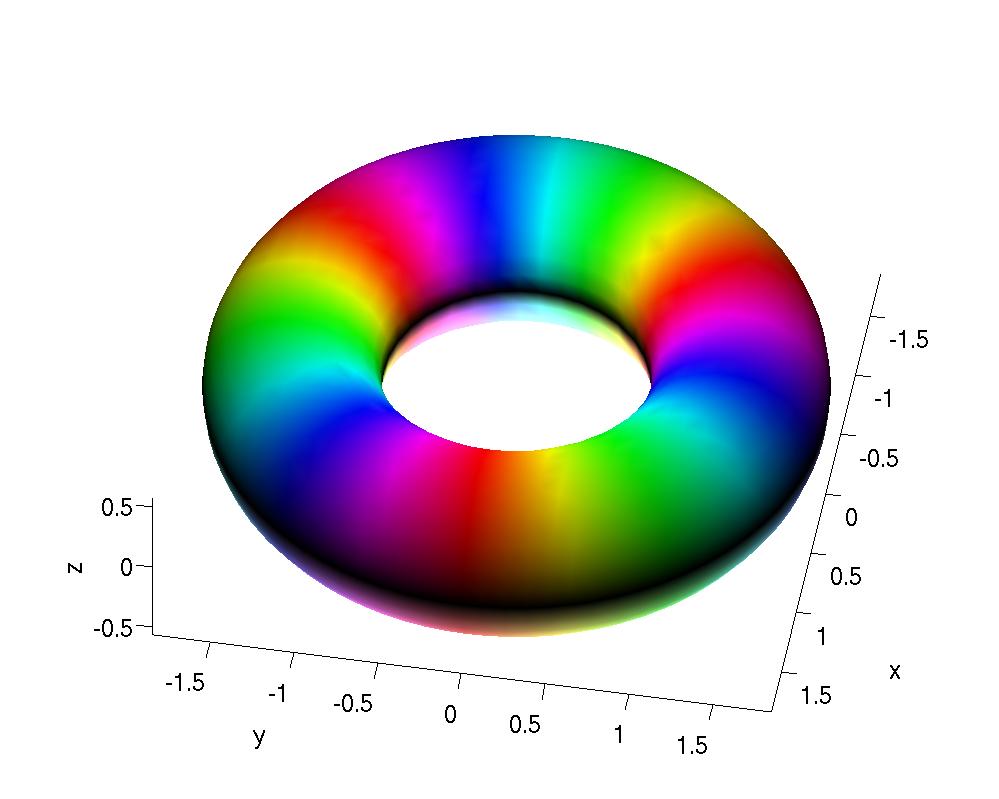}}}
\mbox{
\subfloat[$m=3/2$, $B^{\rm numerical}=2.9992$]{\includegraphics[width=0.33\linewidth]{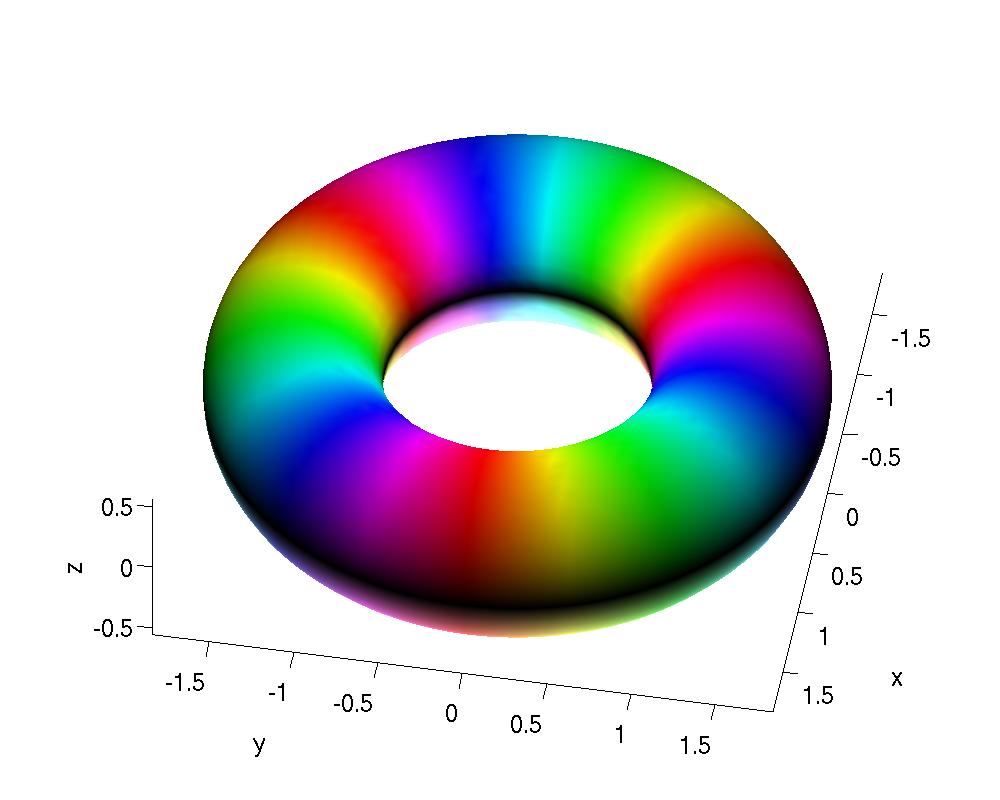}}
\subfloat[$m=2$, $B^{\rm numerical}=2.9989$]{\includegraphics[width=0.33\linewidth]{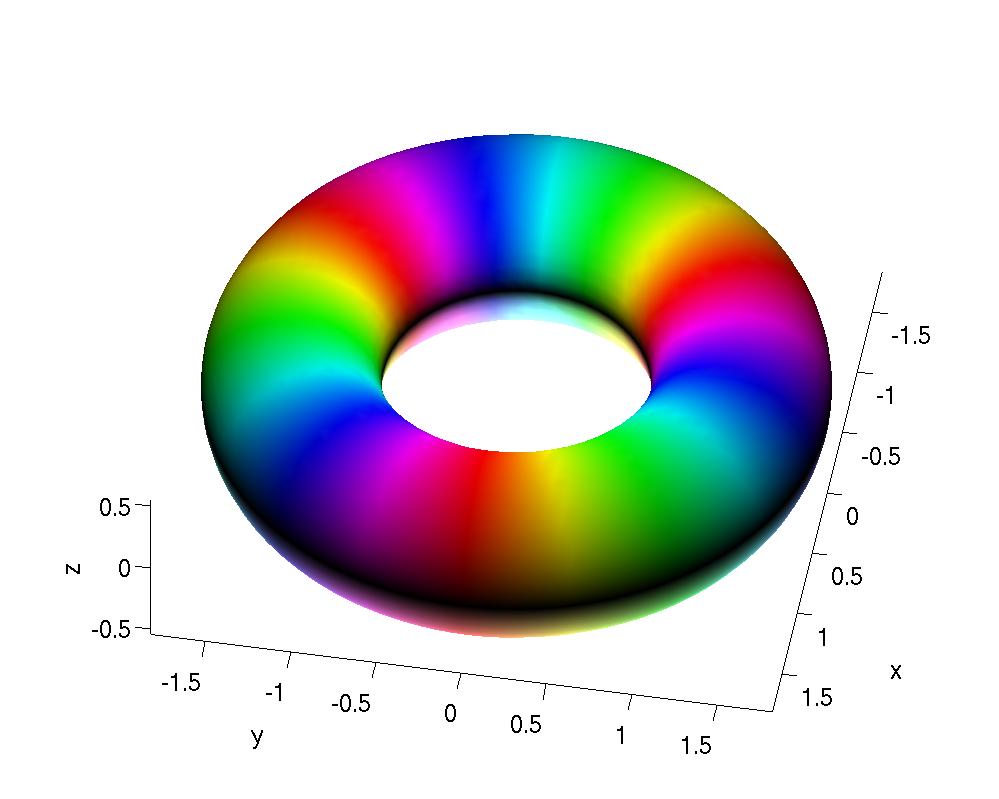}}
\subfloat[$m=3$, $B^{\rm numerical}=2.9994$]{\includegraphics[width=0.33\linewidth]{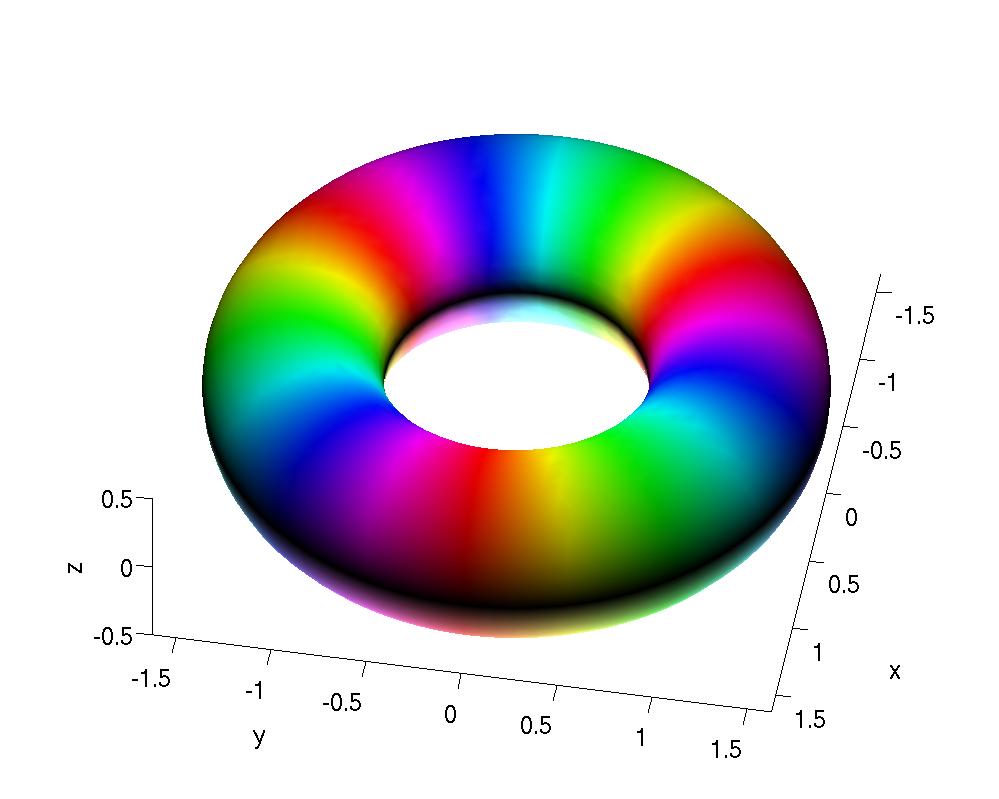}}}
\caption{Isosurfaces showing the solutions for the 2+4 model, 
$B=3$ and various values of the mass parameter, $m$, and the
toroidal Ansatz \eqref{eq:torus_reduced} as initial guess for the
relaxation. 
The color represents the phase of the scalar field $\phi_2$ and the
lightness is given by $|\Im(\phi_1)|$. 
The calculations are done on an $81^3$ cubic lattice
with the relaxation method. }
\label{fig:T4B3_TOIC}
\end{center}
\end{figure}

Figs.~\ref{fig:T4B3_TEIC} and \ref{fig:T4B3_TOIC} show the two series
of numerical solutions starting from the tetrahedral and toroidal
initial guess, respectively. It is observed that for $m\gtrsim 3$ both 
series converge to a flat torus. The difference in the colors is due
to a permutation in the fields $n_3$ and $n_2$. The two flat torii for
$m=4$ are physically the same and are not shown in
Figs.~\ref{fig:T4B3_TEIC} and \ref{fig:T4B3_TOIC}, but can be seen in
Fig.~\ref{fig:t4}. 

\begin{figure}[!htp]
\begin{center}
\includegraphics[width=0.8\linewidth]{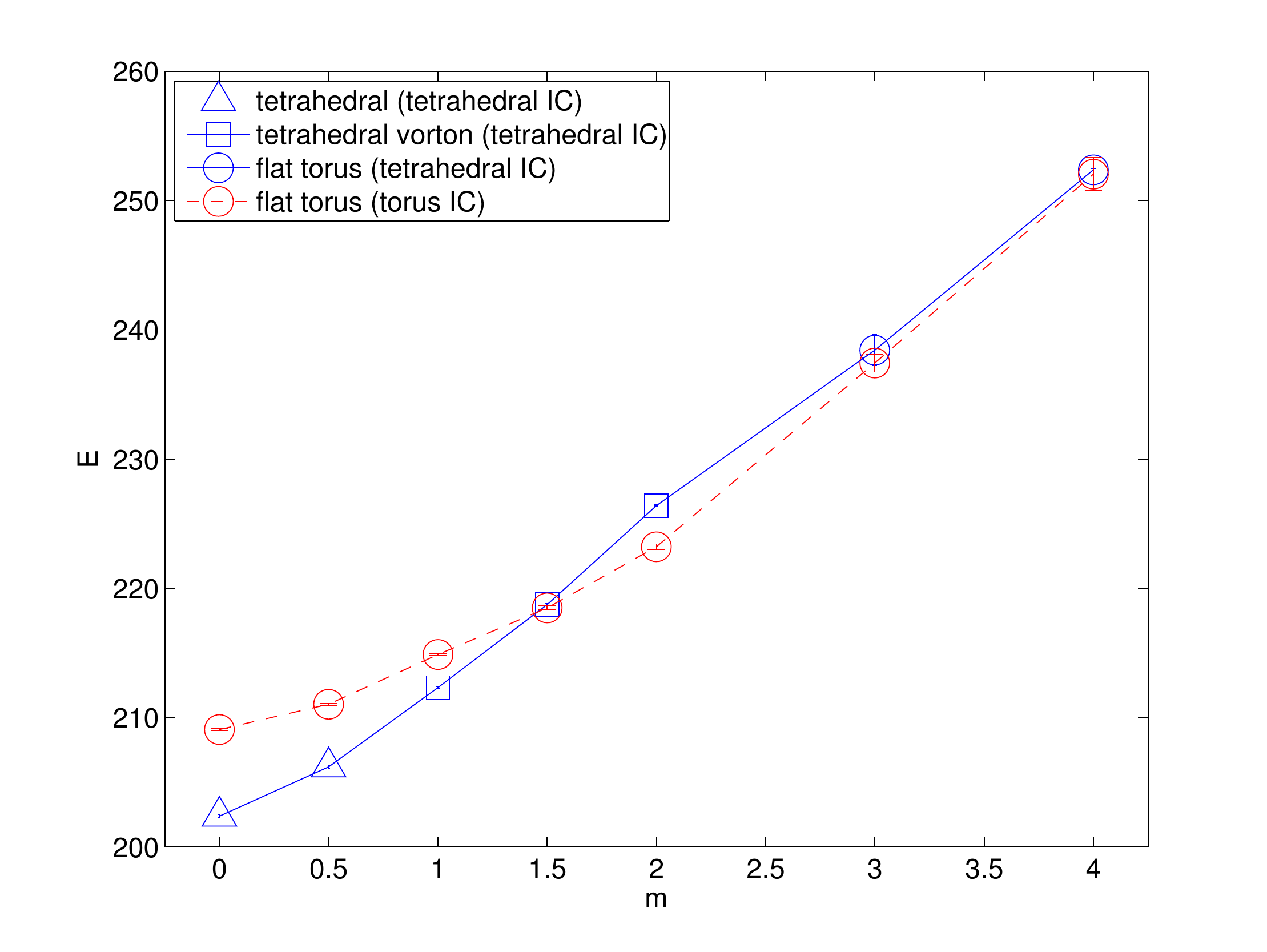}
\caption{Energies of numerical solutions whose initial guesses are
tetrahedrals (blue line) and torii (red line) with varying mass
$m$. For small $m\leq 1$, the tetrahedral Ansatz gives tetrahedral
solutions. At $m\gtrsim 1.5$ the tetrahedral Ansatz gives rise to
solutions that are heavier than that with the toroidal Ansatz,
suggesting a first-order phase transition. For large $m\gtrsim 3$,
both series give torii. } 
\label{fig:energycomparison}
\end{center}
\end{figure}

In order to determine which state is the lowest-energy state, we also
compare the energies for the two series of numerical solutions. 
In Fig.~\ref{fig:energycomparison} we show the energies of the two
series of numerical solutions: the blue solid line shows the solution
whose initial guess is the tetrahedral Ansatz and the red dashed line
has the torus Ansatz as initial guess. We see that for $m\lesssim 1.5$
the tetrahedral is the lowest-energy state. 
Our calculation indicates that a first-order phase transition takes
place around $m=m_{\rm critical}\sim 1.5$, where the tetrahedral
state rises above that of the toroidal one.
For $m\gtrsim 3$ both Ans\"atze give a flat torus after
relaxation has found a solution. Thus the tetrahedral state either
becomes unstable or slowly merges together with that of the toroidal
one. 
The instability sets in for $m$ between 3 and 4. 
In order to check that the phase transition around $m\sim 1.5$ really
takes place, we have run the solutions with an exceptionally long
relaxation time and found solutions with a very high accuracy (the
equations of motion are satisfied at every spatial position better
than $10^{-4}$ and about $10^{-5}$ on average). 
Indicative of the two different states crossing around the critical
mass, $m_{\rm critical}\sim 1.5$, is that the two different solutions
have almost exactly the same energy. 

\begin{figure}[!htp]
\begin{center}
\captionsetup[subfloat]{labelformat=empty}
\mbox{
\subfloat[$m=0$, $B^{\rm numerical}=2.9977$]{\includegraphics[width=0.33\linewidth]{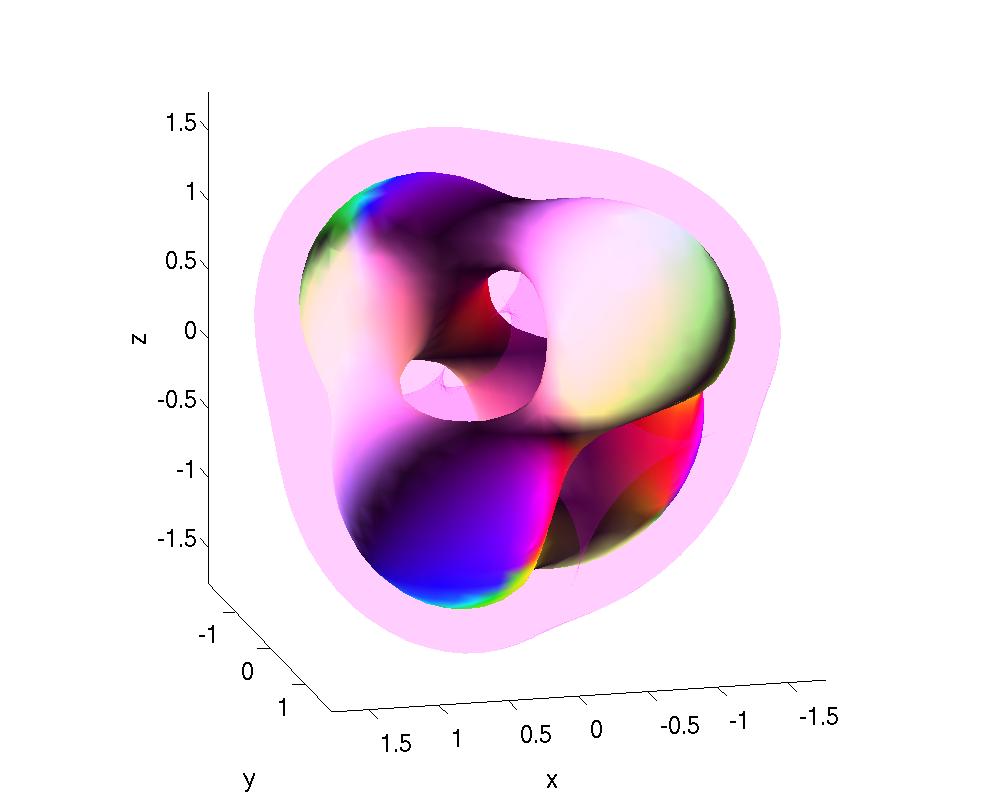}}
\subfloat[$m=1/2$, $B^{\rm numerical}=2.9983$]{\includegraphics[width=0.33\linewidth]{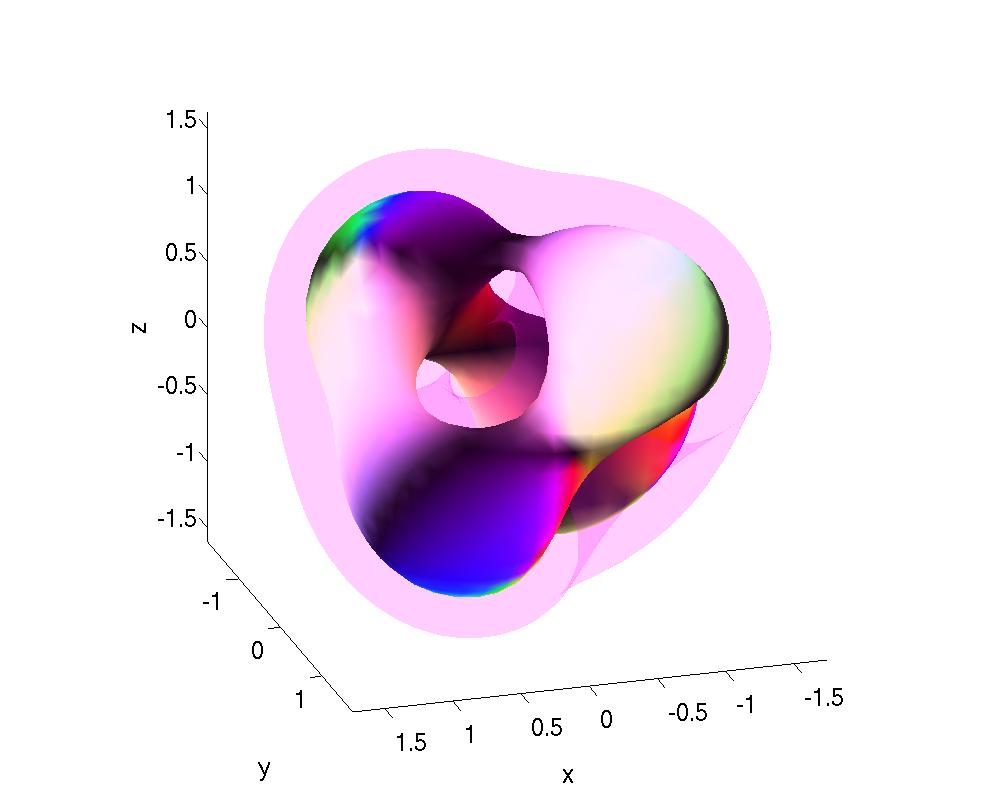}}
\subfloat[$m=1$, $B^{\rm numerical}=2.9986$]{\includegraphics[width=0.33\linewidth]{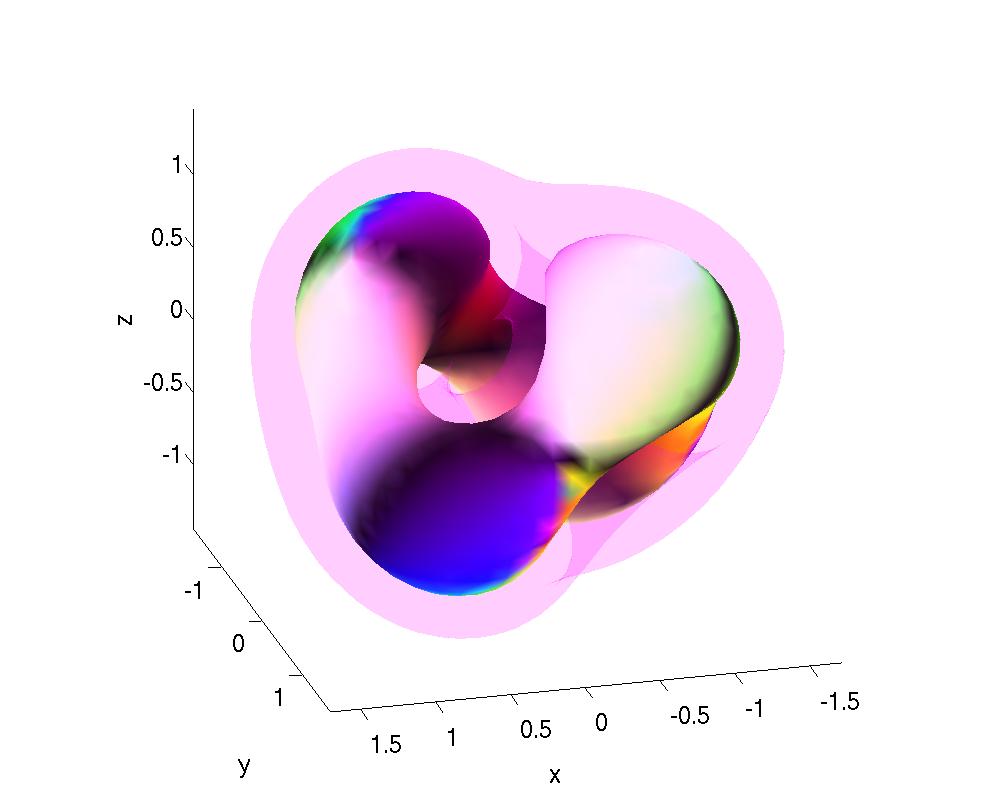}}}
\mbox{
\subfloat[$m=3/2$, $B^{\rm numerical}=2.9993$]{\includegraphics[width=0.33\linewidth]{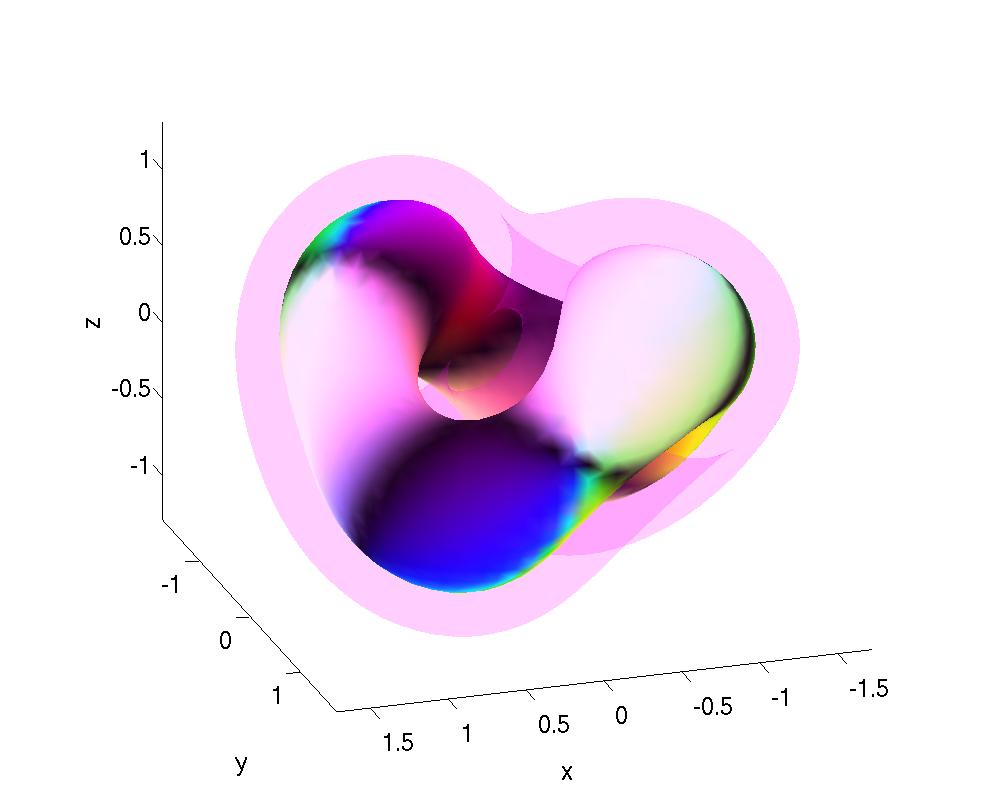}}
\subfloat[$m=2$, $B^{\rm numerical}=2.9986$]{\includegraphics[width=0.33\linewidth]{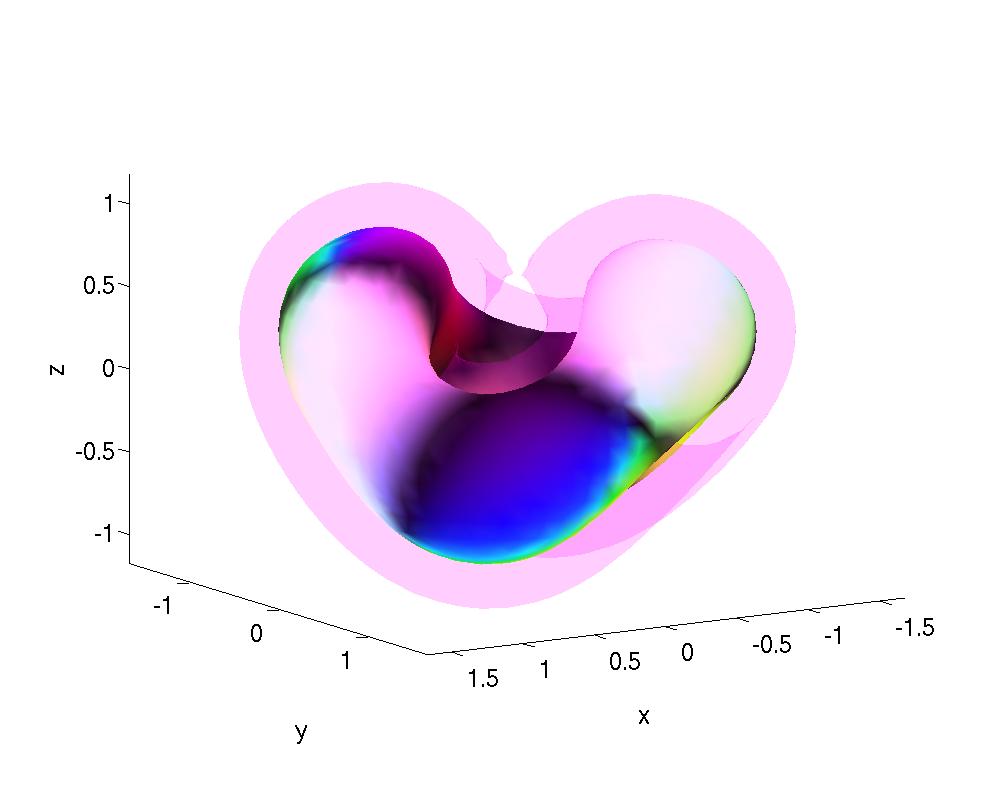}}
\subfloat[$m=3$, $B^{\rm numerical}=2.9982$]{\includegraphics[width=0.33\linewidth]{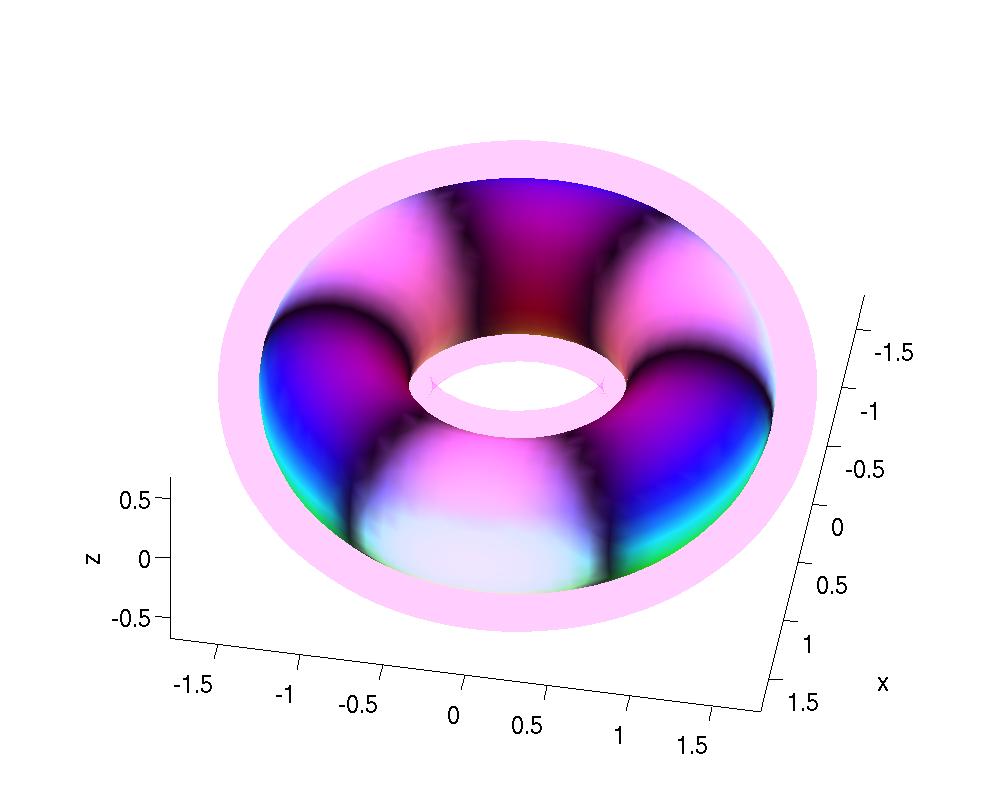}}}
\caption{Isosurfaces showing the solutions for the 2+4 model, for
$B=3$ and various values of the mass parameter, $m$, and the
tetrahedral Ansatz \eqref{eq:tetrahedral_ansatz} as initial guess for
the relaxation. The colored isosurface and the magenta shadow show the
isosurface at half and a quarter of the maximal value of the baryon
charge density, respectively. 
The color represents the phase of the scalar field $\phi_2$ and the
lightness is given by $|\Im(\phi_1)|$. 
The calculations are done on an $81^3$ cubic lattice
with the relaxation method. }
\label{fig:T4B3_TEIC_aura}
\end{center}
\end{figure}

In order to see that the numerical solutions for $m$ between 1 and 2
are actually tetrahedral in nature as opposed to bent torii, we show
the solutions with colored isosurfaces at half the maximal value as
well as at a quarter of the maximal value of the baryon charge density
in Fig.~\ref{fig:T4B3_TEIC_aura}. It is observed that there is a cloud
connecting the solution between the string at antipodal points.
For sufficiently large $m\sim 2m_{\rm critical}$, the tetrahedral
solution becomes unstable and thus for large $m$ only the torus
exists. 

\section{Summary and Discussion \label{sec:summary} }

We have studied Skyrmion solutions in the BEC Skyrme model, 
which is a Skyrme model with the potential 
term motivated by two-component BECs.
We have constructed stable Skyrmion solutions  
for $P=1,2,3,4,5$ and $Q=1$,  
yielding the baryon numbers $B=1,2,3,4,5$ as well as a metastable
solution for $P=6$ and $Q=1$ ($B=6$). We suspect that higher baryon
charged solutions will all be metastable.
The energy and baryon charge distributions of 
the configurations with $P>1$ are all of toroidal shape.
They are vortex rings of the field $\phi_1$, 
with the field $\phi_2$ trapped in their cores,
where the phase of the field $\phi_2$ winds $P$ times 
along the ring.
We have found that configurations with charge $(P,Q)$ decay into $Q$ rings
of charge $(P,1)$. This string splitting can be understood as
repulsion of global vortex strings. 
Finally we have discovered a first-order phase transition between
Skyrmions with a discrete point symmetry and axial (toroidal)
symmetry.

In two-component BECs, 
one can introduce a Rabi oscillation term 
$\gamma(\phi_1(x)^* \phi_2(x)+{\rm c.c.})$, 
known as a Josephson term in superconductors,
in the Lagrangian. 
Introduction of this term deforms the Skyrmions inside a domain wall 
\cite{Nitta:2012wi,Nitta:2012xq,Gudnason:2014nba,Garaud:2012}.
What deformation this term introduces for toroidal Skyrmions in the
BEC Skyrme model remains as a future problem.
On the other hand, if we introduce the potential term 
$V \sim \phi_1 + \phi_1^*$ \cite{Gudnason:2014hsa}, 
our configurations will become $P$ sine-Gordon kinks on a vortex ring, 
which is a $(3+1)$-dimensional analogue of
Ref.~\cite{Kobayashi:2013ju}, 
in which sine-Gordon kinks on a domain wall ring 
were constructed in $2+1$ dimensions.

Two-component BECs are known 
to admit a stable composite soliton, viz.~a D-brane soliton, that is,  
a domain wall on which 
vortices end from both sides \cite{Kasamatsu:2010aq}, 
originally found in the massive $\mathbb{C}P^1$ model
\cite{Gauntlett:2000de,Isozumi:2004vg}.  
The (BEC) Skyrme model discussed in this paper 
has the same potential term and 
is expected to admit such a D-brane soliton.
A configuration made of
a domain wall and an anti-domain wall 
stretched by lump-strings in the massive $\mathbb{C}P^1$ model 
was considered in Ref.~\cite{Nitta:2012kk}, 
in which it was discussed that 
such a configuration is unstable to decay,  
resulting in the creation of Hopfions. 
Therefore, the same mechanism should work 
also in the (BEC) Skyrme model discussed in this paper 
creating Skyrmions from brane annihilation, 
as was discussed for two-component BECs \cite{Nitta:2012hy}.

The Bogomol'nyi-Prasad-Sommerfield (BPS) Skyrme model, proposed
recently \cite{Adam:2010fg}, consists of only the sixth-order
derivative term as well as appropriate potentials. This model admits
exact solutions with compact support. By choosing the potential of the
BEC Skyrme model in this paper, we may be able to construct exact
solutions of Skyrmions with toroidal shape.

The Skyrmions with the charge $(P,Q)$ are related through the Hopf
map to $(P,Q)$ Hopfions 
\cite{Kobayashi:2013bqa,Kobayashi:2013xoa}
in the Ising Faddeev-Skyrme (FS) model \cite{Nitta:2012kk}, 
that is, the FS model 
\cite{Faddeev:1975,Faddeev:1996zj} with an Ising-type
potential term admitting two discrete vacua. 
The domain wall in the BEC Skyrme model 
is mapped to a domain wall 
with a U(1) modulus interpolating between these two vacua
\cite{Abraham:1992vb,Kudryavtsev:1997nw,Arai:2002xa}, 
and a global vortex is mapped to a lump or baby Skyrmion 
\cite{Piette:1994ug,Weidig:1998ii}.
This model also admits 
a twisted domain-wall tube with the U(1) modulus twisted along the
cycle of the tube \cite{Kobayashi:2013ju} as a baby-Skyrmion string. 
The original FS model without said potential term is known to admit
Hopfions, i.e.~solitons with Hopf charge 
$\pi_3(S^2) \simeq \mathbb{Z}$
\cite{deVega:1977rk,Gladikowski:1996mb,Faddeev:1996zj,Battye:1998pe,Hietarinta:2000ci,Sutcliffe:2007ui,Radu:2008pp}, 
and, in particular, Hopfions with Hopf charge 7 or 
higher were found to have knot structures  
\cite{Battye:1998pe,Hietarinta:2000ci,Sutcliffe:2007ui}.
The $(P,Q)$ Hopfions in the Ising FS model are 
not knots but toroidal domain walls 
characterized by two integers $(P,Q)$,  
where the U(1) modulus of the domain wall is twisted $P$ and $Q$ times 
along the toroidal and poloidal cycles of the torus, respectively. 
In this case, some configurations with $Q>1$ were found to be stable 
\cite{Kobayashi:2013xoa}, 
unlike our case of Skyrmions for which 
all configurations for $Q>1$ are unstable. 
This is because there is no repulsion between lumps. 

If we consider compactifying space to $\mathbb{R}^2\times S^1$ we have
another solution in addition to the one studied here, in which the
vortex string extends along the $S^1$ direction and has $P$ twists on
its U(1) modulus. The corresponding solution for the case of the
Hopfion was discussed in Ref.~\cite{Kobayashi:2013aza}. 
Skyrmions in the conventional model on $S^2\times S^1$ were
discussed in Ref.~\cite{Canfora:2014aia}.

\section*{Acknowledgments}

We thank Michikazu Kobayashi for discussions in the early stage 
of this work. 
The work of M.~N.~is supported in part by Grant-in-Aid for Scientific Research 
No.~25400268
and by the ``Topological Quantum Phenomena'' 
Grant-in-Aid for Scientific Research 
on Innovative Areas (No.~25103720)  
from the Ministry of Education, Culture, Sports, Science and Technology 
(MEXT) of Japan. 
S.~B.~G.~thanks Keio University for hospitality during which this
project took shape. 
The authors thank the referee for valuable comments.

\begin{appendix}

\section{String splitting for $Q>1$\label{app:stringsplitting}}

\begin{figure}[!tph]
\begin{center}
\mbox{
\includegraphics[width=0.24\linewidth]{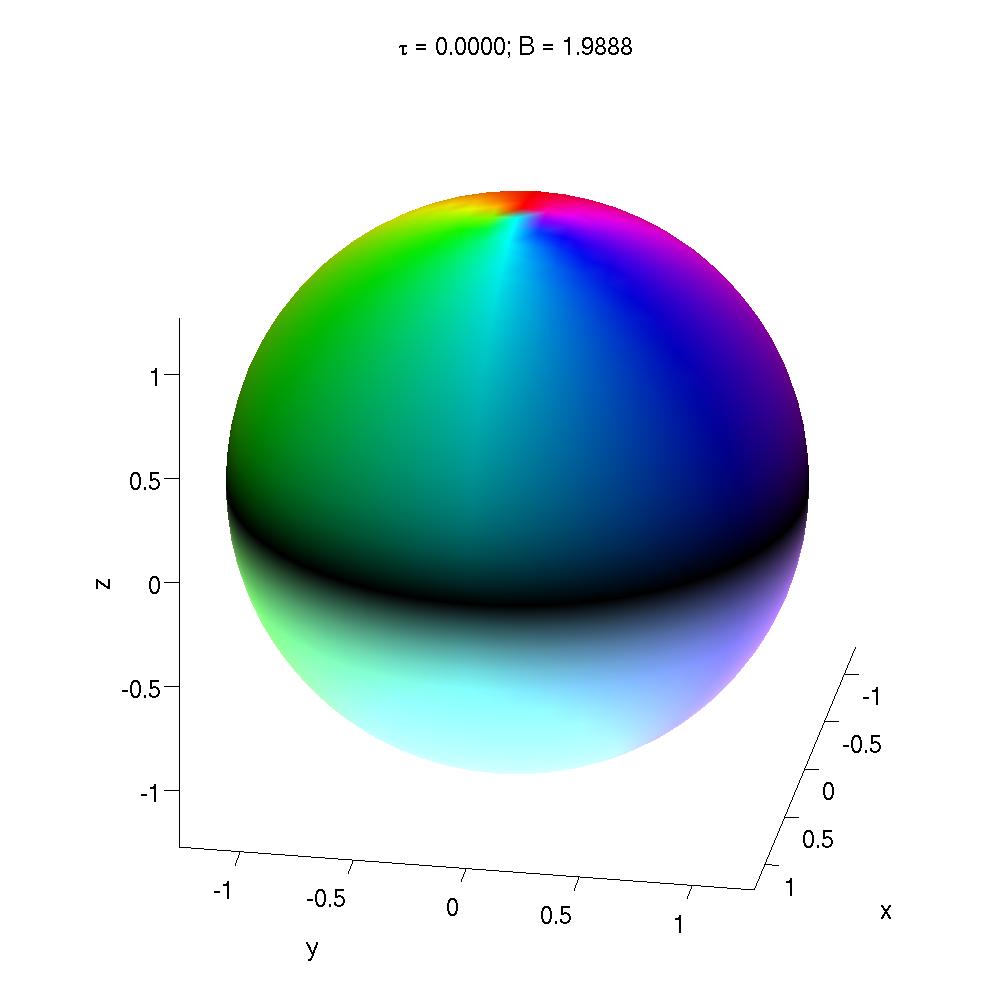}
\includegraphics[width=0.24\linewidth]{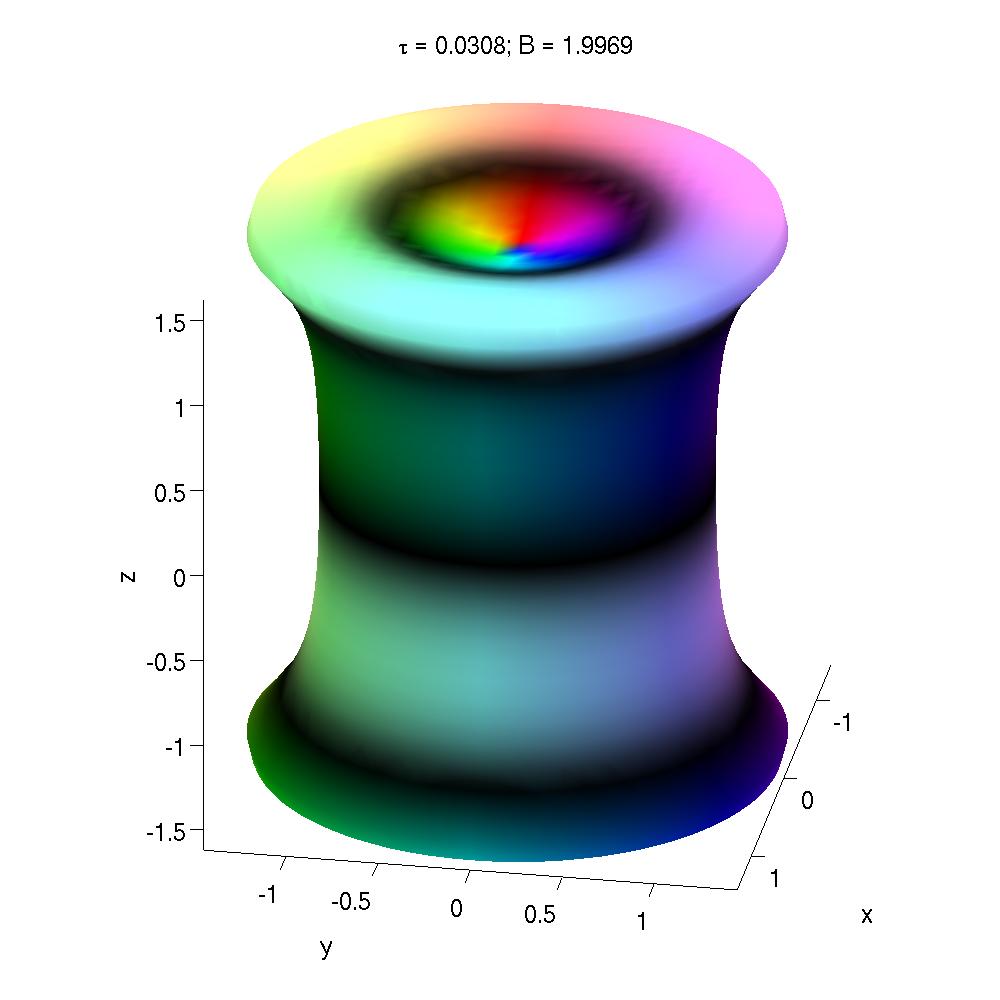}
\includegraphics[width=0.24\linewidth]{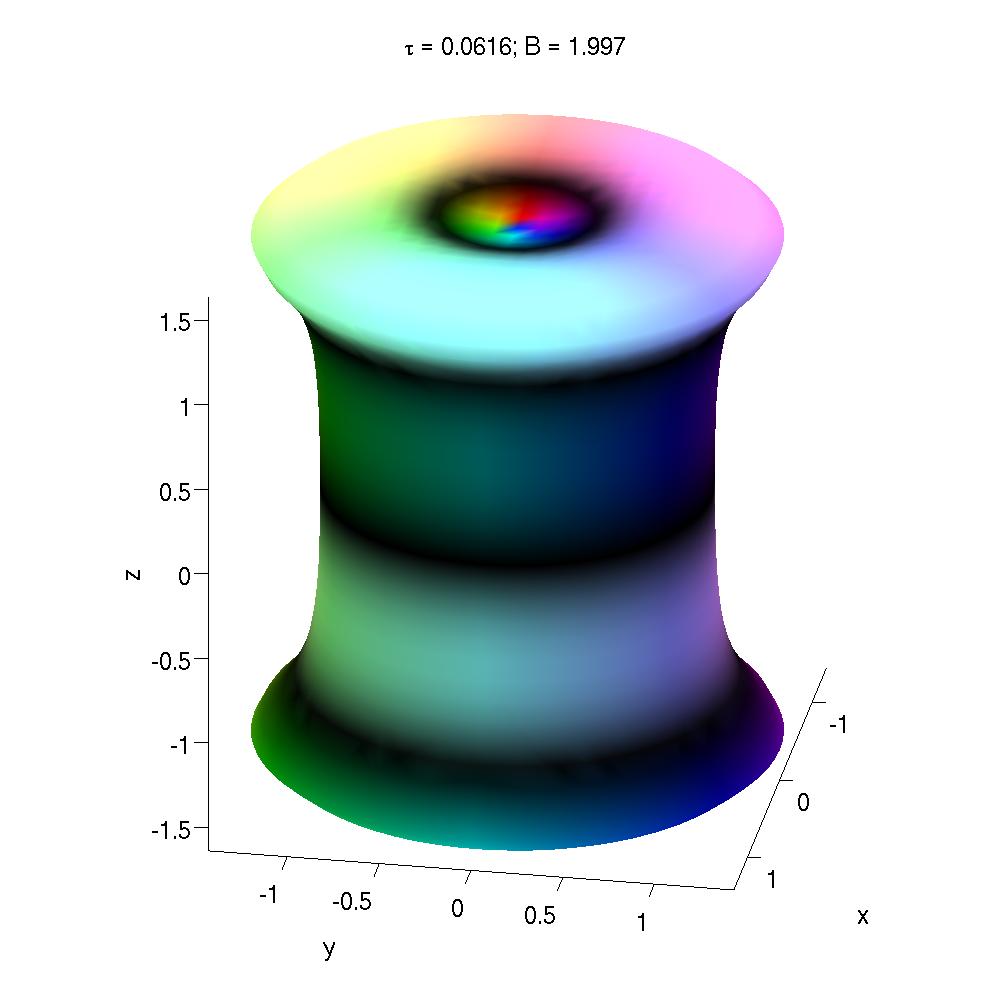}
\includegraphics[width=0.24\linewidth]{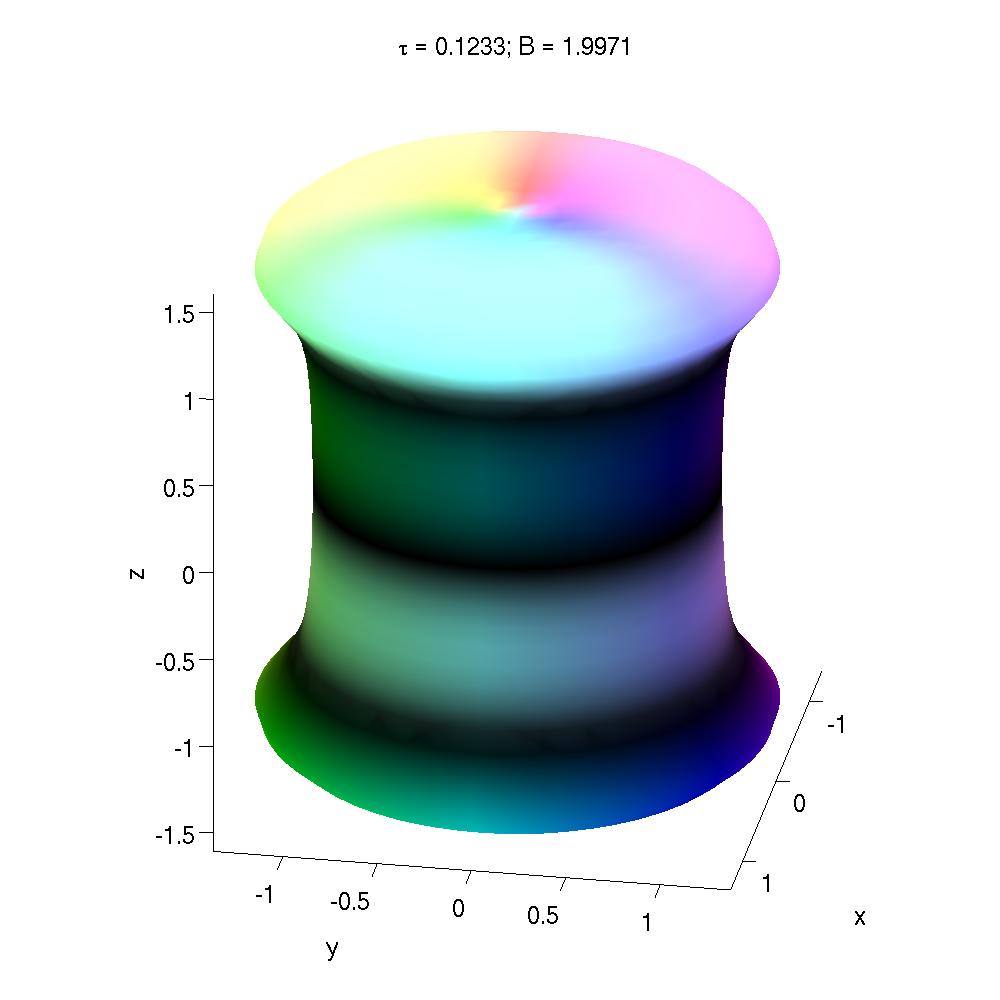}}
\mbox{
\includegraphics[width=0.24\linewidth]{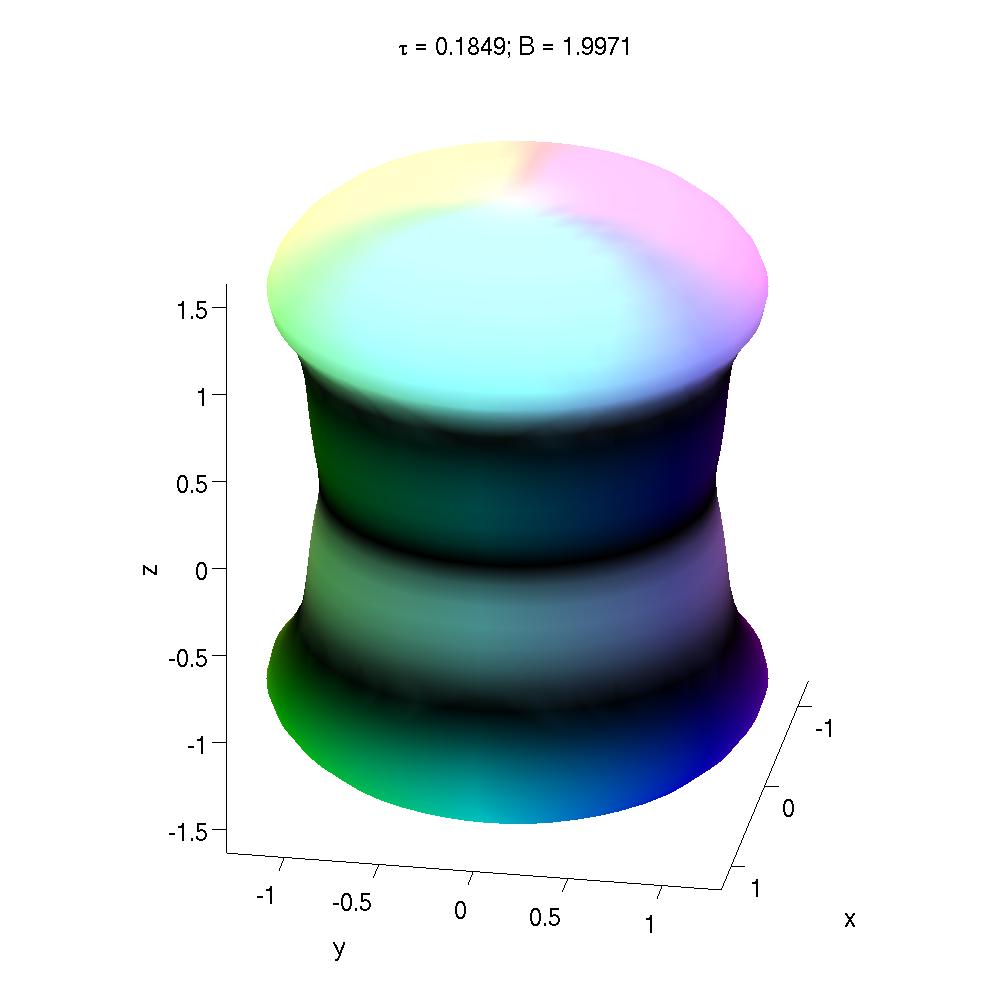}
\includegraphics[width=0.24\linewidth]{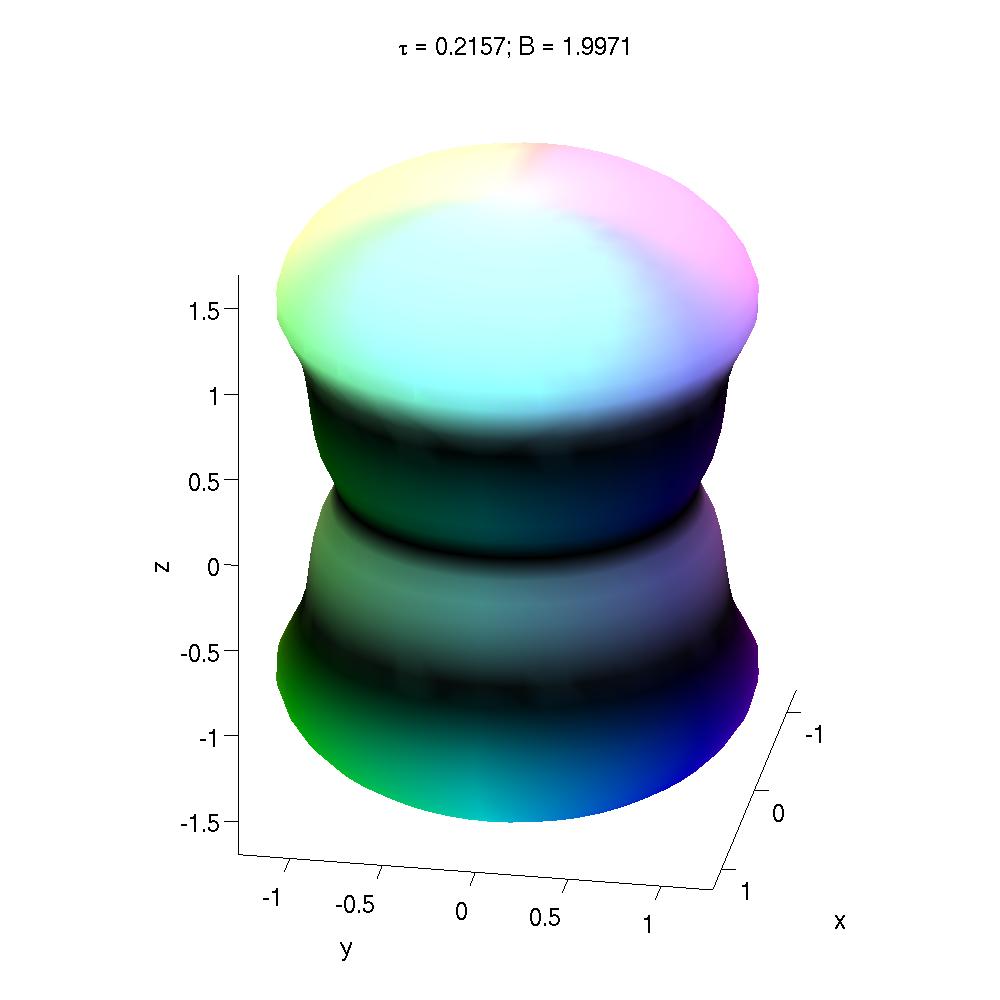}
\includegraphics[width=0.24\linewidth]{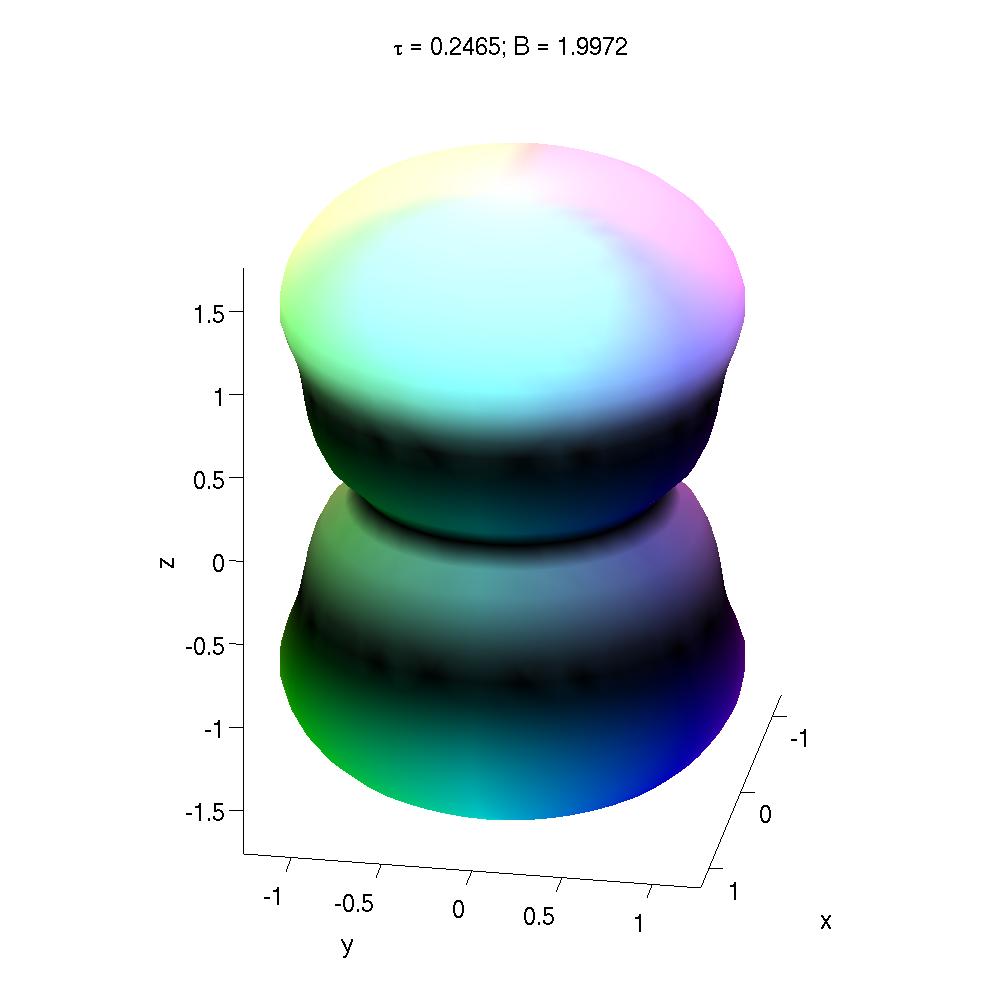}
\includegraphics[width=0.24\linewidth]{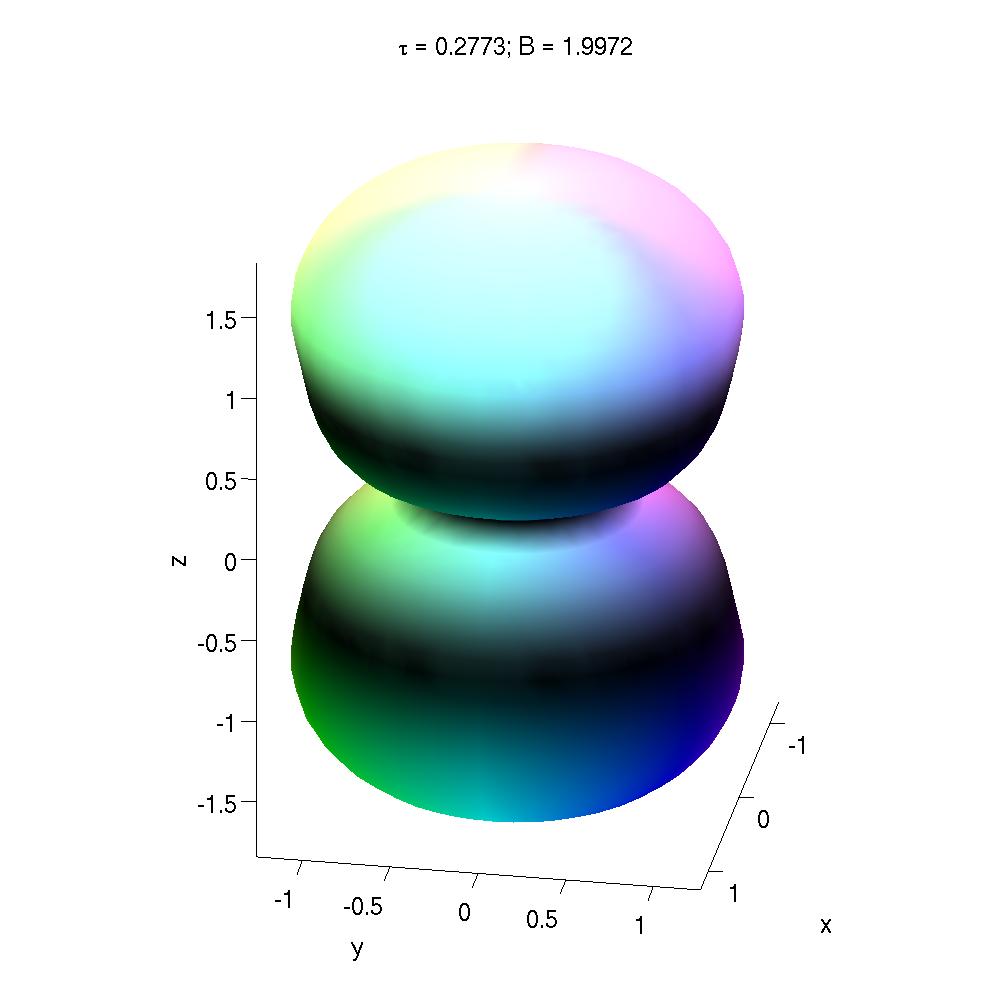}}
\mbox{
\includegraphics[width=0.24\linewidth]{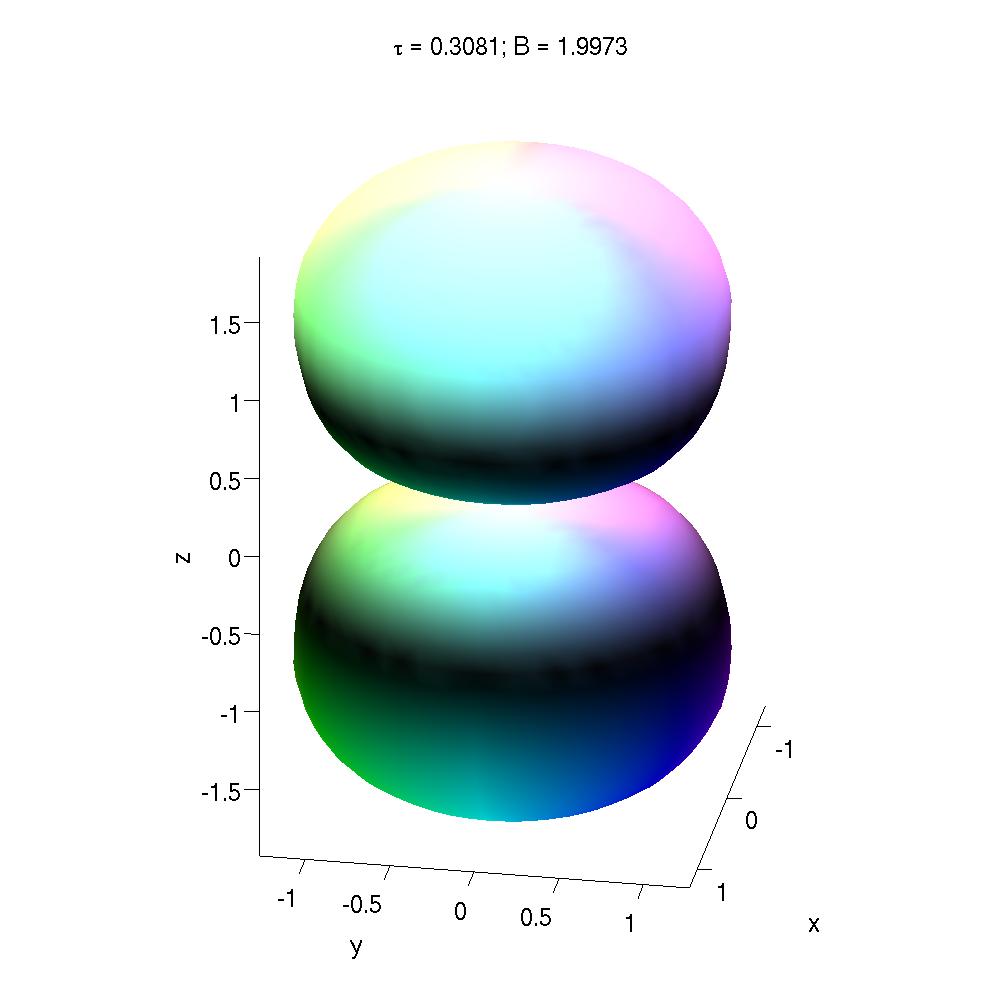}
\includegraphics[width=0.24\linewidth]{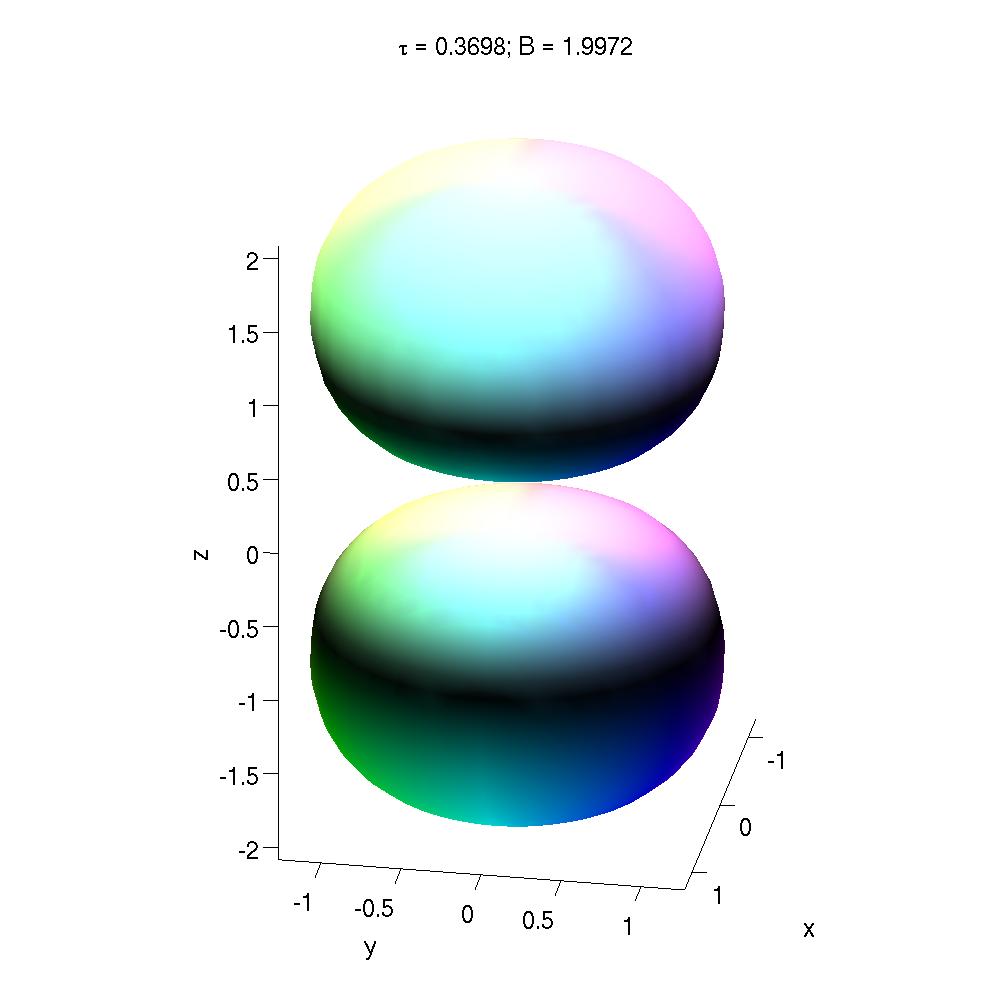}
\includegraphics[width=0.24\linewidth]{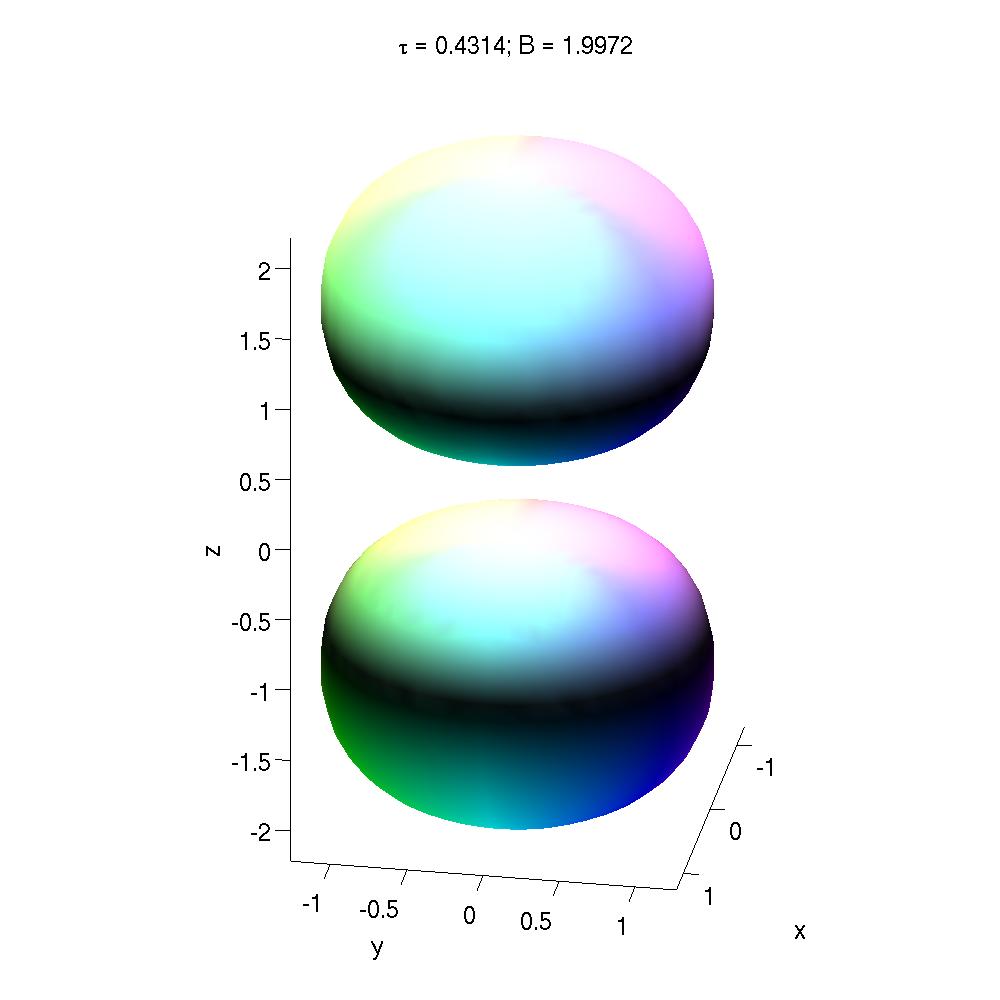}
\includegraphics[width=0.24\linewidth]{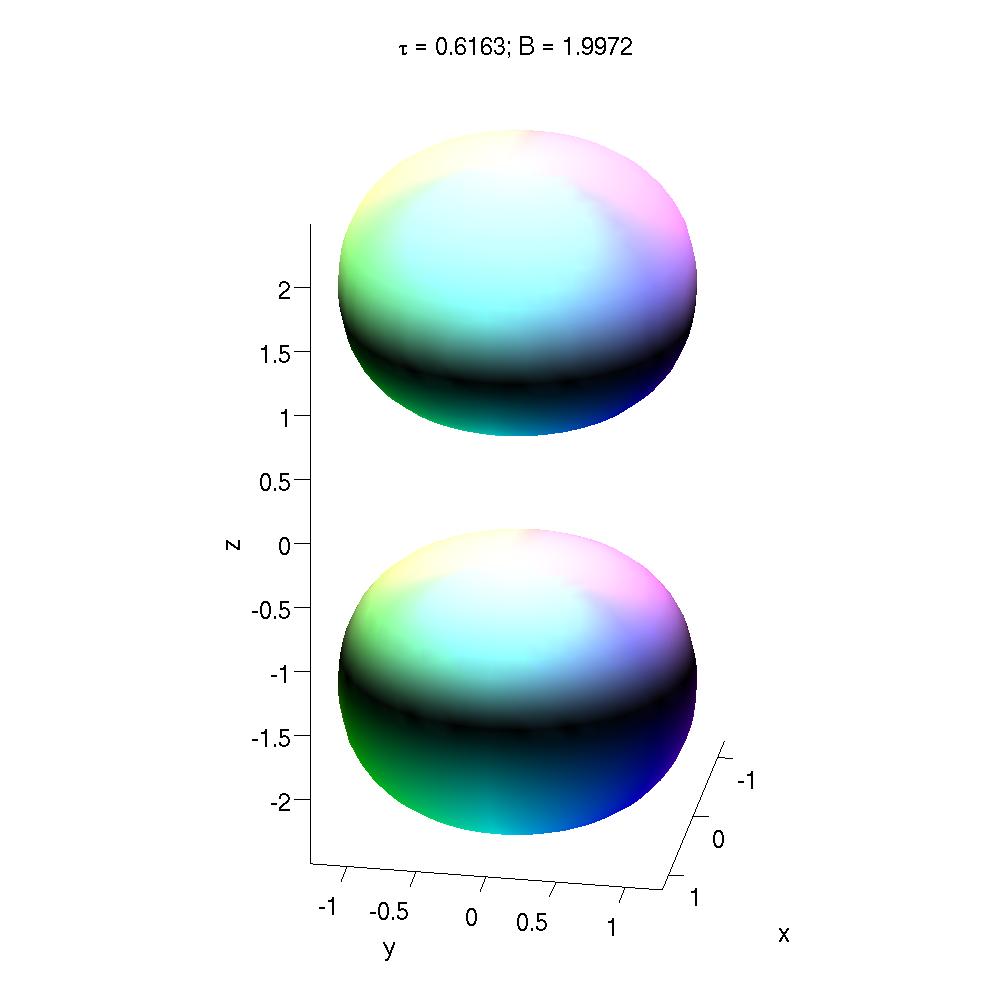}}
\caption{Isosurfaces showing an initial configuration with
  $(P,Q)=(1,2)$ ($B=2$) in 2+6 model ($\kappa=0$, $c_6=1$ and $m=4$)
  which after some finite relaxation time splits the Skyrmion into two
  separate Skyrmions of charge one, i.e.~$(P,Q)=(1,1)$. 
  The isosurfaces show constant baryon charge density equal to half
  its maximum value. 
  The color represents the phase of the scalar field $\phi_2$ and the
  lightness is given by $|\Im(\phi_1)|$. 
  The calculation is carried out on an $81^3$ cubic lattice with the
  relaxation method. }
\label{fig:stringsplitting12}
\end{center}
\end{figure}

\begin{figure}[!tph]
\begin{center}
\mbox{
\includegraphics[width=0.24\linewidth]{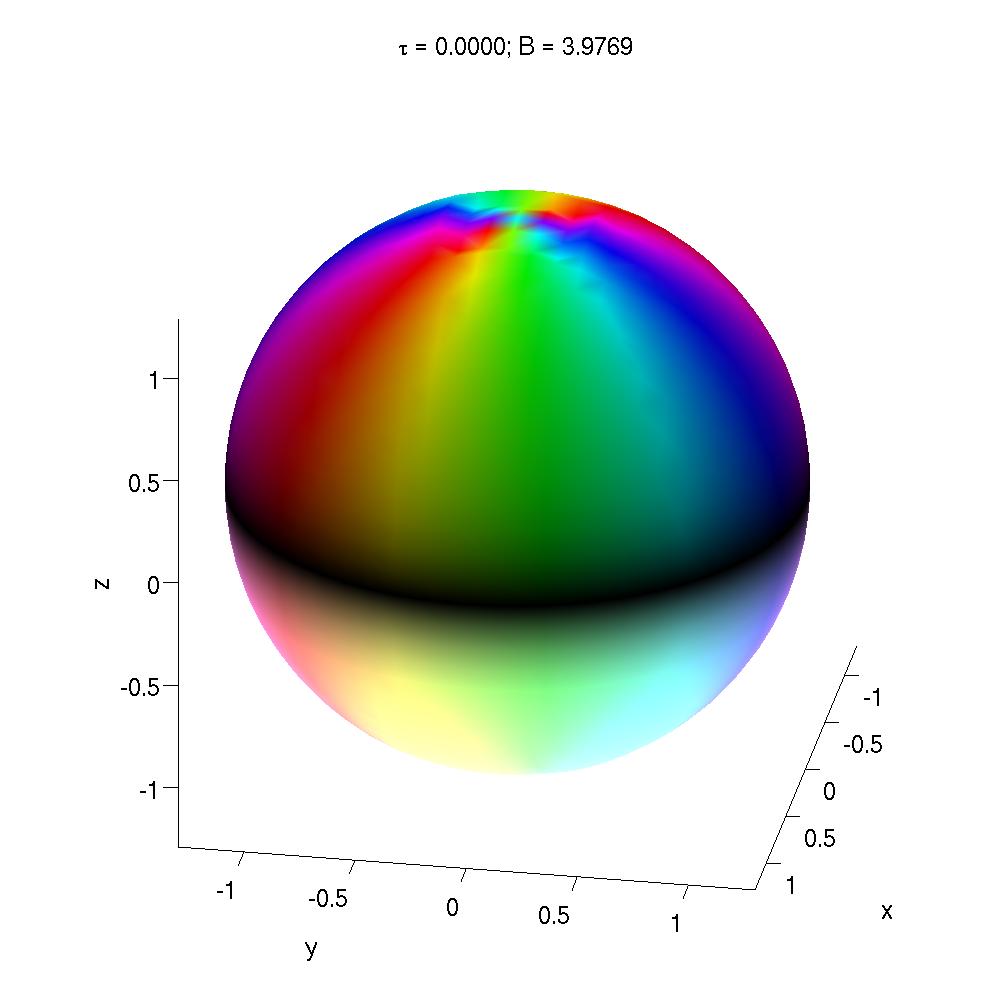}
\includegraphics[width=0.24\linewidth]{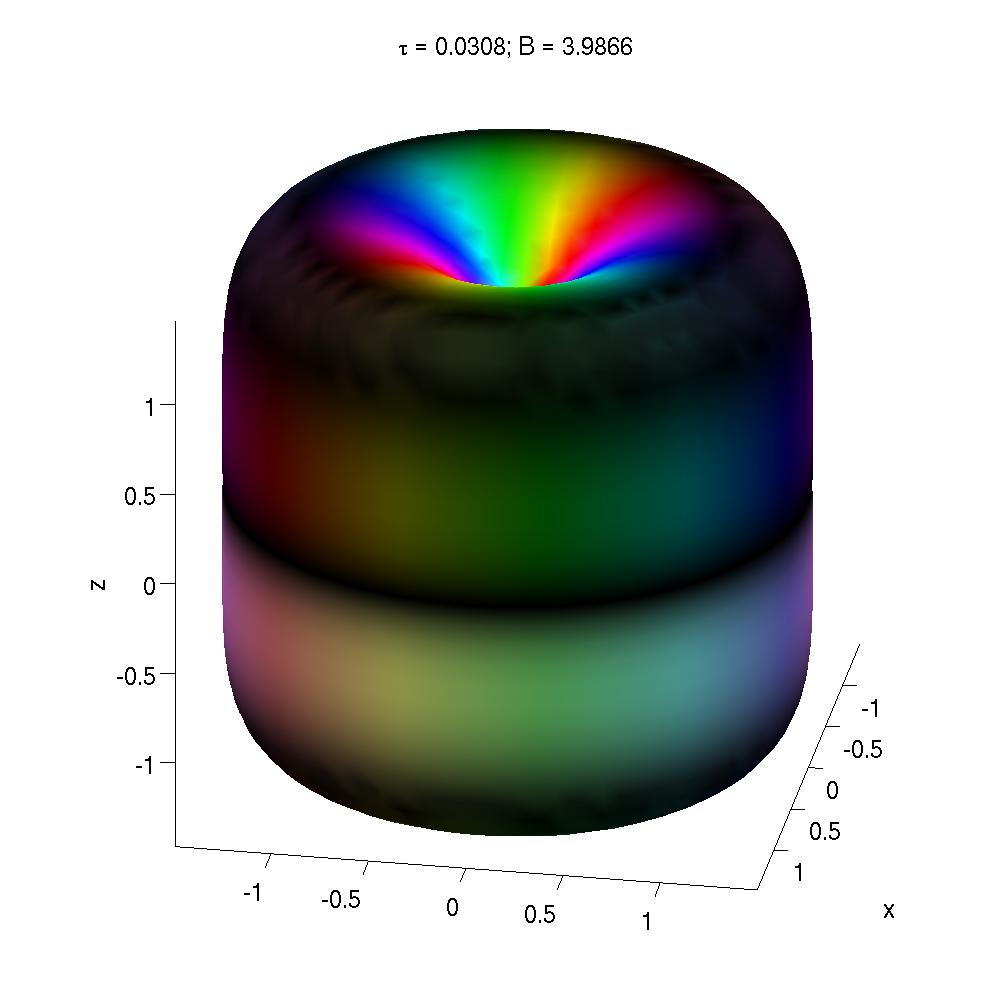}
\includegraphics[width=0.24\linewidth]{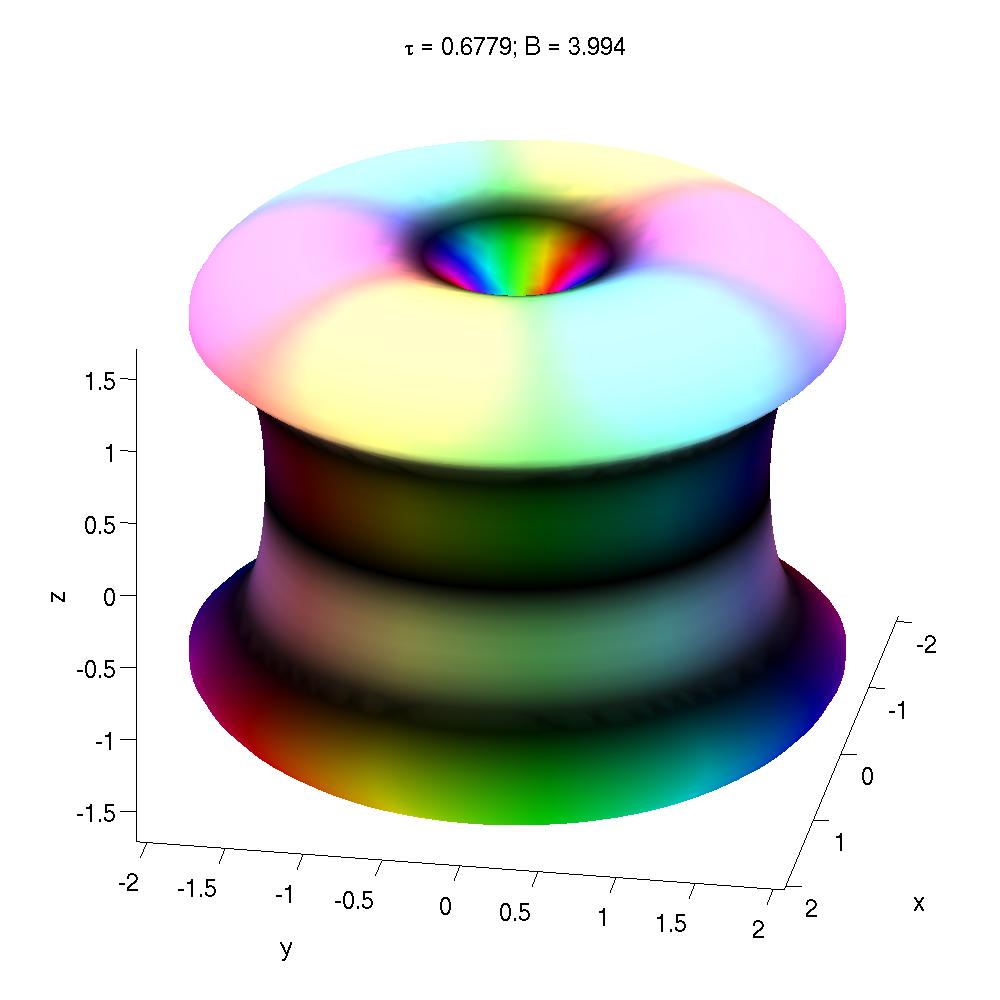}
\includegraphics[width=0.24\linewidth]{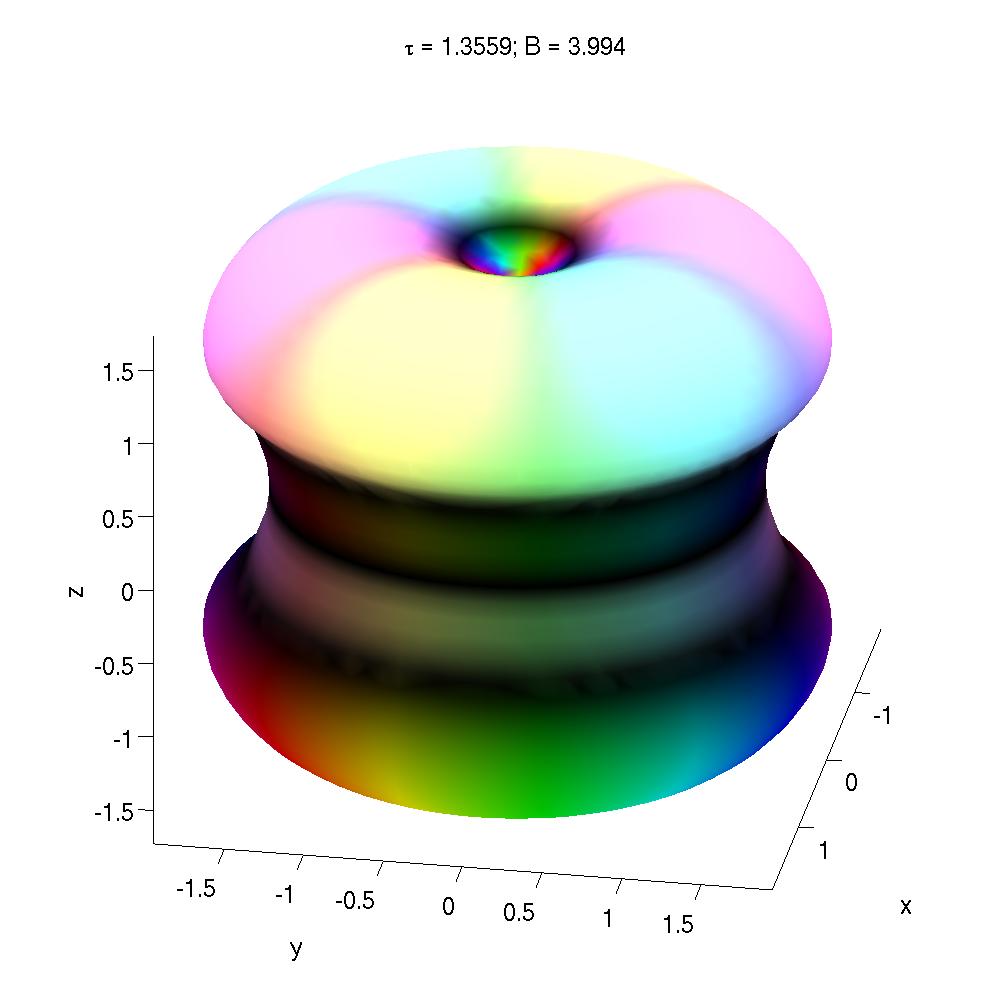}}
\mbox{
\includegraphics[width=0.24\linewidth]{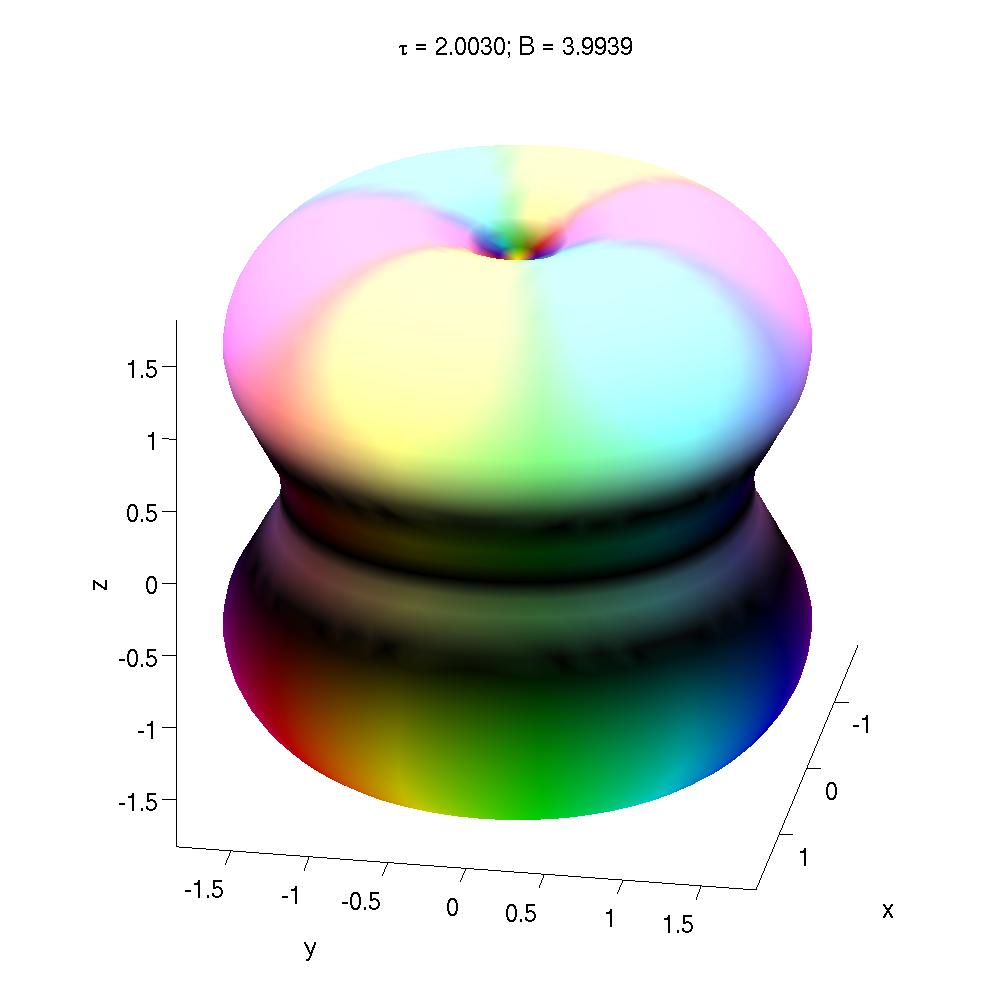}
\includegraphics[width=0.24\linewidth]{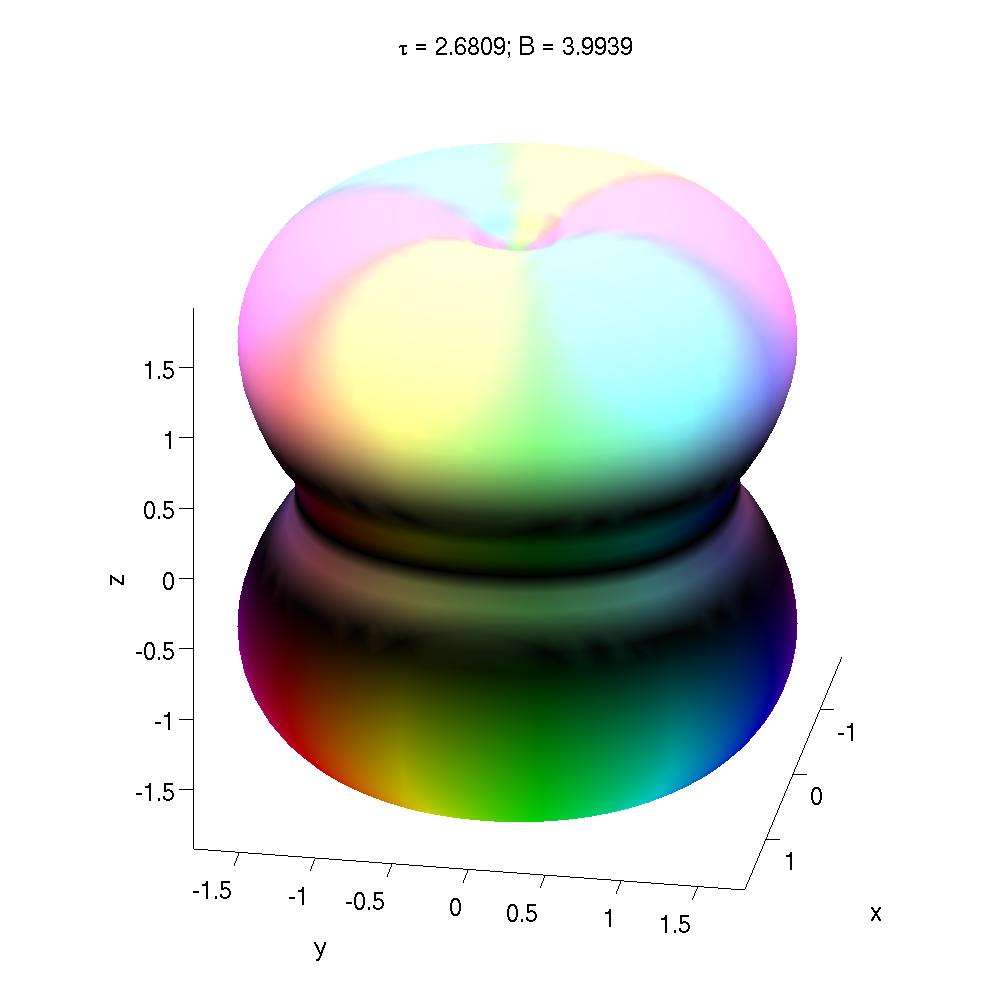}
\includegraphics[width=0.24\linewidth]{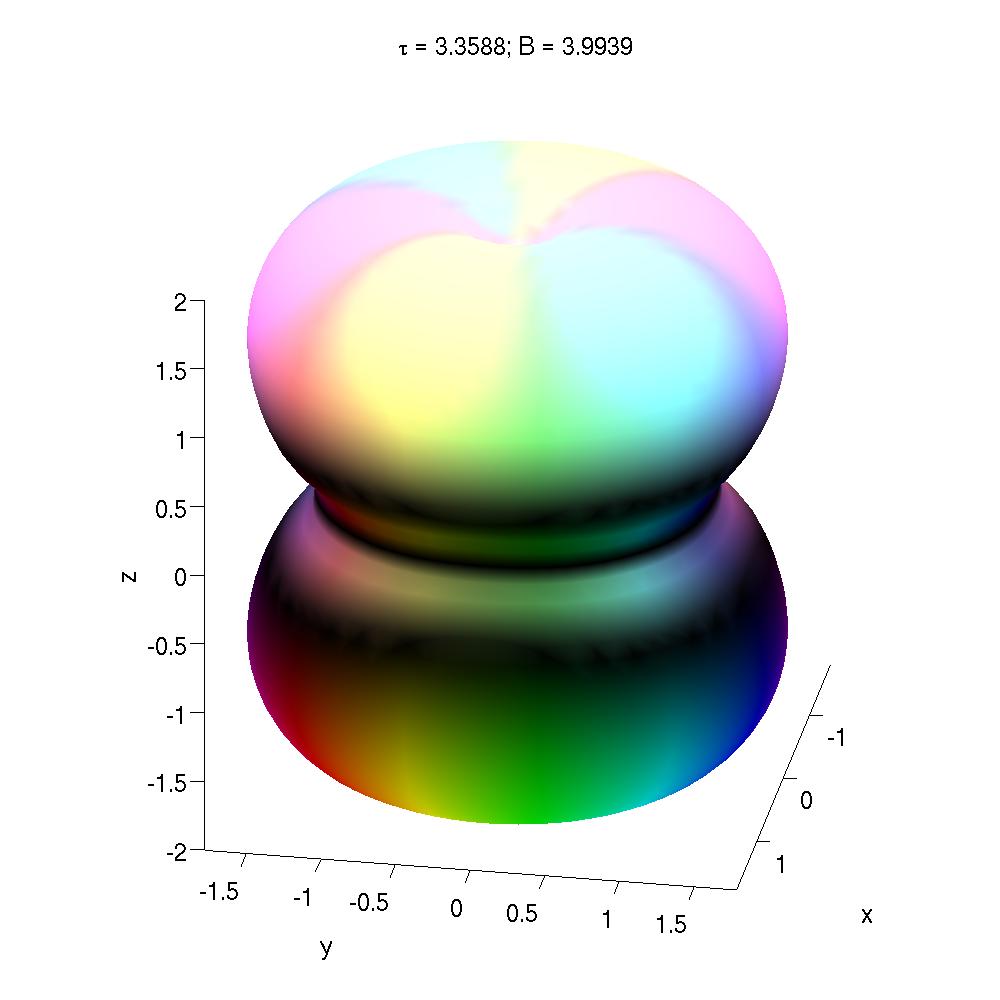}
\includegraphics[width=0.24\linewidth]{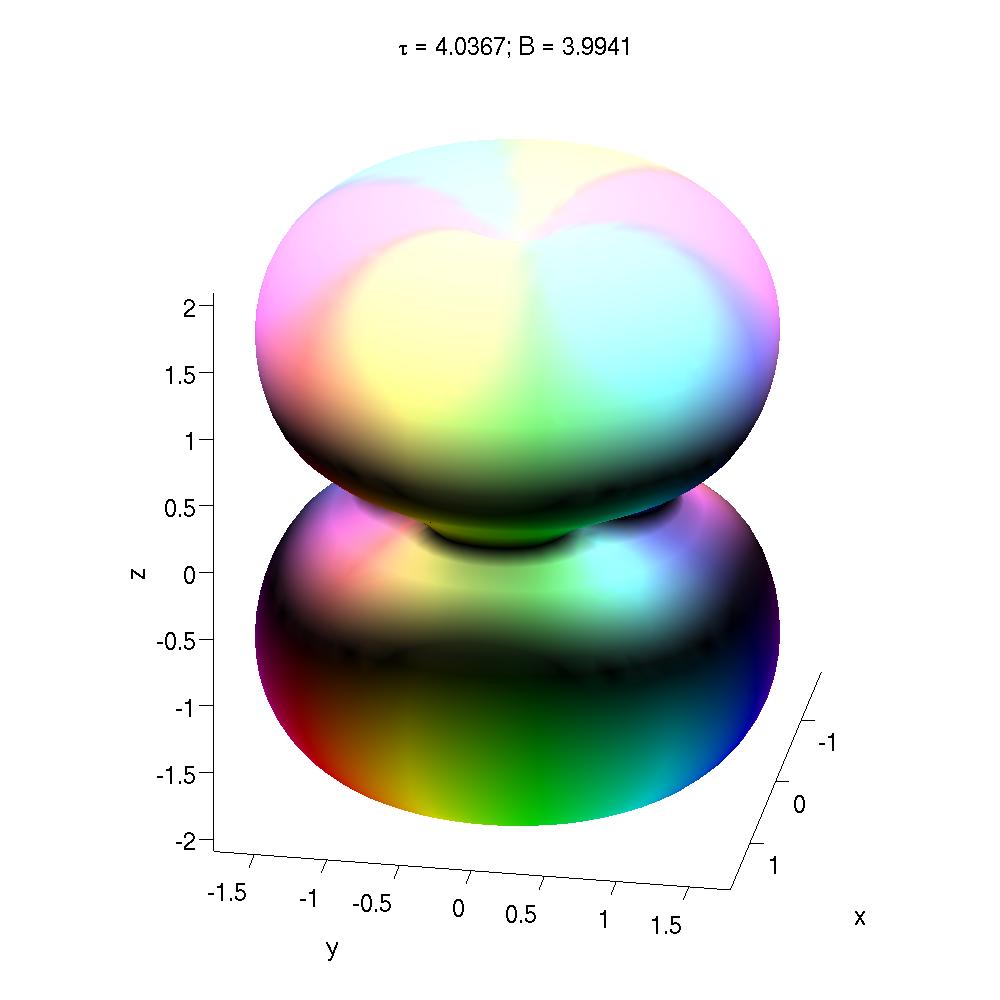}}
\mbox{
\includegraphics[width=0.24\linewidth]{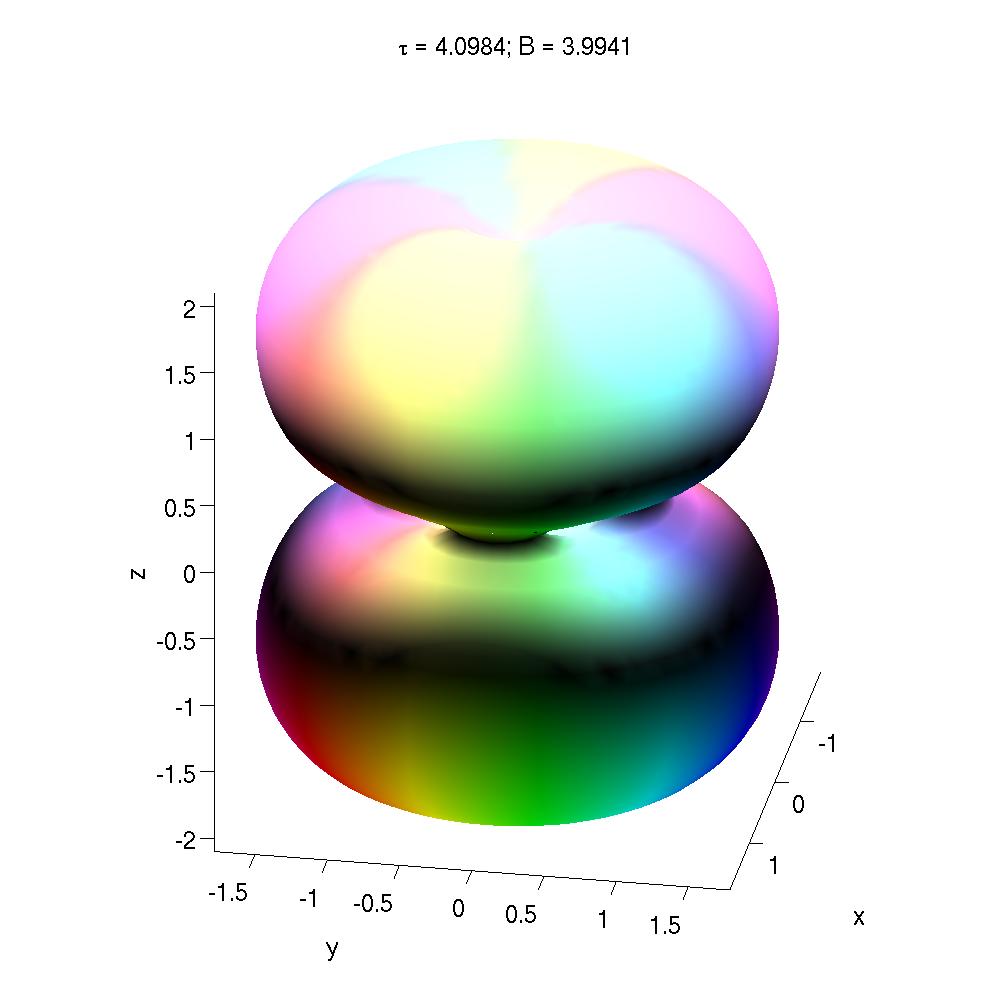}
\includegraphics[width=0.24\linewidth]{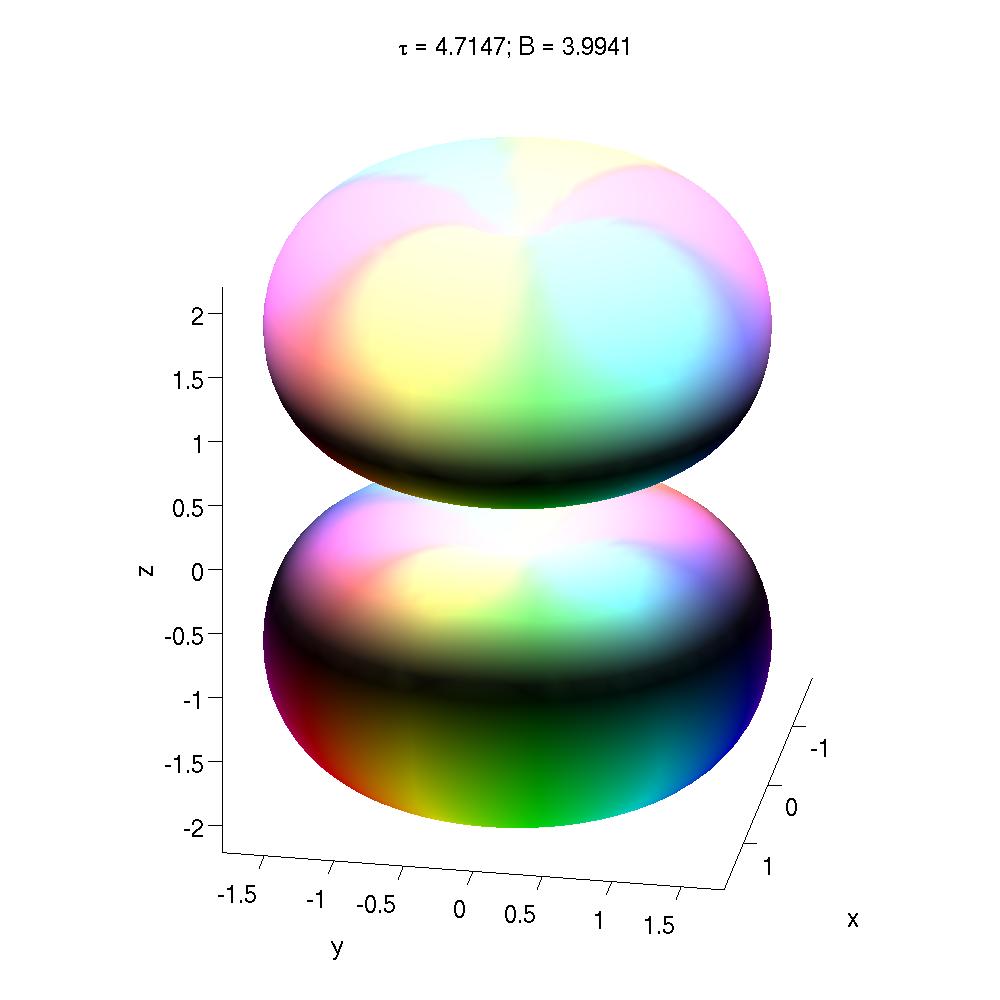}
\includegraphics[width=0.24\linewidth]{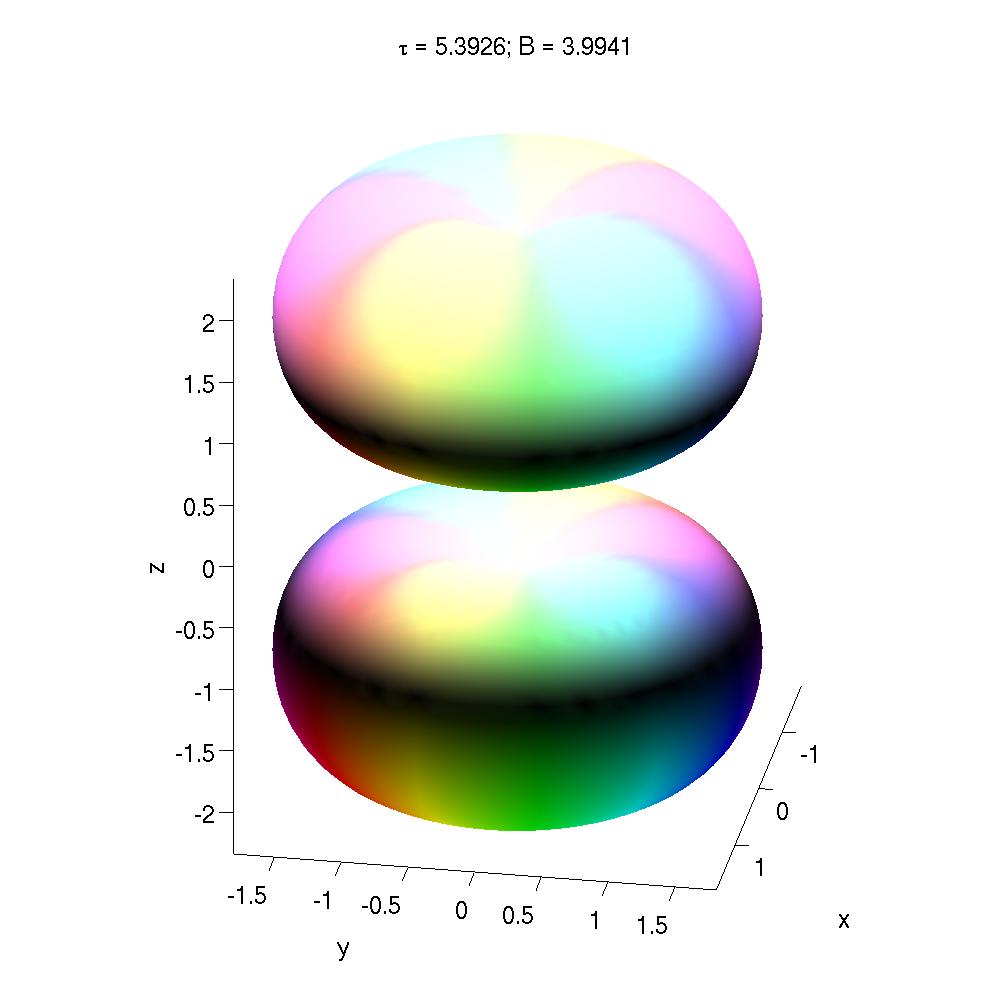}
\includegraphics[width=0.24\linewidth]{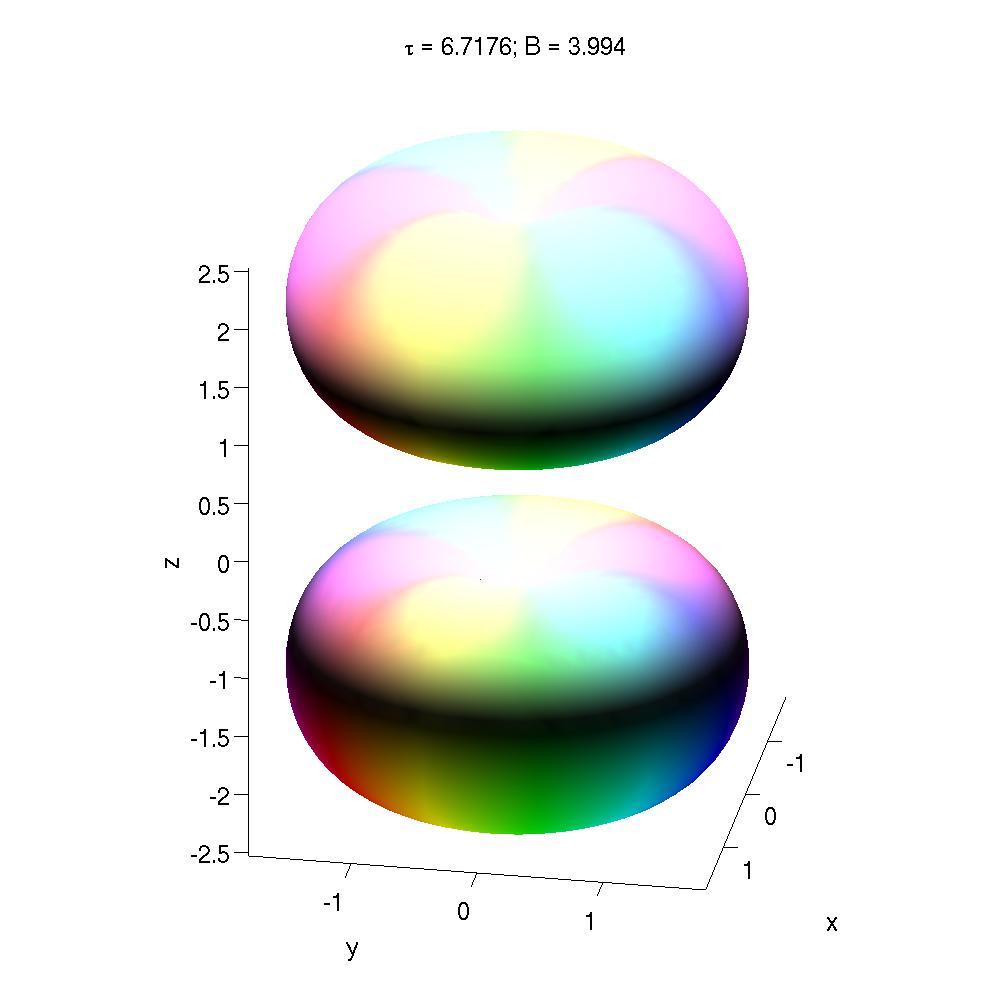}}
\caption{Isosurfaces showing an initial configuration with
  $(P,Q)=(2,2)$ ($B=4$) in 2+6 model ($\kappa=0$, $c_6=1$ and $m=4$)
  which after some finite relaxation time splits the Skyrmion into two 
  separate Skyrmions of charge two, i.e.~$(P,Q)=(2,1)$.
  The isosurfaces show constant baryon charge density equal to half
  its maximum value. 
  The color represents the phase of the scalar field $\phi_2$ and the
  lightness is given by $|\Im(\phi_1)|$. 
  The calculation is carried out on an $81^3$ cubic lattice with the
  relaxation method. }
\label{fig:stringsplitting22}
\end{center}
\end{figure}

In this section we show that the relaxation of the $(P,Q)=(P,2)$ torus
splits into two separate $(P,Q)=(P,1)$ objects for $P=1,2$. For
concreteness we carry out the calculations in the 2+6 model
($\kappa=0$ and $c_6=1$). 
In Figs.~\ref{fig:stringsplitting12} and \ref{fig:stringsplitting22} are
shown the $(1,2)\to 2\times(1,1)$ and $(2,2)\to 2\times(2,1)$ string
splittings as function of relaxation time $\tau$, respectively. 

\section{Comparison of torus and Skyrmion\label{app:comparison}}

\begin{figure}[!thp]
\begin{center}
\captionsetup[subfloat]{labelformat=empty}
\mbox{
\includegraphics[width=0.4\linewidth]{T4B21n}
\includegraphics[width=0.4\linewidth]{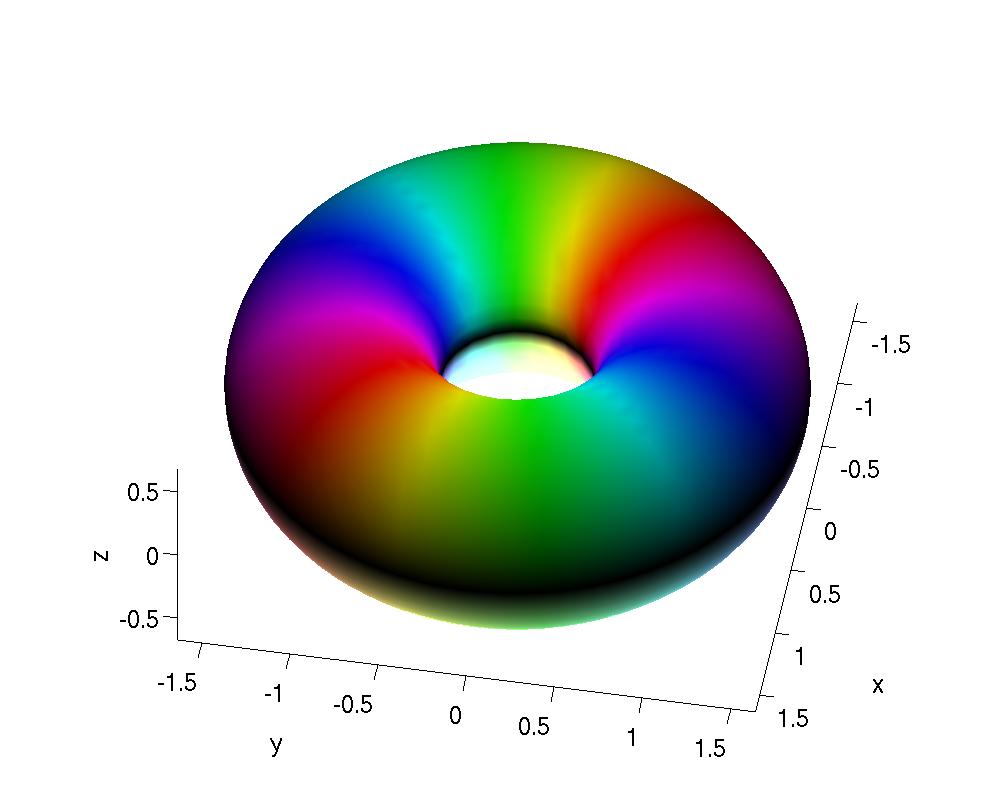}}
\mbox{
\includegraphics[width=0.49\linewidth]{T4B21_baryonslice}
\includegraphics[width=0.49\linewidth]{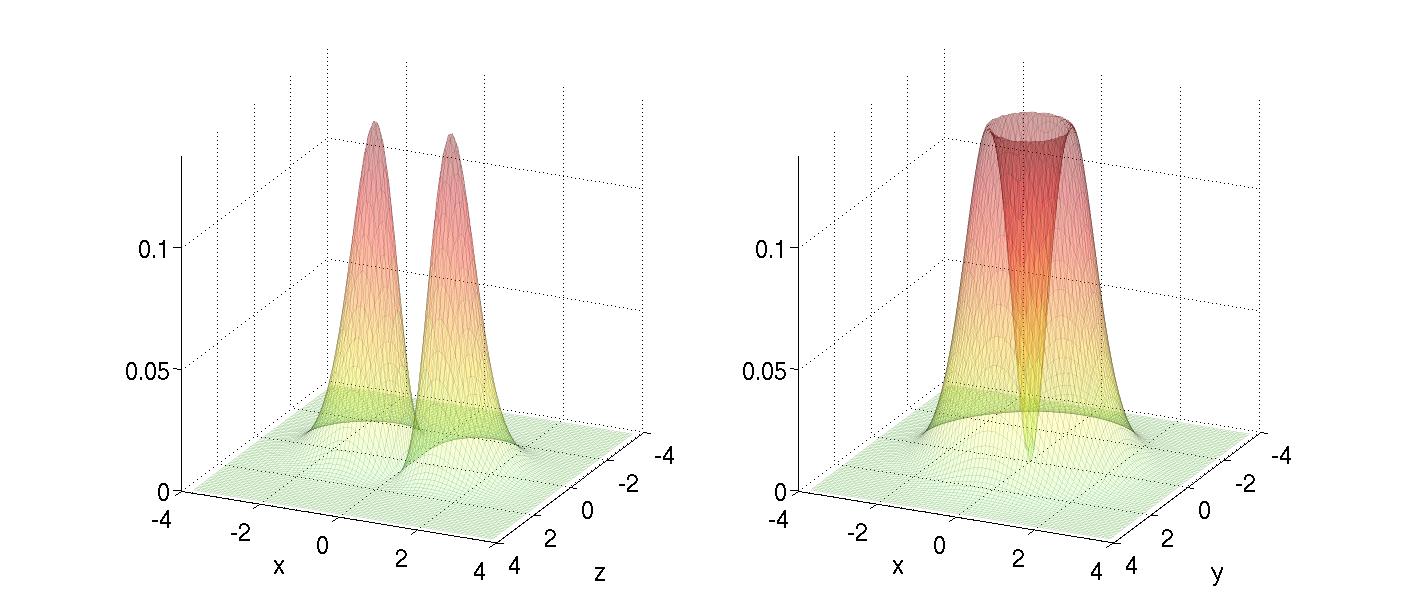}}
\mbox{
\subfloat[$m=4$, $B^{\rm numerical}=1.9994$]{\includegraphics[width=0.49\linewidth]{T4B21_energyslice}}
\subfloat[$m=0$, $B^{\rm numerical}=1.9715$]{\includegraphics[width=0.49\linewidth]{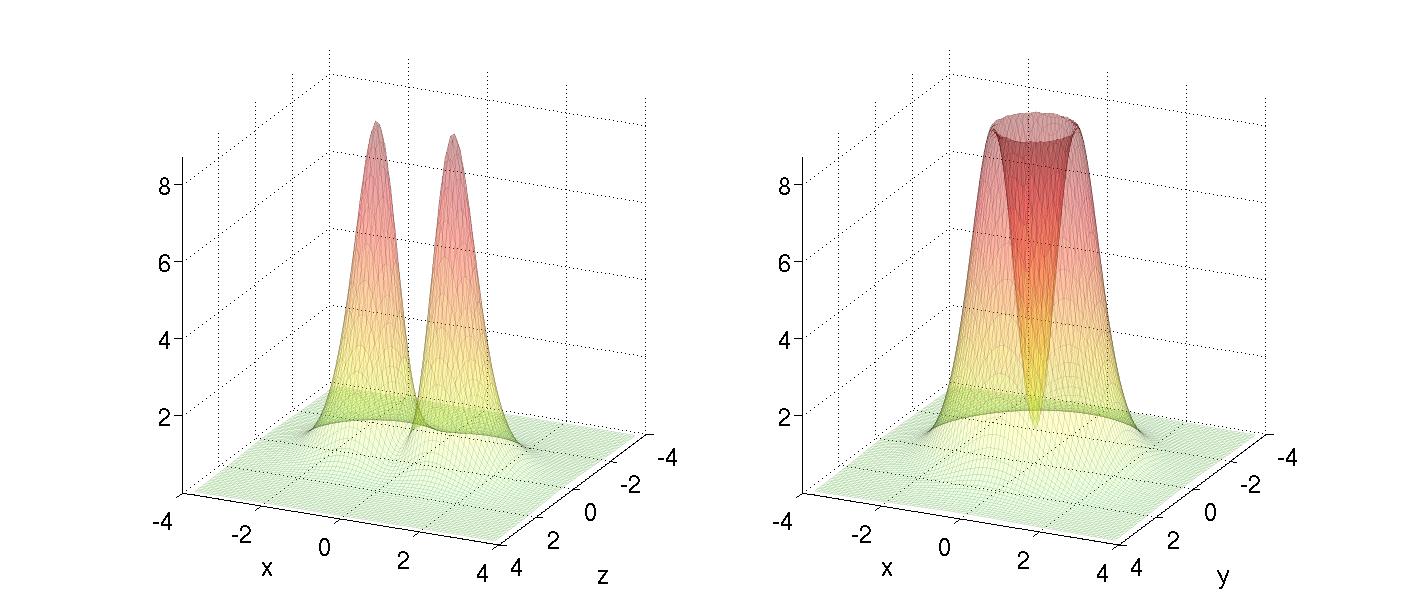}}}
\caption{Comparison between the BEC Skyrmion in the 2+4 model ($m=4$)
  on the left and the normal Skyrmion ($m=0$) on the right. From top
  to bottom is shown the isosurface of the baryon density at half
  maximum, the baryon density at $xz$ and $xy$ slices through the
  origin of the torus, and finally similar energy density slices. }
\label{fig:24comparison}
\end{center}
\end{figure}

\begin{figure}[!thp]
\begin{center}
\captionsetup[subfloat]{labelformat=empty}
\mbox{
\includegraphics[width=0.4\linewidth]{T6B21n}
\includegraphics[width=0.4\linewidth]{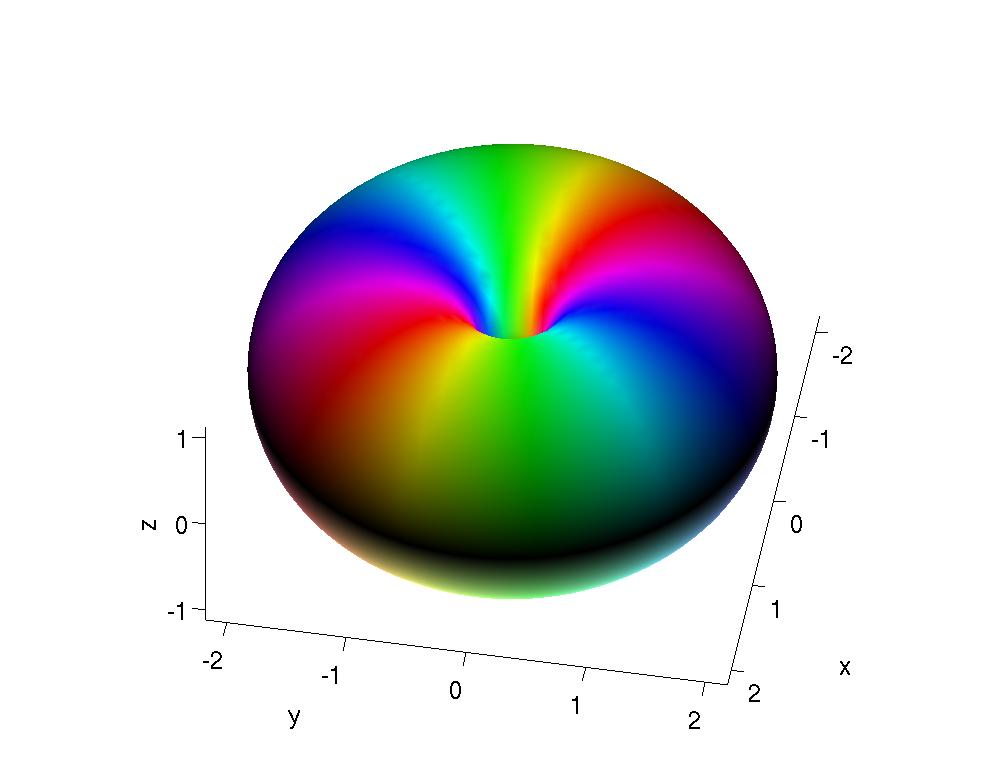}}
\mbox{
\includegraphics[width=0.49\linewidth]{T6B21_baryonslice}
\includegraphics[width=0.49\linewidth]{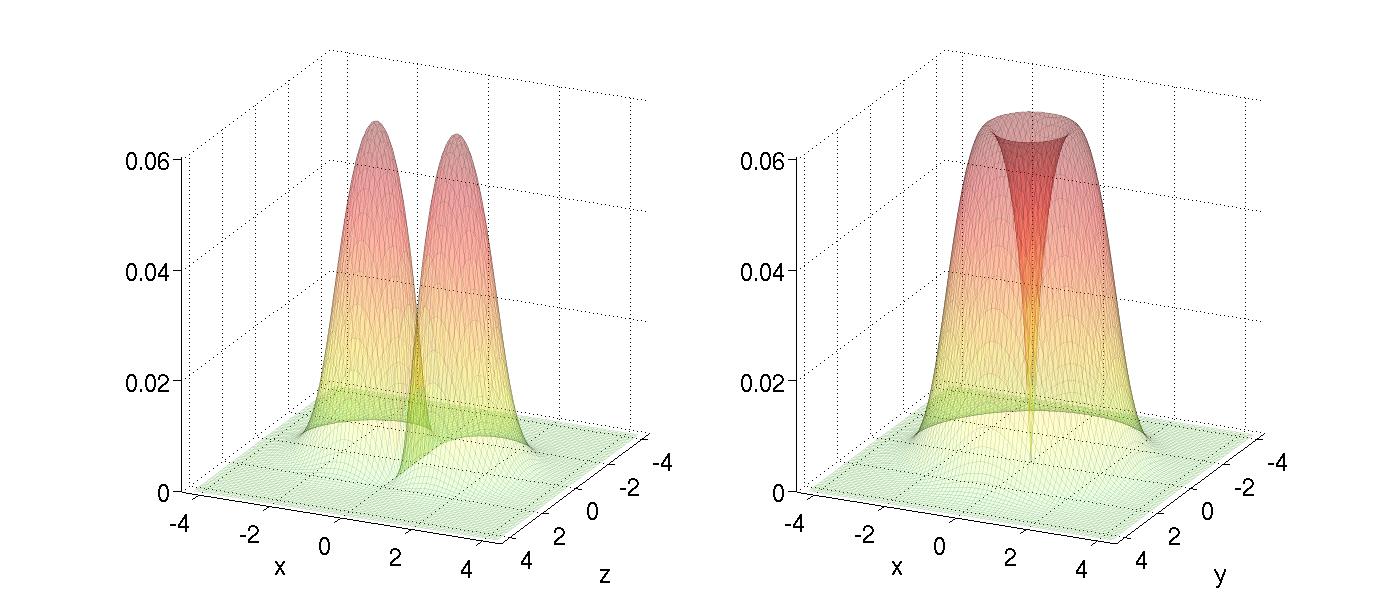}}
\mbox{
\subfloat[$m=4$, $B^{\rm numerical}=1.9998$]{\includegraphics[width=0.49\linewidth]{T6B21_energyslice}}
\subfloat[$m=0$, $B^{\rm numerical}=1.9834$]{\includegraphics[width=0.49\linewidth]{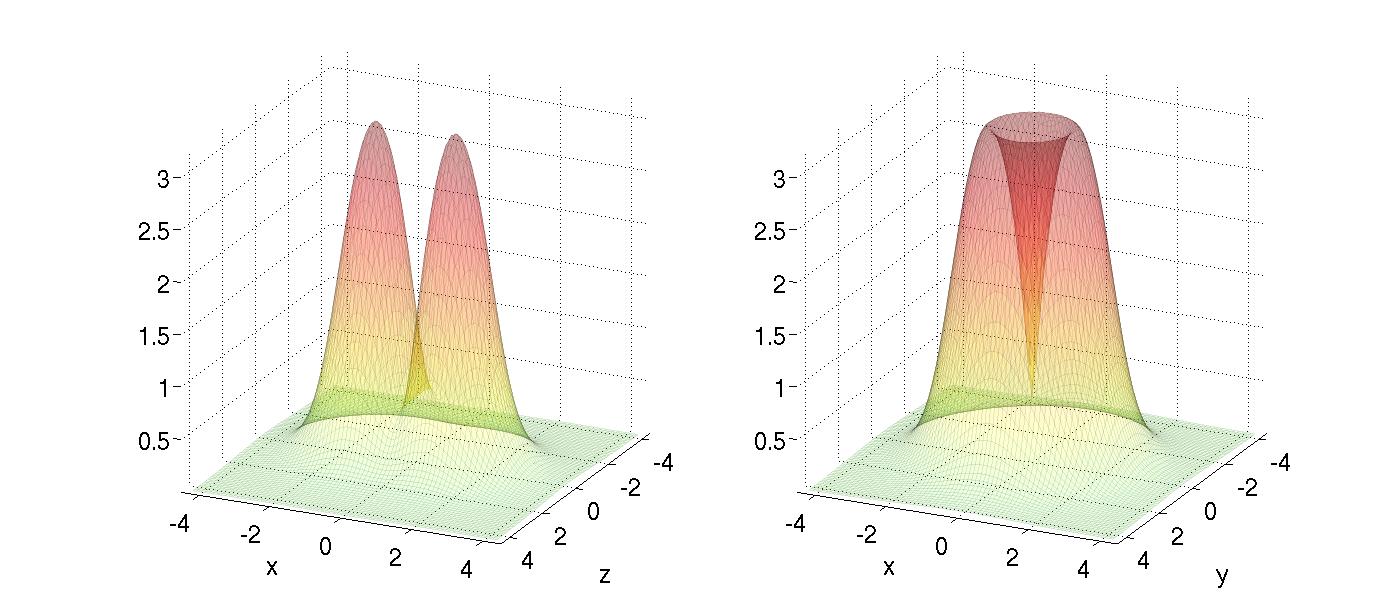}}}
\caption{Comparison between the BEC Skyrmion in the 2+6 model ($m=4$)
  on the left and the ``normal'' Skyrmion in the 2+6 model ($m=0$) on
  the right. From top to bottom is shown the isosurface of the baryon
  density at half maximum, the baryon density at $xz$ and $xy$ slices
  through the origin of the torus, and finally similar energy density
  slices. } 
\label{fig:26comparison}
\end{center}
\end{figure}

In this section we will compare the case of $(P,Q)=(2,1)$ and thus
baryon number $2$ and $m=4$, where the Skyrmion is a torus, with the
case of $m=0$, which is just the normal $B=2$ Skyrmion and also in the
form of a torus. We will make the comparison for both the 2+4 model
and the 2+6 model.
In Figs.~\ref{fig:24comparison} and \ref{fig:26comparison} are shown
the comparison for the 2+4 and 2+6 models, respectively. For the 2+4
model, the main difference is the size (and in turn the total mass) of
the two solutions. For the 2+6 model, differences are evident both in
the baryon charge density slices (middle row) and the energy density
slices (bottom row). For the BEC Skyrmion in the 2+6 model, the torus
is more hollow with respect to its potential-less counterpart. 

\end{appendix}

\newcommand{\J}[4]{{\sl #1} {\bf #2} (#3) #4}
\newcommand{\andJ}[3]{{\bf #1} (#2) #3}
\newcommand{\AP}{Ann.\ Phys.\ (N.Y.)}
\newcommand{\MPL}{Mod.\ Phys.\ Lett.}
\newcommand{\NP}{Nucl.\ Phys.}
\newcommand{\PL}{Phys.\ Lett.}
\newcommand{\PR}{ Phys.\ Rev.}
\newcommand{\PRL}{Phys.\ Rev.\ Lett.}
\newcommand{\PTP}{Prog.\ Theor.\ Phys.}
\newcommand{\hep}[1]{{\tt hep-th/{#1}}}


\begin{thebibliography}{100}


\bibitem{Skyrme:1962vh} 
  T.~H.~R.~Skyrme,
  ``A Unified Field Theory of Mesons and Baryons,''
  Nucl.\ Phys.\  {\bf 31}, 556 (1962);  
  ``A Nonlinear field theory,''  
  Proc.\ Roy.\ Soc.\ Lond.\ A {\bf 260}, 127 (1961).  

\bibitem{Adkins:1983ya} 
  G.~S.~Adkins, C.~R.~Nappi and E.~Witten,
  ``Static Properties of Nucleons in the Skyrme Model,''  
  Nucl.\ Phys.\ B {\bf 228}, 552 (1983).  

\bibitem{Sakai:2004cn} 
  T.~Sakai and S.~Sugimoto,
  ``Low energy hadron physics in holographic QCD,''  
  Prog.\ Theor.\ Phys.\  {\bf 113}, 843 (2005)  [hep-th/0412141]; 
  ``More on a holographic dual of QCD,''  
  Prog.\ Theor.\ Phys.\  {\bf 114}, 1083 (2005)  [hep-th/0507073].  

\bibitem{Hata:2007mb} 
  H.~Hata, T.~Sakai, S.~Sugimoto and S.~Yamato,
  ``Baryons from instantons in holographic QCD,''  
  Prog.\ Theor.\ Phys.\  {\bf 117}, 1157 (2007)  [hep-th/0701280
    [HEP-TH]].  

\bibitem{Ruostekoski:2001fc} 
  J.~Ruostekoski and J.~R.~Anglin,
  ``Creating vortex rings and three-dimensional skyrmions in
  Bose-Einstein condensates,'' 
  Phys.\ Rev.\ Lett.\  {\bf 86}, 3934 (2001)
  [cond-mat/0103310];  
  R.~A.~Battye, N.~R.~Cooper and P.~M.~Sutcliffe,
  ``Stable skyrmions in two-component Bose-Einstein condensates,''  
  Phys.\ Rev.\ Lett.\  {\bf 88}, 080401 (2002)
  [cond-mat/0109448].  

\bibitem{3D-skyrmions}
  U.~A.~Khawaja and H.~T.~C.~Stoof,
  ``Skyrmions in a ferromagnetic Bose-Einstein condensate,''
  Nature (London) {\bf 411}, 918 (2001);
  ``Skyrmion physics in Bose-Einstein ferromagnets,''
  Phys.\ Rev.\  A {\bf 64}, 043612 (2001);
  C.~M.~Savage and J.~Ruostekoski,
  ``Energetically stable particle-like skyrmions in a trapped
  Bose-Einstein condensate,''
  Phys.\ Rev.\ Lett.\  {\bf 91}, 010403 (2003);
  [cond-mat/0306112].  
  J.~Ruostekoski,
  ``Stable particlelike solitons with multiply-quantized vortex lines
  in Bose-Einstein condensates,'' 
  Phys.\ Rev.\ A {\bf 70}, 041601 (2004);
  [cond-mat/0408376].  
  S.~Wuster, T.~E.~Argue, and C.~M.~Savage,
  ``Numerical study of the stability of skyrmions in Bose-Einstein
  condensates,'' 
  Phys.\ Rev.\ A {\bf 72}, 043616 (2005);
  I.~F.~Herbut and M.~Oshikawa, 
  ``Stable Skyrmions in spinor condensates,''
  Phys.\ Rev.\ Lett.\ {\bf 97}, 080403 (2006) 
  [arXiv:cond-mat/0604557]; 
  %
  A.~Tokuno, Y.~Mitamura, M.~Oshikawa, I.~F.~Herbut, 
  ``Skyrmion in spinor condensates and its stability in trap potentials,''
  Phys.\ Rev.\ A {\bf 79}, 053626 (2009) 
  [arXiv:0812.2736]. 

\bibitem{Kawakami:2012zw} 
  T.~Kawakami, T.~Mizushima, M.~Nitta and K.~Machida,
  ``Stable Skyrmions in SU(2) Gauged Bose-Einstein Condensates,''  
  Phys.\ Rev.\ Lett.\  {\bf 109}, 015301 (2012)  
  [arXiv:1204.3177 [cond-mat.quant-gas]].  

\bibitem{Nitta:2012hy} 
  M.~Nitta, K.~Kasamatsu, M.~Tsubota and H.~Takeuchi,
  ``Creating vortons and three-dimensional skyrmions from domain wall
  annihilation with stretched vortices in Bose-Einstein condensates,''
  Phys.\ Rev.\ A {\bf 85}, 053639 (2012)  
  [arXiv:1203.4896 [cond-mat.quant-gas]].  

\bibitem{Kasamatsu:2005}
  K.~Kasamatsu, M.~Tsubota and M.~Ueda, 
  ``Vortices in multicomponent Bose-Einstein condensates,''
  Int.\ J.\ Mod.\ Phys.\ {\bf B} 19, 1835 (2005).

\bibitem{Takeuchi:2012ee} 
  H.~Takeuchi, K.~Kasamatsu, M.~Tsubota and M.~Nitta,
  ``Tachyon Condensation Due to Domain-Wall Annihilation in
  Bose-Einstein Condensates,''  
  Phys.\ Rev.\ Lett.\  {\bf 109}, 245301 (2012)  
  [arXiv:1205.2330 [cond-mat.quant-gas]];  
  H.~Takeuchi, K.~Kasamatsu, M.~Nitta and M.~Tsubota,
  ``Vortex Formations from Domain Wall Annihilations in Two-component
  Bose-Einstein Condensates,''  
  J.\ Low.\ Temp.\ Phys.\  {\bf 162}, 243 (2011)  
  [arXiv:1205.2328 [cond-mat.quant-gas]];  
  H.~Takeuchi, K.~Kasamatsu, M.~Tsubota and M.~Nitta,
  ``Tachyon Condensation and Brane Annihilation in Bose-Einstein
  Condensates: Spontaneous Symmetry Breaking in Restricted
  Lower-dimensional Subspace,''   
  J.\ Low.\ Temp.\ Phys.\ {\bf 171}, 443-454 (2013) 
  [arXiv:1211.3952 [cond-mat.other]].  

\bibitem{Metlitski:2003gj} 
  M.~A.~Metlitski and A.~R.~Zhitnitsky,
  ``Vortex rings in two-component Bose-Einstein condensates,''  
  JHEP {\bf 0406}, 017 (2004)
  [arXiv:cond-mat/0307559].  

\bibitem{Davis:1988jq} 
  R.~L.~Davis and E.~P.~S.~Shellard,
  ``The Physics Of Vortex Superconductivity. 2,''  
  Phys.\ Lett.\ B {\bf 209}, 485 (1988);  
  ``Cosmic Vortons,''  
  Nucl.\ Phys.\ B {\bf 323}, 209 (1989).  

\bibitem{Vilenkin:2000}
  A.~Vilenkin and E.~P.~S.~Shellard, 
  {\it Cosmic Strings and Other Topological Defects}, 
  (Cambridge Monographs on Mathematical Physics), 
  Cambridge University Press (July 31, 2000).

\bibitem{Radu:2008pp} 
  E.~Radu and M.~S.~Volkov,
  ``Existence of stationary, non-radiating ring solitons in field
  theory: knots and vortons,''  
  Phys.\ Rept.\  {\bf 468}, 101 (2008)  [arXiv:0804.1357 [hep-th]].  

\bibitem{Garaud:2013iba} 
  J.~Garaud, E.~Radu and M.~S.~Volkov,
  ``Stable Cosmic Vortons,''
  Phys.\ Rev.\ Lett.\  {\bf 111}, 171602 (2013)
  [arXiv:1303.3044 [hep-th]].

\bibitem{Gudnason:2014hsa} 
  S.~B.~Gudnason and M.~Nitta,
  ``Incarnations of Skyrmions,''
  Phys.\ Rev.\ D {\bf 90}, 085007 (2014)
  [arXiv:1407.7210 [hep-th]].

\bibitem{Gudnason:2014gla} 
  S.~B.~Gudnason and M.~Nitta,
  ``Effective field theories on solitons of generic shapes,''
  arXiv:1407.2822 [hep-th].

\bibitem{Adam:2010fg} 
  C.~Adam, J.~Sanchez-Guillen and A.~Wereszczynski,
  ``A Skyrme-type proposal for baryonic matter,''
  Phys.\ Lett.\ B {\bf 691}, 105 (2010)
  [arXiv:1001.4544 [hep-th]];
  ``A BPS Skyrme model and baryons at large $N_c$,''
  Phys.\ Rev.\ D {\bf 82}, 085015 (2010)
  [arXiv:1007.1567 [hep-th]].

\bibitem{Gudnason:2013qba} 
  S.~B.~Gudnason and M.~Nitta,
  ``Baryonic sphere: a spherical domain wall carrying baryon number,''
  Phys.\ Rev.\ D {\bf 89}, 025012 (2014)
  [arXiv:1311.4454 [hep-th]].

\bibitem{Kobayashi:2014xua} 
  M.~Kobayashi and M.~Nitta,
  ``Non-relativistic Nambu-Goldstone modes associated with
  spontaneously broken space-time and internal symmetries,'' 
  Phys.\ Rev.\ Lett.\  {\bf 113}, 120403 (2014)
  [arXiv:1402.6826 [hep-th]].

\bibitem{Abraham:1992vb} 
  E.~R.~C.~Abraham and P.~K.~Townsend,
  ``Q kinks,''  
  Phys.\ Lett.\ B {\bf 291}, 85 (1992);  
  ``More on Q kinks: A (1+1)-dimensional analog of dyons,''  
  Phys.\ Lett.\ B {\bf 295}, 225 (1992).  

\bibitem{Arai:2002xa} 
  M.~Arai, M.~Naganuma, M.~Nitta and N.~Sakai,
  ``Manifest supersymmetry for BPS walls in N=2 nonlinear sigma
  models,''  
  Nucl.\ Phys.\ B {\bf 652}, 35 (2003)  [hep-th/0211103];  
  ``BPS wall in N=2 SUSY nonlinear sigma model with Eguchi-Hanson
  manifold,''  
  In *Arai, A. (ed.) et al.: A garden of quanta* 299-325
  [hep-th/0302028].  

\bibitem{Nitta:2012xq} 
  M.~Nitta,
  ``Josephson vortices and the Atiyah-Manton construction,''  
  Phys.\ Rev.\ D {\bf 86}, 125004 (2012)  [arXiv:1207.6958 [hep-th]].  

\bibitem{Nitta:2012kj} 
  M.~Nitta,
  ``Defect formation from defect--anti-defect annihilations,''
  Phys.\ Rev.\ D {\bf 85}, 101702 (2012)
  [arXiv:1205.2442 [hep-th]].

\bibitem{Kudryavtsev:1997nw}
  A.~E.~Kudryavtsev, B.~M.~A.~Piette and W.~J.~Zakrzewski,
  ``Skyrmions and domain walls in (2+1) dimensions,''
  Nonlinearity {\bf 11}, 783 (1998)
  [arXiv:hep-th/9709187].
  D.~Harland and R.~S.~Ward,
  ``Walls and chains of planar skyrmions,''
  Phys.\ Rev.\  D {\bf 77}, 045009 (2008)
  [arXiv:0711.3166 [hep-th]].

\bibitem{Nitta:2012wi} 
  M.~Nitta,
  ``Correspondence between Skyrmions in 2+1 and 3+1 Dimensions,''  
  Phys.\ Rev.\ D {\bf 87}, 025013 (2013)  [arXiv:1210.2233 [hep-th]];  
  ``Matryoshka Skyrmions,''  
  Nucl.\  Phys.\ B {\bf 872}, 62 (2013)  [arXiv:1211.4916 [hep-th]].  

\bibitem{Witten:1984eb} 
  E.~Witten,
  ``Superconducting Strings,''  
  Nucl.\ Phys.\ B {\bf 249}, 557 (1985).  

\bibitem{Battye:1997qq} 
  R.~A.~Battye and P.~M.~Sutcliffe,
  ``Symmetric skyrmions,''
  Phys.\ Rev.\ Lett.\  {\bf 79}, 363 (1997)
  [hep-th/9702089].

\bibitem{Houghton:1997kg} 
  C.~J.~Houghton, N.~S.~Manton and P.~M.~Sutcliffe,
  ``Rational maps, monopoles and Skyrmions,''
  Nucl.\ Phys.\ B {\bf 510}, 507 (1998)
  [hep-th/9705151].

\bibitem{Gudnason:2014nba} 
  S.~B.~Gudnason and M.~Nitta,
  ``Domain wall Skyrmions,''
  Phys.\ Rev.\ D {\bf 89}, 085022 (2014)
  [arXiv:1403.1245 [hep-th]].

\bibitem{Garaud:2012}
  J.~Garaud and E.~Babaev, 
  ``Skyrmionic state and stable half-quantum vortices in chiral p-wave
  superconductors,'' 
  Phys.\ Rev.\ B {\bf 86}, 060514 (2012) [arXiv:1201.2946 [cond-mat.supr-con]]. 

\bibitem{Kobayashi:2013ju} 
  M.~Kobayashi and M.~Nitta,
  ``Jewels on a wall ring,''  
  Phys.\  Rev.\ D {\bf 87}, 085003 (2013)  [arXiv:1302.0989 [hep-th]].  

\bibitem{Kasamatsu:2010aq} 
  K.~Kasamatsu, H.~Takeuchi, M.~Nitta and M.~Tsubota,
  ``Analogues of D-branes in Bose-Einstein condensates,''  
  JHEP {\bf 1011}, 068 (2010)  [arXiv:1002.4265 [cond-mat.quant-gas]]; 
  K.~Kasamatsu, H.~Takeuchi and M.~Nitta,
  ``D-brane solitons and boojums in field theory and Bose-Einstein condensates,''
  J.\ Phys.\ Condens.\ Matter {\bf 25}, 404213 (2013)
  [arXiv:1303.4469 [cond-mat.quant-gas]].
  K.~Kasamatsu, H.~Takeuchi, M.~Tsubota and M.~Nitta,
  ``Wall-vortex composite solitons in two-component Bose-Einstein condensates,''
  Phys.\ Rev.\ A {\bf 88}, no. 1, 013620 (2013)
  [arXiv:1303.7052 [cond-mat.quant-gas]].

\bibitem{Gauntlett:2000de} 
  J.~P.~Gauntlett, R.~Portugues, D.~Tong and P.~K.~Townsend,
  ``D-brane solitons in supersymmetric sigma models,''  
  Phys.\ Rev.\ D {\bf 63}, 085002 (2001) 
  [hep-th/0008221];  
  M.~Shifman and A.~Yung,
  ``Domain walls and flux tubes in N=2 SQCD: D-brane prototypes,''  
  Phys.\ Rev.\ D {\bf 67}, 125007 (2003)  [hep-th/0212293].  

\bibitem{Isozumi:2004vg} 
  Y.~Isozumi, M.~Nitta, K.~Ohashi and N.~Sakai,
  ``All exact solutions of a 1/4 Bogomol'nyi-Prasad-Sommerfield equation,''
  Phys.\ Rev.\ D {\bf 71}, 065018 (2005)
  [hep-th/0405129];
  M.~Eto, Y.~Isozumi, M.~Nitta, K.~Ohashi and N.~Sakai,
  ``Solitons in the Higgs phase: The Moduli matrix approach,''  
  J.\ Phys.\ A {\bf 39}, R315 (2006)  [hep-th/0602170];  
  M.~Eto, Y.~Isozumi, M.~Nitta and K.~Ohashi,
  ``1/2, 1/4 and 1/8 BPS equations in SUSY Yang-Mills-Higgs systems:
  Field theoretical brane configurations,''   
  Nucl.\ Phys.\ B {\bf 752}, 140 (2006)  [hep-th/0506257].  

\bibitem{Nitta:2012kk} 
  M.~Nitta,
  ``Knots from wall--anti-wall annihilations with stretched strings,''  
  Phys.\ Rev.\ D {\bf 85}, 121701 (2012)  [arXiv:1205.2443 [hep-th]].  

\bibitem{Kobayashi:2013bqa} 
  M.~Kobayashi and M.~Nitta,
  ``Toroidal domain walls as Hopfions,''  
  arXiv:1304.4737 [hep-th].  

\bibitem{Kobayashi:2013xoa} 
  M.~Kobayashi and M.~Nitta,
  ``Torus knots as Hopfions,''
  Phys.\ Lett.\ B {\bf 728}, 314 (2014)
  [arXiv:1304.6021 [hep-th]].

\bibitem{Faddeev:1975}
  L.~D.~Faddeev, Princeton preprint IAS-75-QS70.

\bibitem{Faddeev:1996zj} 
  L.~D.~Faddeev and A.~J.~Niemi,
  ``Knots and particles,''  
  Nature {\bf 387}, 58 (1997)  [hep-th/9610193].  

\bibitem{Piette:1994ug}
  B.~M.~A.~Piette, B.~J.~Schroers and W.~J.~Zakrzewski,
  ``Multi - Solitons In A Two-Dimensional Skyrme Model,''
  Z.\ Phys.\  C {\bf 65}, 165 (1995);
  [arXiv:hep-th/9406160];
  ``Dynamics of baby skyrmions,''
  Nucl.\ Phys.\  B {\bf 439}, 205 (1995).
  [arXiv:hep-ph/9410256].

\bibitem{Weidig:1998ii}
  T.~Weidig,
  ``The baby Skyrme models and their multi-skyrmions,''
  Nonlinearity {\bf 12}, 1489-1503 (1999). 
  [arXiv:hep-th/9811238].

\bibitem{deVega:1977rk} 
  H.~J.~de Vega,
  ``Closed Vortices and the HOPF Index in Classical Field Theory,''  
  Phys.\ Rev.\ D {\bf 18}, 2945 (1978);  
  A.~Kundu and Y.~P.~Rybakov,
  ``Closed Vortex Type Solitons With Hopf Index,''  
  J.\ Phys.\ A {\bf 15}, 269 (1982).  

\bibitem{Gladikowski:1996mb} 
  J.~Gladikowski and M.~Hellmund,
  ``Static solitons with nonzero Hopf number,''  
  Phys.\ Rev.\ D {\bf 56}, 5194 (1997)  [hep-th/9609035].  

\bibitem{Battye:1998pe} 
  R.~A.~Battye and P.~M.~Sutcliffe,
  ``Knots as stable soliton solutions in a three-dimensional classical
  field theory,''   
  Phys.\ Rev.\ Lett.\  {\bf 81}, 4798 (1998)  [hep-th/9808129];  
  ``Solitons, links and knots,''  
  Proc.\ Roy.\ Soc.\ Lond.\ A {\bf 455}, 4305 (1999)  [hep-th/9811077].  

\bibitem{Hietarinta:2000ci} 
  J.~Hietarinta and P.~Salo,
  ``Ground state in the Faddeev-Skyrme model,'' 
  Phys.\ Rev.\ D {\bf 62}, 081701 (2000).  

\bibitem{Sutcliffe:2007ui} 
  P.~Sutcliffe,
  ``Knots in the Skyrme-Faddeev model,''  
  Proc.\ Roy.\ Soc.\ Lond.\ A {\bf 463}, 3001 (2007)  [arXiv:0705.1468 [hep-th]].  

\bibitem{Kobayashi:2013aza} 
  M.~Kobayashi and M.~Nitta,
  ``Winding Hopfions on R$^{2} \times S^{1}$,''
  Nucl.\ Phys.\ B {\bf 876}, 605 (2013)
  [arXiv:1305.7417 [hep-th]].

\bibitem{Canfora:2014aia} 
  F.~Canfora, F.~Correa and J.~Zanelli,
  ``Exact multi-soliton solutions in the four dimensional Skyrme model,''
  arXiv:1406.4136 [hep-th].

\end{thebibliography}
\end{document}